\documentclass[twocolumn,times,astrosym,iop,useAMS,usenatbib,savetrees]{aastex63} 

\hypersetup{linkcolor=red,citecolor=blue,filecolor=cyan,urlcolor=magenta}

\usepackage{txfonts} 
\usepackage{textgreek}
\newcommand{\mic}{\hbox{\textmu m}}
\newcommand{\lsun}{\hbox{$L_\odot$}}
\newcommand{\msun}{\hbox{$M_\odot$}}

\newcommand{\tdust}{\hbox{$T_\mathrm{dust}$}}
\newcommand{\ldust}{\hbox{$L_\mathrm{dust}$}}
\newcommand{\mdust}{\hbox{$M_\mathrm{dust}$}}
\newcommand{\lthick}{\hbox{$\lambda_\mathrm{thick}$}}

\newcommand{\sn}{\hbox{S/N}}
 
 \DeclareMathSymbol{\la}{3}{AMSa}{46}
 \DeclareMathSymbol{\ga}{3}{AMSa}{38}

\submitjournal{The Astrophysical Journal}

\shorttitle{Measurements of the dust properties in $z\simeq1-3$ SMGs with ALMA}
\shortauthors{E. da Cunha et al.}

\begin{document}



\title{\bf Measurements of the dust properties in $z\simeq1-3$ sub-millimeter galaxies with ALMA}


\author{E. da Cunha}
\affiliation{International Centre for Radio Astronomy Research, University of Western Australia, 35 Stirling Hwy, Crawley, WA 6009, Australia}
\affiliation{Research School of Astronomy and Astrophysics, Australian National University, Canberra, ACT 2611, Australia}
\affiliation{ARC Centre of Excellence for All Sky Astrophysics in 3 Dimensions (ASTRO 3D)}
\email{E-mail: Elisabete.daCunha@uwa.edu.au}

\author{J. A. Hodge}
\affiliation{Leiden Observatory, Leiden University, P.O. Box 9513, 2300 RA Leiden, The Netherlands}
\author{C. M. Casey}
\affiliation{The University of Texas at Austin, 2515 Speedway Boulevard Stop C1400, Austin, TX 78712, USA}
\author{H. S. B. Algera}
\affiliation{Leiden Observatory, Leiden University, P.O. Box 9513, 2300 RA Leiden, The Netherlands}
\author{M. Kaasinen}
\affiliation{Max-Planck Institut f{\"u}r Astronomie, Konigstuhl 17, D-69117 Heidelberg, Germany}
\affiliation{Universit{\"a}t Heidelberg, Zentrum f{\"u}r Astronomie, Institut f{\"u}r Theoretische Astrophysik, Albert-Ueberle-Strasse 2, D-69120 Heidelberg, Germany}
\author{I. Smail}
\affiliation{Centre for Extragalactic Astronomy, Department of Physics, Durham University, South Road, Durham, DH1 3LE, UK}
\author{F. Walter}
\affiliation{Max-Planck Institut f{\"u}r Astronomie, Konigstuhl 17, D-69117 Heidelberg, Germany}
%
\author{W. N. Brandt}
\affiliation{Department of Astronomy \& Astrophysics, The Pennsylvania State University, 525 Davey Lab, University Park, PA 16802, USA}
\affiliation{Institute for Gravitation and the Cosmos, The Pennsylvania State University, University Park, PA 16802, USA}
\affiliation{Department of Physics, The Pennsylvania State University, University Park, PA 16802, USA}
\author{H. Dannerbauer}
\affiliation{Instituto de Astrof\'isica de Canarias (IAC), E-38205 La Laguna, Tenerife, Spain}
\affiliation{Universidad de La Laguna, Dpto. Astrof\'isica, E-38206 La Laguna, Tenerife, Spain}
\author{R. Decarli}
\affiliation{INAF -- Osservatorio di Astrofisica e Scienza dello Spazio, via Gobetti 93/3, I-40129, Bologna, Italy}
\author{B. A. Groves}
\affiliation{International Centre for Radio Astronomy Research, University of Western Australia, 35 Stirling Hwy, Crawley, WA 6009, Australia}
\affiliation{Research School of Astronomy and Astrophysics, Australian National University, Canberra, ACT 2611, Australia}
\author{K. K. Knudsen}
\affiliation{Department of Space, Earth and Environment, Chalmers University of Technology, Onsala Space Observatory, SE-43992 Onsala, Sweden}
\author{A. M. Swinbank}
\affiliation{Centre for Extragalactic Astronomy, Department of Physics, Durham University, South Road, Durham, DH1 3LE, UK}
\author{A. Weiss}
\affiliation{Max-Planck-Institut f{\"u}r Radioastronomie, Auf dem H{\"u}gel 69, D-53121 Bonn, Germany}
\author{P. van der Werf}
\affiliation{Leiden Observatory, Leiden University, P.O. Box 9513, 2300 RA Leiden, The Netherlands}
\author{J. A. Zavala}
\affiliation{The University of Texas at Austin, 2515 Speedway Boulevard Stop C1400, Austin, TX 78712, USA}

\date{\today}

\begin{abstract}
We present Atacama Large Millimetre Array (ALMA) 2mm continuum observations of a complete and unbiased sample of 99 870\mic-selected sub-millimeter galaxies (SMGs) in the Extended {\it Chandra} Deep Field South (ALESS). Our observations of each SMG reach average sensitivities of 53\,\textmu Jy~beam$^{-1}$. We measure the flux densities for 70 sources, for which we obtain a typical 870\mic-to-2mm flux ratio of $14\pm5$. We do not find a redshift dependence of this flux ratio, which would be expected if the dust emission properties of our SMGs were the same at all redshifts. By combining our ALMA measurements with existing {\it Herschel}/SPIRE observations, we construct a (biased) subset of 27 galaxies for which the cool dust emission is sufficiently well sampled to obtain precise constraints on their dust properties using simple isothermal models. Thanks to our new 2mm observations, the dust emissivity index is well-constrained and robust against different dust opacity assumptions. The median dust emissivity index of our SMGs is $\beta\simeq1.9\pm0.4$, consistent with the emissivity index of dust in the Milky Way and other local and high-redshift galaxies, as well as classical dust grain model predictions. We also find a negative correlation between the dust temperature and $\beta$, similar to low-redshift observational and theoretical studies. Our results indicate that $\beta\simeq2$ in high-redshift dusty star-forming galaxies, implying little evolution in dust grain properties between our SMGs and local dusty galaxy samples, and suggesting these high-mass and high-metallicity galaxies have dust reservoirs driven by grain growth in their ISM.
\end{abstract}

\section{Introduction}
Sub-millimeter galaxies (SMGs) are the most dust-rich galaxies in the Universe, with large ($\gtrsim10^{12} L_\sun$) infrared (IR) luminosities that are powered by high star formation rates (SFR $\gtrsim100 M_\odot\,\mathrm{yr}^{-1}$; .e.g., \citealt{Blain2002,Barger2012,Casey2014,Swinbank2014, daCunha2015,Ugne2020}). This combination makes SMGs the ideal targets for studies of dust formation and the interplay between gas, dust, and stars (e.g., \citealt{Hodge2012,Hodge2015,Hodge2016,Hodge2019,Swinbank2015,Chen2017}; see \citealt{Hodge2020} for a recent review). Although SMGs are relatively rare (e.g., \citealt{Weiss2009}), they contribute significantly ($\gtrsim 20\%$) to the SFR density at $z>1$ (e.g., \citealt{Chapman2005,Sargent2012,Swinbank2014,Ugne2020}), and they are likely progenitors of the most massive galaxies in the local Universe (e.g., \citealt{Blain2002}, \citealt{Casey2014}, \citealt{Simpson2014})

The first SMGs were detected using SCUBA at 850\mic\ \citep{Smail1997, Hughes1998, Barger1998}, which remains one of the prime wavelengths to detect these galaxies (e.g., \citealt{Geach2017}), thanks to a combination of available instruments, spectral window, and the negative $k$-correction at that wavelength. Other single-dish samples of SMGs have also been obtained at 1.1-1.3mm using MAMBO (e.g., \citealt{Eales2003,Bertoldi2007,Greve2008}) and AzTEC (e.g., \citealt{Aretxaga2011,Yun2012}), at 1.4mm/2mm with the SPT \citep{Vieira2010} and at 2mm with GISMO \citep{Staguhn2014,Magnelli2019}.
Selecting SMGs from observations at longer wavelengths is thought to favour galaxies at higher redshifts (e.g., \citealt{Smolcic2012,Vieira2013,Staguhn2014,Magnelli2019,Hodge2020}), although it is difficult to compare the redshift distributions in an unbiased way (see e.g., \citealt{Zavala2014} for a discussion), and account for intrinsic variations of galaxy far-IR spectral energy distributions (SEDs). Nevertheless, the 2mm band has been put forth as a potential candidate to detect high-redshift ($z>3$) galaxies (e.g., \citealt{Casey2018a,Casey2018b,Casey2019,Zavala2021}). The negative $k$-correction is stronger at 2mm than at 850\mic, thus, for a fixed SED, the 2-mm band should pick up more high-redshift galaxies than at 870\mic. In addition, better atmospheric transmission and larger fields of view can be achieved at 2mm (but corresponding poorer resolution).
Such an effort is currently ongoing (see \citealt{Zavala2021} for first results). To understand the relationship between the populations detected at 850\mic\ and at 2mm we require a detailed characterization of the (sub-)millimeter SEDs of these sources. Multi-wavelength sub-millimeter observations are still rare, with most observations focusing on a single wavelength. Only a handful of sources observed at 2mm have complementary shorter wavelength detections \citep{Staguhn2014,Magnelli2019}. Thus, a more systematic multi-wavelength dust-continuum investigation is warranted in order to reveal the dust properties of (sub-)millimeter-detected sources.

In recent years, there has been much debate over the origin of the large dust masses of SMGs (typically $\gtrsim10^8~M_\odot$, i.e., a few percent of their stellar mass; e.g., \citealt{Ugne2020}). However, such measurements have mostly been performed for biased samples (that favour the dustiest sources) and until recently relied solely on single-dish observations (e.g., \citealt{Rowlands2014,Ugne2021}). The build-up of such large dust masses over timescales $\sim 0.5 - 2$~Gyr is extremely difficult to explain with models of dust production and growth relying solely on stellar sources (e.g., \citealt{Morgan2003, Dwek2007}). Additional physical mechanisms such as ISM dust growth and/or non-standard initial mass functions (IMFs) may be required (the so-called `dust budget crisis'; e.g.~\citealt{Rowlands2014}). More precise dust mass constraints for unbiased samples of SMGs are needed to further investigate this issue.
Another matter of debate in the community is what drives the intense star-formation activity in SMGs: whether it is a mode of enhanced star-formation efficiency, driven by major mergers (e.g., \citealt{Hayward2011}), and/or, a more modest star-formation efficiency driven by secular evolution in large disks with high gas fractions (e.g., \citealt{Dave2011}). One way to disentangle these two evolutionary modes proposed for SMGs is to compare the observed SFRs with the mass of gas available to form stars, which, until we have CO observations for large samples, can be roughly inferred from the dust mass (assuming $M_\mathrm{gas}/M_\mathrm{dust}$; \citealt{Scoville2014,Scoville2016,Groves2015,Scoville2017,Kaasinen2019}).
The estimations of both dust and gas masses rely on the assumption that for SMGs the dust emissivity index ($\beta$, which describes the wavelength-dependence of the dust emissivity per unit mass) is similar to what is measured in the Milky Way and other local galaxies, i.e., $\beta\simeq1.5-2.0$ (e.g., \citealt{Galliano2018}). However, $\beta$ has only been directly measured in small samples of SMGs using observations at the long enough wavelengths needed to break the intrinsic degeneracy between $\beta$ and the cold dust temperature (e.g., \citealt{Birkin2021}; and see previous efforts with {\it Herschel} and AzTEC; \citealt{Chapin2009,Magnelli2012}).

To provide constraints on the dust emissivity index and temperature, we use ALMA to perform a systematic study of the 2mm emission of a complete sample of 870\mic-selected SMGs. First detected as part of the APEX LABOCA 870\,\mic\ survey of the Extended {\em Chandra} Deep Field South, LESS \citep{Weiss2009}, our sample is taken from the ALMA Cycle 0 follow-up program (ALESS), in which we observed 122 of the LESS sources at high sensitivity and spatial resolution through snapshot observations at 870\mic\ in Band 7 \citep{Karim2013,Hodge2013}. The high resolution of the ALMA observations de-blended multiple sources that were previously misidentified as single sources and located the SMGs to within 0.3~arcsec \citep{Hodge2013}. These ALESS observations yield a sample of 99 robustly identified SMGs, a sample large and reliable enough to enable a complete and unbiased multi-wavelength study of the properties of this galaxy population (e.g., \citealt{Simpson2014,daCunha2015}; see also recent similar efforts in the UDS and COSMOS fields, \citealt{Simpson2017,Stach2018,Ugne2020,Simpson2020}). In this paper, we present the first systematic (i.e., resolved) comparison of the 2mm emission of a 870\mic-selected sample, with which we characterize the long-wavelength SEDs and derive robust dust properties for individual SMGs.

This paper is organized as follows. In Section~\ref{sec:observations}, we describe our ALMA Band 4 imaging of the ALESS SMGs. In Section~\ref{sec:2mm_prop}, we obtain and analyze 2\,mm flux measurements and compare them with the 870\,\mic\ properties of our sources. In Section~\ref{sec:dustprop}, we derive the dust properties of our sources by fitting their observed SEDs using simple dust models. In Section~\ref{discussion} we discuss the robustness of our constraints, selection effects, and we compare the dust emissivities derived for our SMGs with other measurements and theoretical predictions. We provide a summary and conclusions in Section~\ref{conclusions}. Throughout the paper, we use a concordance $\Lambda$CDM cosmology with $H_0=70$~km s$^{-1}$ Mpc$^{-1}$, $\Lambda=0.7$, and $\Omega_m=0.3$ \citep[e.g.,][]{Planck2018}.

\section{ALMA Band 4 observations of SMGs}
\label{sec:observations}

\subsection{Observations}

Our ALMA Band 4 continuum observations were carried out between December 26th, 2015, and January 1st, 2016, as part of the Cycle 3 Project \#2015.1.00948.S (PI: E. da Cunha).

We targeted the 69 LESS fields that contain at least one {\sc main} catalog (i.e., most reliable) source from the Cycle 0 ALESS observations at 870~\mic\ \citep{Hodge2013}, and centered each pointing on the brightest source in each field. Thanks to the multiplicity of the single-dish detected LESS sources, 24 of our 69 LESS fields contain multiple SMGs identified in the Cycle 0 observations by \cite{Hodge2013}, resulting in a total of 99 {\sc main} ALESS sources in our target fields, as well as 32 additional {\sc supplementary} catalog sources. At the frequency of our observations, the primary beam of ALMA is 40.7~arcsec, ensuring that, when centering each field on the brightest ALESS source, the remaining ALESS sources within the $\simeq18$-arcsec 870\mic\ primary beam were covered by our pointings.

Our observations were taken in Band 4 (at a representative frequency of 152~GHz) using the total 7.5-GHz bandwidth available for continuum observations. Between 34 and 41 12-meter antennas were used in the most compact array configuration in Cycle 3 (C34-1), with baselines ranging from 15 to 310 meters. This antenna configuration was sufficient to achieve our desired resolution of 2.3\,arcsec, which allows us to separate the different sources in fields where there are multiple ALESS SMGs, while not resolving out each individual source (based on their typical sizes of $\lesssim0.5$\,arcsec; \citealt{Simpson2015,Hodge2016}). The weather conditions were adequate for Band 4 observations (precipitable water vapor between 1.35 and 3.82 mm). The quasar J0334-4008 was used for atmospheric, bandpass, flux, and pointing calibration, and J0348-2749 was used as a phase calibrator. ALESS045.1 was also used as an atmospheric calibrator.
Each of our 69 target fields was observed for 160 seconds.

\subsection{Data reduction and imaging}

The observations were processed using the ALMA automated data reduction pipeline in the Common Astronomy Software Application ({\sc casa}) version 4.5.1, and checked by the ALMA data quality assessment team. We verified that the pipeline produced high-quality data, and therefore use the data as delivered by ALMA.

\begin{figure*}
\centering
\begin{minipage}{\linewidth}
\includegraphics[trim={1cm 1.3cm 1cm 0.75cm},width=0.33\textwidth]{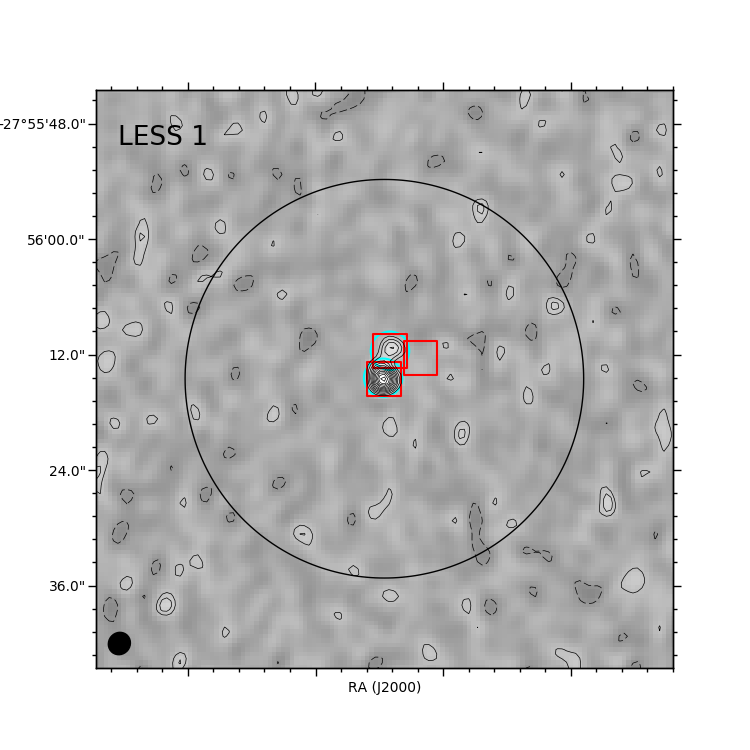}
\includegraphics[trim={1cm 1.3cm 1cm 0.75cm},width=0.33\textwidth]{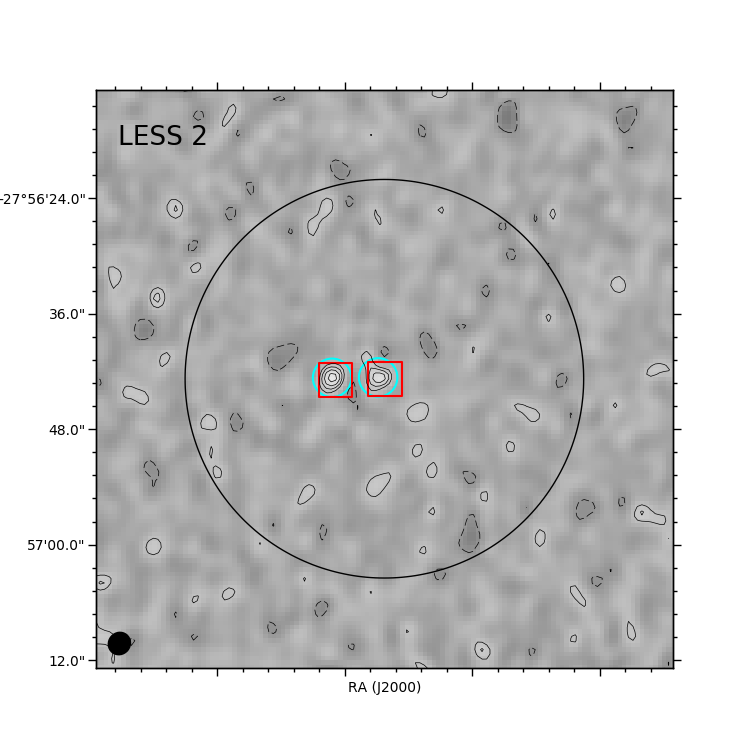}
\includegraphics[trim={1cm 1.3cm 1cm 0.75cm},width=0.33\textwidth]{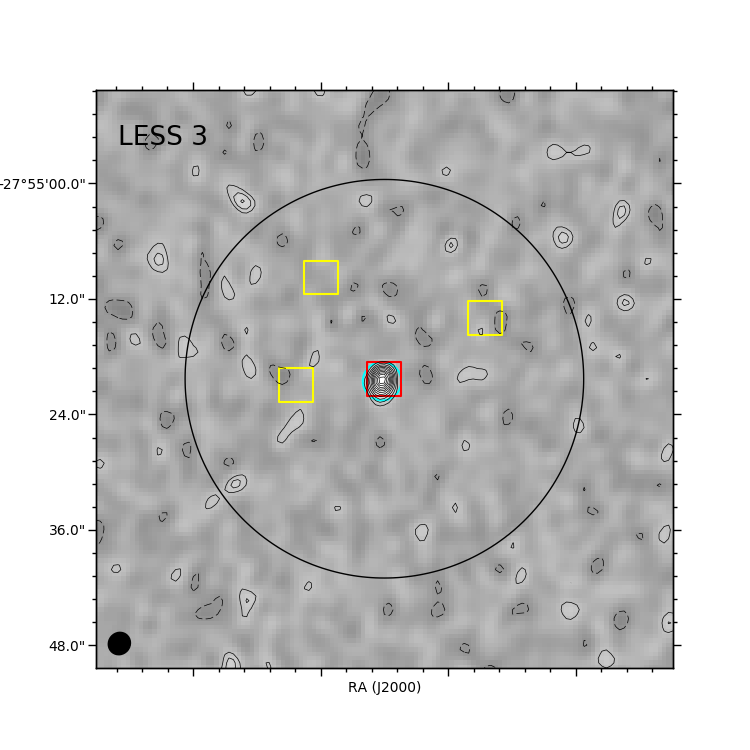}
\includegraphics[trim={1cm 1.3cm 1cm 0.75cm},width=0.33\textwidth]{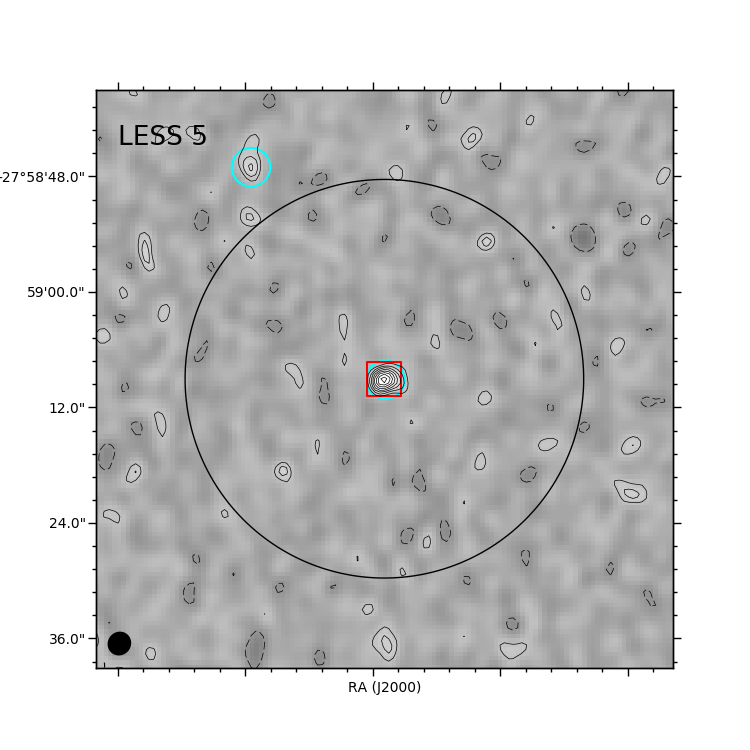}
\includegraphics[trim={1cm 1.3cm 1cm 0.75cm},width=0.33\textwidth]{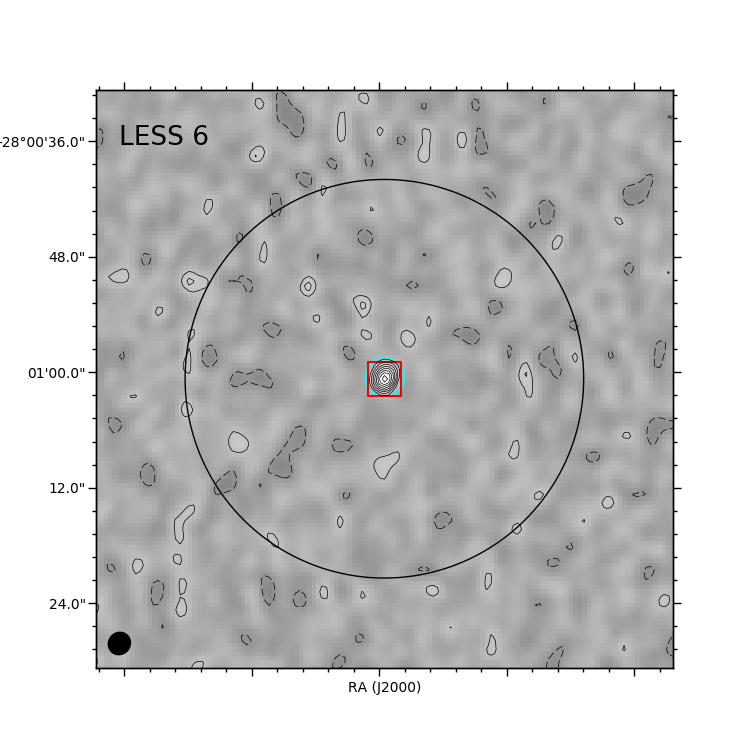}
\includegraphics[trim={1cm 1.3cm 1cm 0.75cm},width=0.33\textwidth]{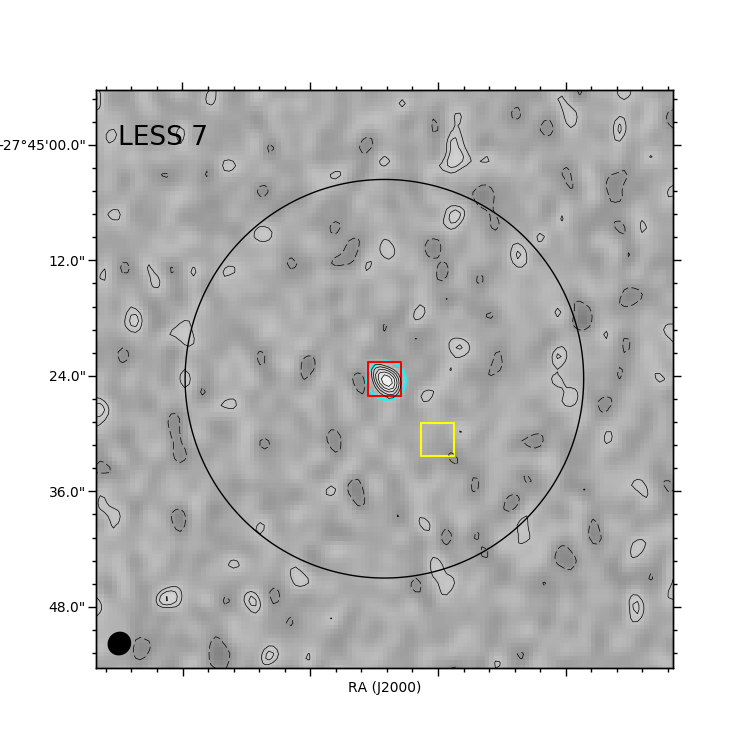}
\end{minipage}
 \caption{ALMA Band 4 (2\,mm) maps of our first six fields. The contours show flux density levels starting at $\pm2\sigma$ and increasing in steps
of $1\sigma$, where $\sigma$ is the rms noise measured in that map. The large black circle shows the primary beam FWHM ($\sim40$\,arcsec), and the small black filled circle in the bottom-left corner shows the synthesized beam ($\sim2.3$\,arcsec). Cyan circles indicate detections with S/N$\ge4$, and squares indicate the positions of known ALESS sources detected at 870\mic\ by \citealt{Hodge2013} (red for main catalog sources, yellow for supplementary sources).} 
 \label{band4_maps}
\end{figure*}

We generate images from the ALMA visibilities using the {\sc clean} task in {\sc casa}. {\sc clean} performs a Fourier transform to map the $uv$ visibilities onto the image plane on the sky, producing a `dirty image'. This image is then deconvolved from the point spread function (i.e. the synthesized `dirty' beam) using the {\sc clean} algorithm \citep{Hogbom1974} with robust (Briggs) weighting of the visibilities; we adopt {\tt robust}$=$0.5. The average rms obtained in our clean images is $\sigma=53\pm2$\,\textmu Jy~beam$^{-1}$ (with the error representing the standard deviation of the noise among all the maps), and the average beam is $2.4\times2.3$\,arcsec. This corresponds to a physical resolution of $\sim18$~kpc at $z\sim1-3$, the typical redshift range of our sample \citep{Danielson2017,daCunha2015}.
In Fig.~\ref{band4_maps}, we show the final cleaned ALESS Band 4 continuum images obtained using this procedure for the first six fields. Each image is $166\arcsec\times166\arcsec$, with a pixel scale of $0.46\arcsec$.
The noise and beam properties of all 69 maps are uniform. All the maps are good quality with rms below our 60\,\textmu Jy~beam$^{-1}$ request and fairly circular beam (the beam axis ratio varies between 1.05 and 1.11), therefore we use all the maps in a common source extraction step in the next section.

\section{2mm properties and comparison with 870\mic\ fluxes}
\label{sec:2mm_prop}

\subsection{Source extraction and catalog matching}
\label{extraction}

We detect sources in our cleaned continuum images using {\sc casa}. First, we identify sources with fluxes above $4\sigma$ in the maps using the {\sc boxit} task, which searches the images to find contiguous sets of pixels above the given threshold, and defines a rectangular box containing those pixels. Then, we use the {\sc imfit} task to fit one or more two-dimensional Gaussian functions to the sources detected in each region defined by {\sc boxit}. For each source, {\sc imfit} returns the peak intensity, the location of the peak pixel on the image, the major and minor axis and position angle, and the total (integrated) flux. For unresolved sources, the peak flux and the total flux are the same, and the size and shape of the source match the synthesized beam. 

Our source extraction procedure yields a total of 53 detections of ALESS main catalog sources with $\sn\ge4$ within the primary beam FWHM (i.e., sensitivity $>50\%$ of the maximum). For these bright, `blindly'-extracted sources, the positions in Band 4 agree very precisely with the previous Band 7 positions cataloged in \cite{Hodge2013}, with only minor offsets of $\delta\mathrm{RA}=0.07\pm0.05^{\prime\prime}$, and $\delta\mathrm{Dec}=0.07\pm0.04^{\prime\prime}$.
We define a `well-sampled subset' of 27 sources for which we detect the 2mm emission at $\sn\ge4$ which also have high-S/N detections at 870\mic\ and at 250\mic\ from {\it Herschel}/SPIRE \citep{Swinbank2014}, as well as spectroscopic redshifts from near-IR spectroscopy \citep{Danielson2017}, ALMA CO \citep{Birkin2021} and [CII] \citep{Swinbank2012}. As we will see in Section~\ref{results}, these are the sources for which we can constrain dust parameters robustly.

For sources below $4\sigma$, since we have prior information based on the Band 7 data, we extract fluxes on their 870\mic\ positions. 17 of those sources are detected with $1.5\le\sn<4$, allowing us to measure their fluxes.
In Table~\ref{table_sources} (Appendix~\ref{app_sources}), we list the prior positions and 870\,\mic\ fluxes of our ALESS sources obtained with ALMA Band 7 \citep{Hodge2013}, along with our measured 2\,mm Band 4 fluxes. All the fluxes are primary-beam corrected (though note that in most cases the correction is unity because we centered our maps on the ALESS sources). Our sources are unresolved at our $\sim2.3$-arcsec resolution. This is expected since most of the ALESS sources were unresolved in Band 7 with higher average spatial resolution \citep{Hodge2013}.

\subsubsection{Flux deboosting}
\label{boosting}

We correct both our measured 2mm flux densities as well as the \cite{Hodge2013} 870\mic\ flux densities for flux boosting due to Eddington bias and noise (note that we only correct the 2mm flux densities of our blindly detected $\ge4\sigma$ sources; sources for which we extracted fluxes at the prior 870\mic\ positions are not affected by flux boosting) . We run simulations of this effect by injecting random sources into maps that have the same noise and synthesized beam properties as our Band 4 and Band 7 observations, and then extracting their fluxes using the same method as for the data, and comparing the original (deboosted) flux densities of our artificial sources with the recovered flux densities.
We find that the flux boosting as a function of signal-to-noise is similar for both bands, and it follows the power law obtained by \cite{Geach2017}. Therefore, we correct (i.e., deboost) the measured flux densities in both bands based on their signal-to-noise using
$S_\nu^\mathrm{deboosted} = S_\nu^\mathrm{measured} / B$,
where the boosting factor $B$ is given by equation 5 in \cite{Geach2017}:
$B=1+0.2\,([\mathrm{S/N}]/5)^{-2.3}$.
We provide the deboosted flux densities in Table~\ref{table_sources}, and adopt those values for all blindly extracted sources throughout the remainder of this paper.

\subsubsection{Undetected sources}

29 of our 99 870\mic-selected ALESS sources are undetected in the 2mm-maps, with no measurable flux at the position of the Band 7 detection ($\sn<1.5$). Of those undetected sources (flagged with an asterisk in Table~\ref{table_sources}), 7 lack photometric counterparts at any other wavelength except for 870\mic\ \citep{Simpson2014,Swinbank2014}, and they are typically low-significance detections at that wavelength (most are $\lesssim4\sigma$ sources except for ALESS099.1 which is detected at $\sim5\sigma$; \citealt{Hodge2013}). These sources might be spurious sources in the Band 7 data, however \cite{Karim2013} estimate that only 1.6\% (i.e., 1 or 2) of the $\ge3.5\sigma$ 870\,\mic\ sources should be spurious. They might instead be high-redshift ($z>3$), high dust optical depth galaxies \citep{daCunha2015} with peculiar dust SEDs. 

To further investigate the properties of our 29 undetected sources, we median-stack the 2mm fluxes at the positions of the 870\,\mic\ detections. The stack produces no  detection, however it allows us to place a $3\sigma$ flux density upper limit of 0.045 mJy.

\subsubsection{2\,mm sources with no counterpart in the ALESS main catalog}

Our source extraction procedure finds an additional sample of eight sources that are detected with $\sn\ge4$ within PB$>0.3$, that we list on Table~\ref{extra_sources} (Appendix~\ref{app_sources}). Since these are blindly-extracted sources, we correct their measured fluxes for boosting using the method described in Section~\ref{boosting}. All but two of these sources fall outside the primary beam of the ALMA 870\,\mic\ observations. Following the method described in \cite{Karim2013,Hodge2013}, we estimate the fraction of spurious sources in our data with $\sn\ge4$ to be less than 10\%, and the corresponding completeness is 96\%. We find that four of these sources have a robust {\it Spitzer}/IRAC 3.6\mic\ counterpart (within 0.5~arcsec) in the ECDFS IRAC-selected catalog of \cite{Damen2011}, confirming they are likely to be real galaxies. The remainder are still within the footprint of that catalog, but have no counterpart within 1 arcsec, meaning that they would have an IRAC 3.6\mic\ magnitude fainter than the $5\sigma$ detection limit of the catalog, $m_\mathrm{AB}=23.8$. They could be spurious or optically-faint SMGs with very low 870\mic-to-2mm flux ratios, either because they are at very high redshifts, and/or or they have very cold dust or unusual dust properties such as low dust emissivity indexes (Section~\ref{colours}; see also \citealt{Wardlow2018}).

\subsection{Comparison with predicted fluxes and 870\,\mic\ fluxes: why are so many sources undetected at 2mm?}

\begin{figure*}
\centering
\begin{minipage}{\linewidth}
\includegraphics[trim={1cm 0.2cm 0.5cm 0cm},width=0.5\textwidth]{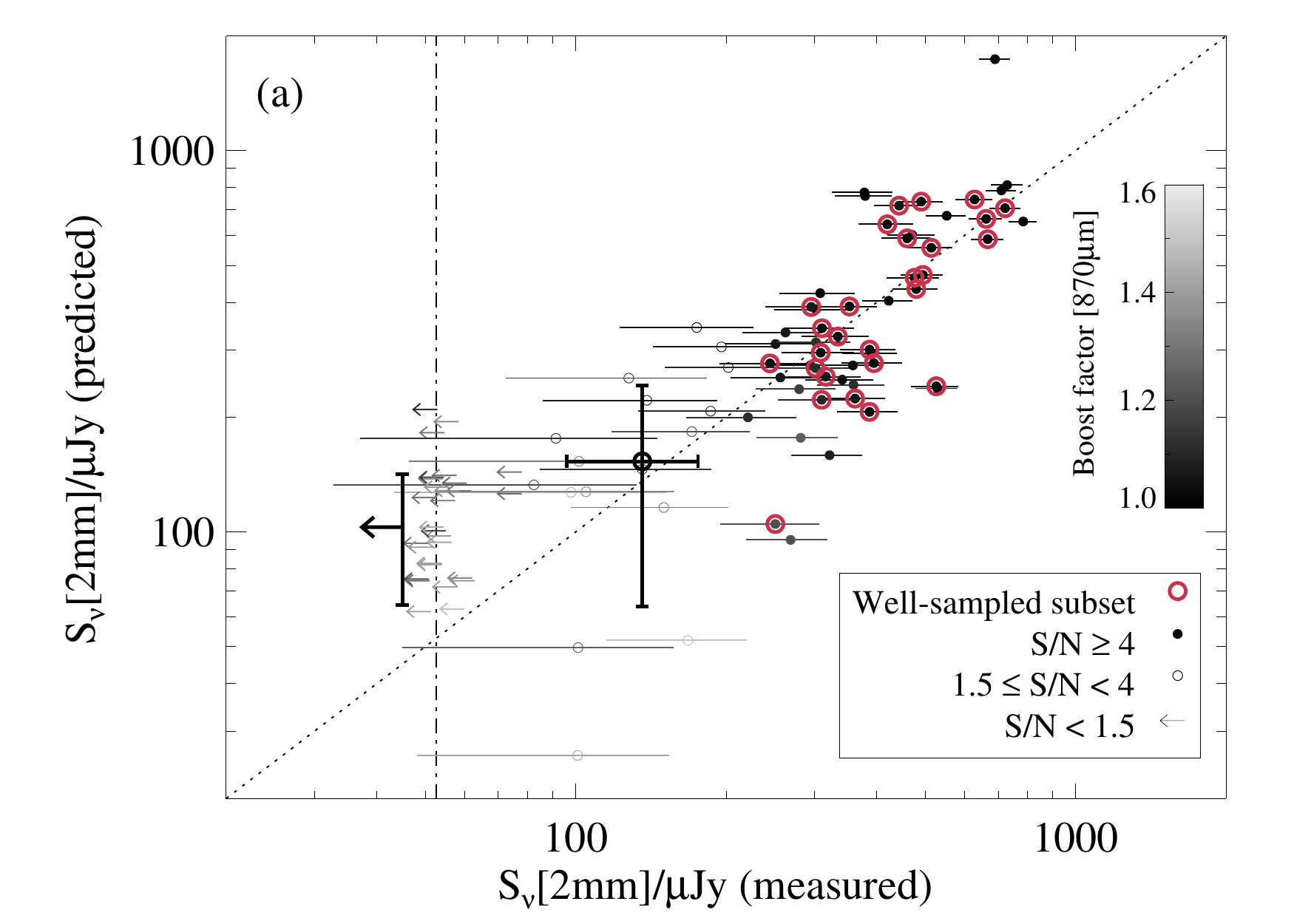}
\includegraphics[trim={1cm 0.2cm 0.5cm 0cm},width=0.5\textwidth]{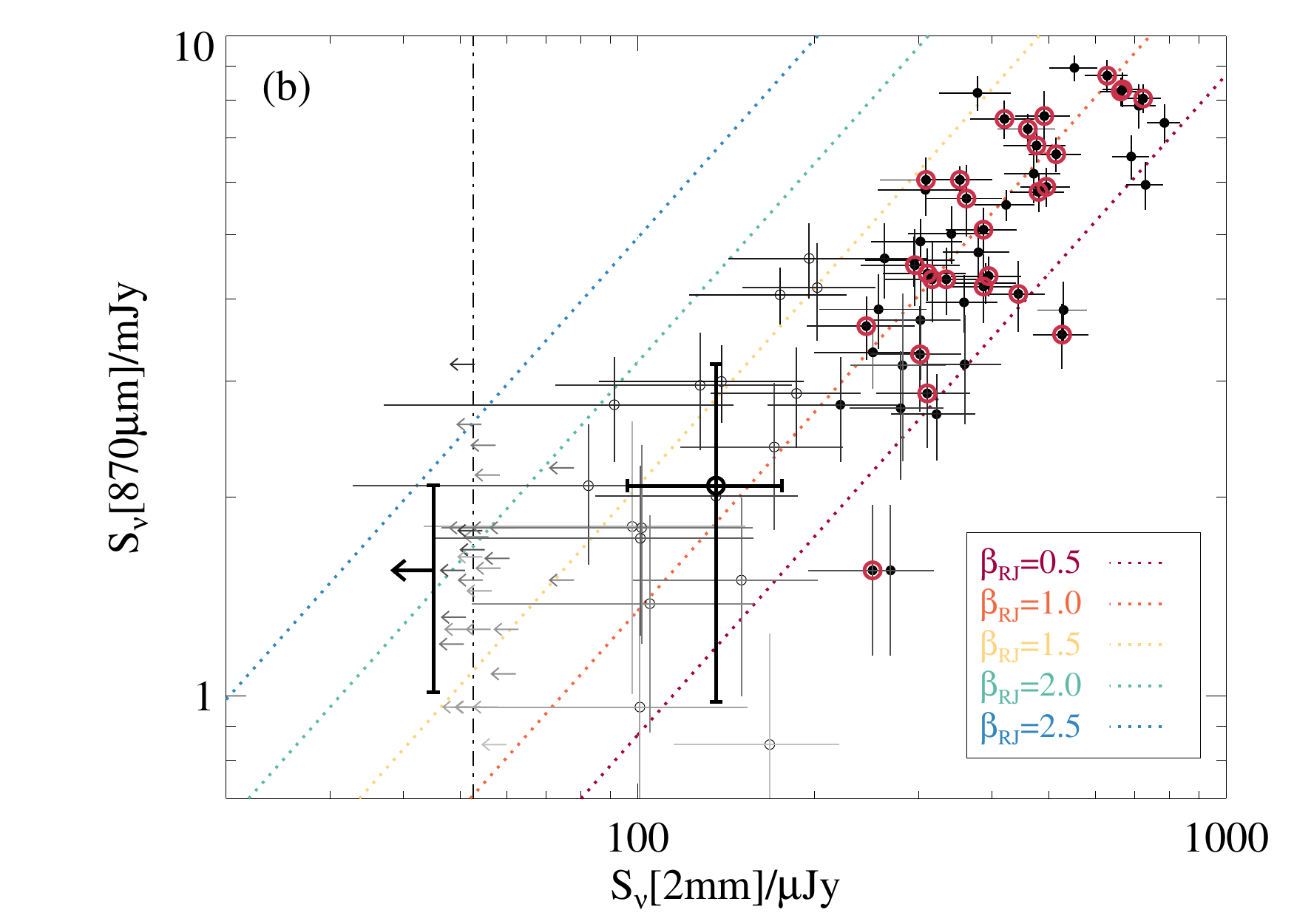}
\end{minipage}
 \caption{(a) Comparison between the measured 2\,mm fluxes of our sources and the predicted MAGPHYS fluxes \citep{daCunha2015}. The dotted line shows the identity. Each data point is color-coded according to the boosting factor of the 870\mic\ observations (Section~\ref{boosting}). The predicted and observed fluxes are well correlated, and the vast majority of our sources have measured fluxes within a factor of 2 of the predictions. We find that the largest deviations occur for faint sources, which had the largest boosting factors in the original 870\mic\ observations. This explains, at least in part, why we over-predicted their 2\,mm fluxes. (b) Measured 2\,mm versus 870\,\mic\ flux densities. The lines show the relation between the fluxes in those two bands {\em assuming the Rayleigh-Jeans approximation}, for different values for the dust emissivity index $\beta_\mathrm{RJ}$ (see Appendix~\ref{appendix_rj} for why that approximation is not appropriate at our observed frequencies). In both panels, the vertical dashed line shows the average sensitivity limit of our observations. The filled black circles represent our individual $\ge4\sigma$ detections (of those, we highlight our `well-sampled subset'; see Section~\ref{extraction} for details), the thin open circles show marginal detections ($1.5 \le \mathrm{S/N} < 4$), and the thin arrows show upper limits for undetected sources. The thick open circle shows the median value for all the 17 marginal detections (error bars are the standard deviation) and the thick arrow shows the upper limit for all 29 undetected sources derived from stacking.} 
 \label{b4_pred_b7}
\end{figure*}

In \cite{daCunha2015}, we developed and used the `high-z extension' of the MAGPHYS SED modeling code \citep{daCunha2008} to fit the observed ultraviolet-to-radio emission of all 99 ALESS main catalog sources. The best-fit model SED of each galaxy was used to estimate the expected 2\,mm flux in preparing our ALMA observations. In Fig.~\ref{b4_pred_b7}(a), we compare those predicted 2\,mm fluxes, scaled down by the 870\mic\ boosting factor (Section~\ref{boosting}), with our measured fluxes. Overall, the predicted and measured fluxes are well correlated, and for the vast majority of sources for which we can extract a flux, they agree within a factor of two. Our 2mm non-detections deviate the most from the predicted fluxes and had the largest 870\mic\ boosting factors ($B=1.36\pm0.10$ on average). Therefore, flux boosting in the original 870\mic\ fluxes can explain our non-detections at least in part. Accounting for 870\,\mic\ flux boosting, the median predicted flux of our 29 non-detections drops from 152\,\textmu Jy to 103\,\textmu Jy, still about a factor of two higher than our $3\sigma$ stack upper limit.

In Fig.~\ref{b4_pred_b7}(b), we compare the 2\,mm and 870\mic\ flux densities. The dispersion is significant at low \sn, but for the brightest sources they are very well correlated. However, our 2\,mm non-detections seem to have a deficit in 2\,mm flux density compared with the 870\,\mic\ flux density (even after accounting for 870\,\mic\ flux boosting). This may indicate different dust emission properties, redshift distributions, or selection effects. We analyze the 870\mic-to-2mm flux ratios in more detail in Section~\ref{colours}.

We note that the flux ratio between two (sub-)millimeter bands is sometimes used to infer the dust emissivity index directly (e.g., \citealt{Aravena2016,Gonzalez2019}). A constant flux ratio between two bands (independent of redshift and dust temperature) is predicted if the two bands sample the low-frequency Rayleigh-Jeans (RJ) tail of the dust emission. If we use RJ approximation, the flux ratio depends on the dust emissivity index $\beta_\mathrm{RJ}$ alone (eq.~\ref{beta_rj}). In Fig.~\ref{b4_pred_b7}(b), we show the correlations between the 870\mic\ and 2mm fluxes predicted in that regime, for different values of $\beta_\mathrm{RJ}$. Those correlations are roughly parallel to our observed correlations and a $\beta_\mathrm{RJ}\simeq1$ seems to match our observations, however that does not necessarily mean that $\beta_\mathrm{RJ}$ is the true emissivity index of our sources. In Appendix~\ref{appendix_rj}, we demonstrate that those values are unlikely to correspond to the real emissivity index in our sources because the Rayleigh Jeans approximation is not valid for our observed bands at the redshifts of the ALESS SMGs, and more sophisticated modeling of the dust emission is needed (Section~\ref{modeling}).

\subsection{Properties of detections vs non-detections}

\begin{figure*}
\centering
\includegraphics[trim={12cm 0cm 0cm 0cm},angle=90,width=\textwidth]{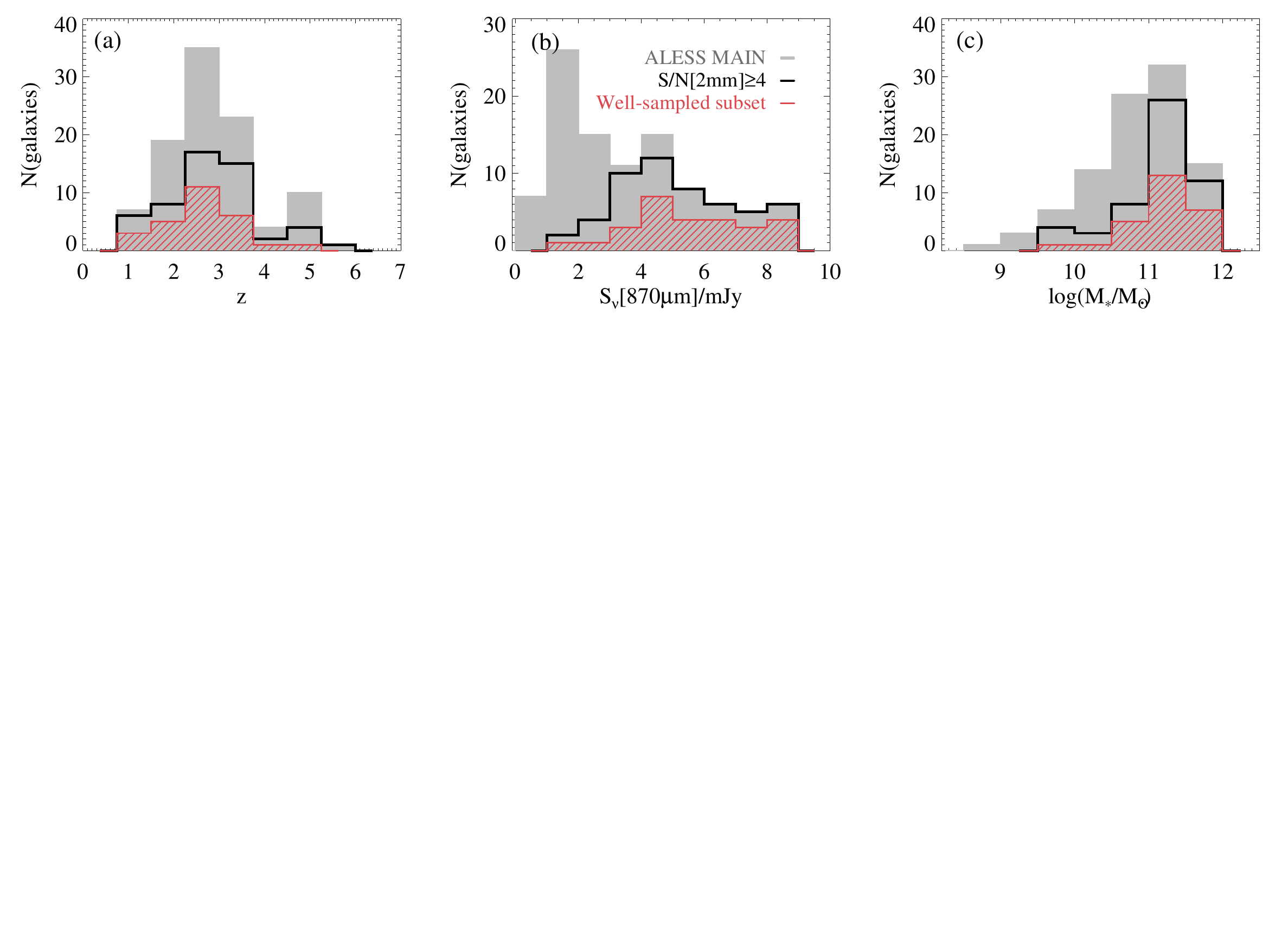}
\caption{Properties of ALESS sources for which we measure a robust 2\,mm flux with S/N$\geq4$ (in black), and of those in our `well-sampled subset' (in red), compared with the properties of our full 870$\mic$-selected sample (in grey): (a) redshift; (b) 870\,\mic\ flux density; (c) median-likelihood estimates of stellar mass derived by fitting the full SEDs of the sources at fixed redshift (we choose the best available redshift for each target; 50/99 are spectroscopic redshifts from \citealt{Swinbank2012}, \citealt{Danielson2017}, and \citealt{Birkin2021}), using the high-redshift extension of the MAGPHYS code \citep{daCunha2015}. }
\label{fig:histograms}
\end{figure*}

The properties of the 99 ALESS main SMGs are described in detail in \citet{Hodge2013,Swinbank2014,Simpson2014,daCunha2015,Danielson2017}. Here we use some of those known properties to investigate what kind of sources are most likely to be detected at 2mm.
Out of the 99 ALESS main catalog SMGs targeted in our 69 Band 4 fields, 53 (i.e., 54\%) are detected above 4$\sigma$, which we consider to be very robust detections.
In Fig.~\ref{fig:histograms}, we plot the distribution of physical properties of our full 870\mic-selected sample and of the subsample of targets for which we achieved robust ($\ge4\sigma$) 2\,mm detections.
We find that the brightest 870\,\mic\ sources are all detected at 2mm, with the detection rate falling steeply for sources below 4~mJy at 870\mic. This also means that we detect the sources with the highest dust masses, dust luminosities, star formation rates, and stellar masses \citep{daCunha2015}. Interestingly, the redshift distribution of 2mm-bright sources follows the parent sample distribution closely, i.e., our detections do not seem to prefer a specific redshift range.

\subsection{870\mic-to-2mm flux density ratios}
\label{colours}

\begin{figure*}
\centering
\includegraphics[width=0.65\textwidth,angle=90]{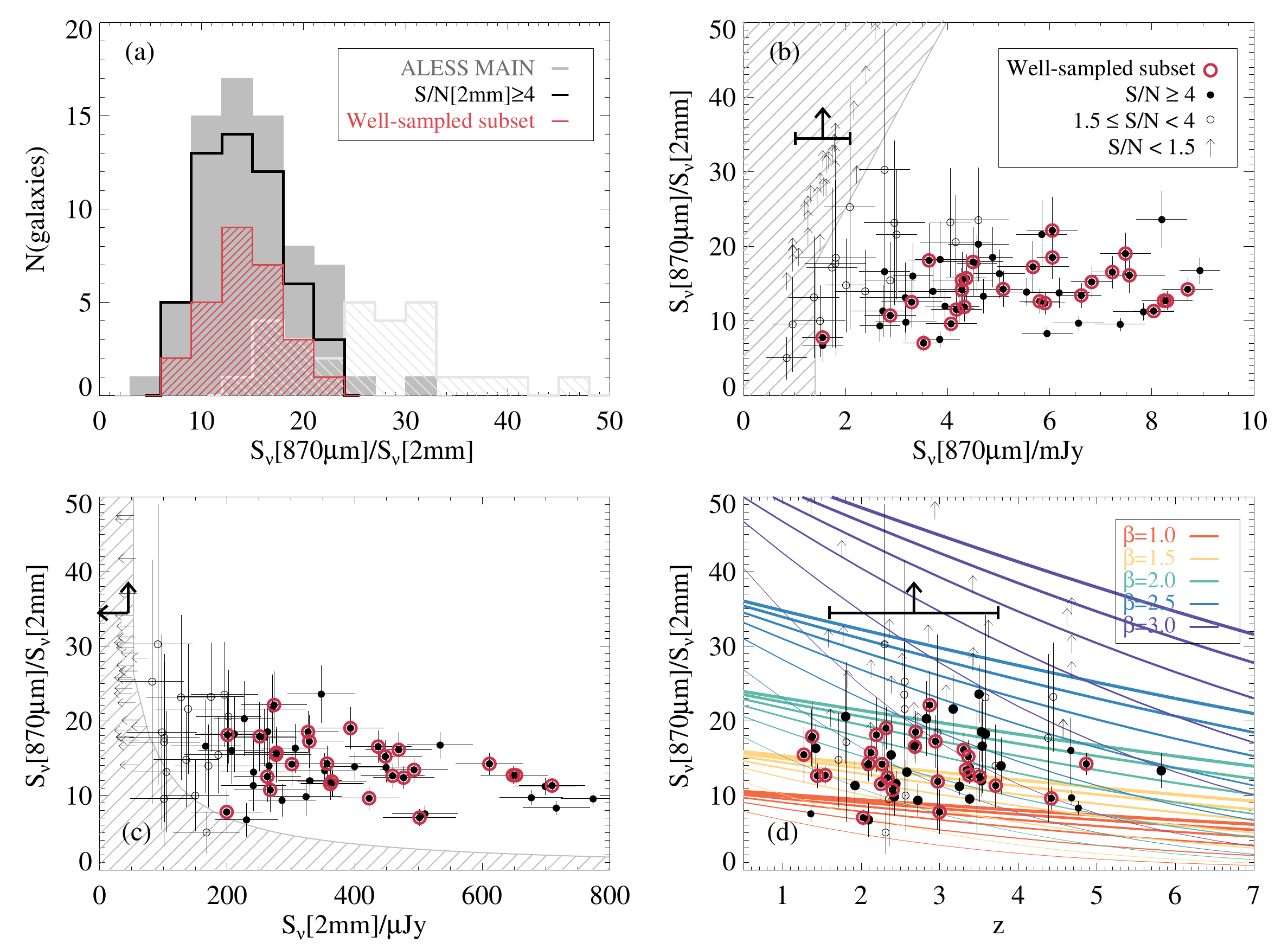}
\caption{The 870\mic-to-2mm flux density ratios, $S_\nu\mathrm{[870\mic]}/S_\nu\mathrm{[2mm]}$, of our 870\mic-selected SMGs. In panels (b) to (d) the symbols have the same meaning as in Fig.~\ref{b4_pred_b7}. (a) Distribution of the flux ratios; the grey histogram shows the full ALESS main sample, including the distribution of upper limits for non-detections as the line-shaded histogram (b) $S_\nu\mathrm{[870\mic]}/S_\nu\mathrm{[2mm]}$ against 870\,\mic\ flux density; (c) $S_\nu\mathrm{[870\mic]}/S_\nu\mathrm{[2mm]}$ against 2\,mm flux density; (d) $S_\nu\mathrm{[870\mic]}/S_\nu\mathrm{[2mm]}$ against redshift; the lines show the predicted ratios as a function of redshift for a set of isothermal, optically-thin dust emission models with temperatures $T_\mathrm{dust}$ varying between 20 and 70~K (thin to thick lines), and emissivity indexes $\beta$ varying between 1.0 and 3.0 (orange to purple). In panels (b) and (c) the shaded region shows the unobservable regions due to the pre-selection at 870\mic\ and the depth of our 2mm observations. The $870\mic/2$mm colors span a relatively narrow range (their median 870\mic-to-2mm flux ratio is $14\pm5$), and are weakly correlated with the 870\mic\ and 2mm fluxes (though this could be due to selection effects), but uncorrelated with redshift. Panel (d) shows that we would expect a negative correlation with redshift for fixed dust temperature and emissivity index. Non-detections ($\mathrm{S/N}<1.5$) seem to have higher $870\mic/2$mm ratios, which would be consistent with higher \tdust\ and/or $\beta$.}
\label{fig:colours}
\end{figure*}

In Fig.~\ref{fig:colours}, we plot the distribution of 870\mic-to-2mm flux density ratios for our ALESS sources. The 870\mic-to-2mm ratios of our strongly detected SMGs span a relatively narrow range. The median flux density ratio for the 70 SMGs for which we measure a 2\,mm flux density is $S_\nu\mathrm{[870\mic]}/S_\nu\mathrm{[2mm]}=14\pm5$ (with the error indicating the standard deviation range; for the well-sampled subset, we find a median of $14\pm4$). For non-detections ($\sn<1.5$), our $3\sigma$ stack upper limit implies ratios $>34.5$. Accounting for the upper limits for non-detections, the median flux ratio of the full ALESS sample increases to $S_\nu\mathrm{[870\mic]}/S_\nu\mathrm{[2mm]}=17\pm9$. We find weak correlations between $S_\nu\mathrm{[870\mic]}/S_\nu\mathrm{[2mm]}$ and the flux densities at 870\mic\ and 2mm, with Spearman rank correlation coefficients $r_{\rm{S}}=0.32$ ($2.3\sigma$) and $r_{\rm{S}}=-0.36$ ($2.6\sigma$), respectively, although these correlations could be due to selection effects.

To put the 870\mic-to-2mm flux density ratios into a more physical context, we compare them, in Fig.~\ref{fig:colours}(d), with the ratios predicted by simple, isothermal and optically-thin dust emission models with varying dust temperatures $T_\mathrm{dust}$ and emissivity indexes $\beta$, as a function of redshift. The models predict that $S_\nu\mathrm{[870\mic]}/S_\nu\mathrm{[2mm]}$ should decrease with redshift, \tdust, and $\beta$, but the specific ratio for a given source is due to a combination of all these parameters. Our sources seem to span a broad model parameter space, however, this is sensitive to errors in the derived 2\,mm flux density for the lower significance detections, as well as uncertainties introduced by photometric redshifts. Nevertheless, if we focus on only strong detections and our well-sampled subset (for which we have spectroscopic redshifts), we find that (i) we do not recover a strong redshift dependence ($r_{\rm{S}}=-0.19$, $<1\sigma$), and (ii) the location of these sources seems to favor models with $\beta\simeq1-2$. The first finding indicates that perhaps the intrinsic dust properties of our SMGs depend on redshift (e.g., dust temperatures could be increasing at high redshift; see also Section~\ref{ratio_evolution}), or more likely that selection effects are playing an important role (see Section~\ref{sec:selection}). The latter finding appears to be in agreement with what is often assumed for the dust properties of galaxies, although the degeneracy between $\beta$ and \tdust\ makes it difficult to estimate the actual dust emissivity index more precisely on a source-by-source basis using this ratio alone.

\section{The dust properties of ALESS SMGs}
\label{sec:dustprop}

To break the degeneracy between the dust temperature and emissivity index for our SMGs, we require additional observations sampling the dust spectral energy distribution closer to its peak\footnote{In Appendix~\ref{appendix_rj} we demonstrate why simply assuming the Rayleigh-Jeans approximation is not appropriate for our data (and for other high-redshift ALMA observations).}. In this section, we describe our dust emission models and fits to the observed ALESS far-infrared/(sub-)millimeter SEDs\footnote{We note that here we will only focus on the thermal dust emission in the far-infrared to millimetre, i.e., we do not include mid-infrared or radio emission from AGN, as they are not expected to contaminate our observations. Furthermore, only three of the sources in our `well-sampled' subset are identified as AGN \citep{Wang2013}; removing those from our analysis would not affect our conclusions.}.

\subsection{Modeling the dust emission}
\label{modeling}

The dust emission of a population of dust grains with equilibrium temperature \tdust\ is described by the general solution to the radiative transfer equation:
\begin{equation}
S_\nu \propto [1-\exp(-\tau_\nu)] B_\nu(T_\mathrm{dust})\,,
\label{rt_solution}
\end{equation}
where $B_\nu(T_\mathrm{dust})$ is the Planck function, and the optical depth $\tau_\nu$ can be written as:
\begin{equation}
\tau_\nu=\kappa_\nu \Sigma_\mathrm{dust}\,,
\label{taunu}
\end{equation}
where $\Sigma_\mathrm{dust}$ is the dust mass surface density, and $\kappa_\nu$ is the frequency-dependent dust opacity, generally described by a power law:
\begin{equation}
\kappa_\nu=\kappa_0 \Big( \frac{\nu}{\nu_0} \Big)^\beta\,,
\label{kappa}
\end{equation}
where $\beta$ is the dust emissivity index, and $\kappa_0$ is the emissivity of dust grains per unit mass at a reference frequency $\nu_0$. This function depends on the chemical and optical properties of dust grains (e.g., \citealt{Draine1984,Draine2007,Galliano2018}). In this paper, we adopt $\kappa_0=0.77\,\mathrm{cm}^2\,\mathrm{g}^{-1}$ at $\nu_0=353$~GHz (i.e., $\lambda_0=850\mic$), to be consistent with \cite{daCunha2008,daCunha2015}.

For simplicity, throughout the remainder of this paper we assume isothermal dust models. Realistically, dust grains in galaxies are not at a single temperature, however (optically-thin) modified black body models have been shown to be a good approximation to the long-wavelength emission ($\lambda\gtrsim100\mic$) caused by dust grains in thermal equilibrium, producing dust masses that are very close to the ones produced by more complex modeling that includes a distribution of temperatures \citep{Draine2007}, provided consistent values for the dust emissivity are used (\citealt{Bianchi2013}; see also, e.g., \citealt{Lianou2019}). Moreover, given the relatively low number of points sampling the dust emission of our sources, such simple isothermal models are the most complexity that can be afforded. Modelling the emission with multiple temperature components with varying emissivity indexes would result in a much larger number of free parameters than observational constraints, and the multiple dust parameters would be very difficult to constrain. This is also true for many other high-redshift sources observed in the dust continuum with ALMA, where often observations only in one or a few bands are available.
Therefore, we adopt isothermal dust models, and explore two cases: an optically-thin approximation and a more general opacity scenario.

\subsubsection{Optically-thin approximation}
\label{optically_thin}

\begin{figure*}
\centering
\includegraphics[width=0.7\textwidth,angle=90]{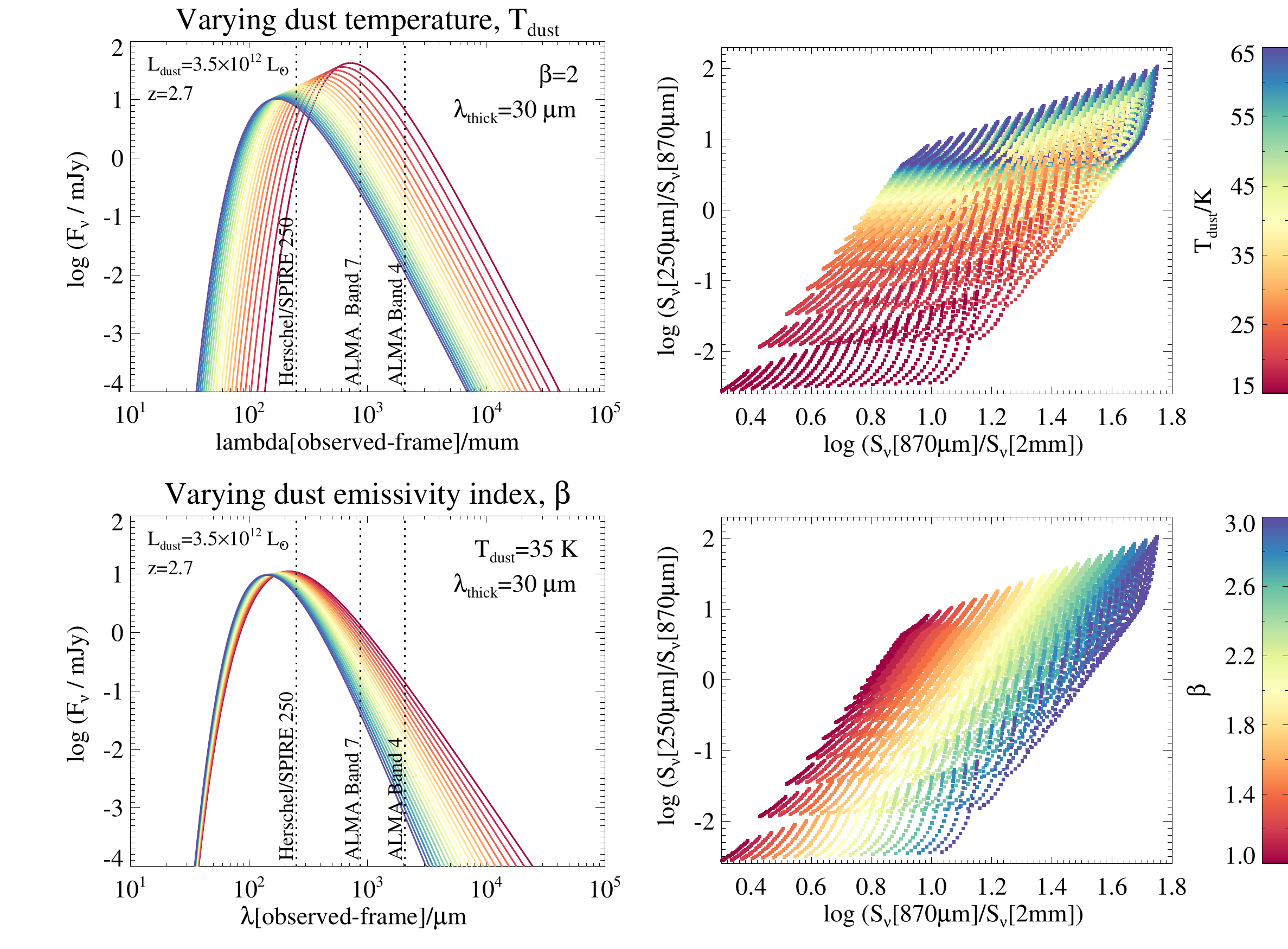}
\includegraphics[width=0.7\textwidth,angle=90]{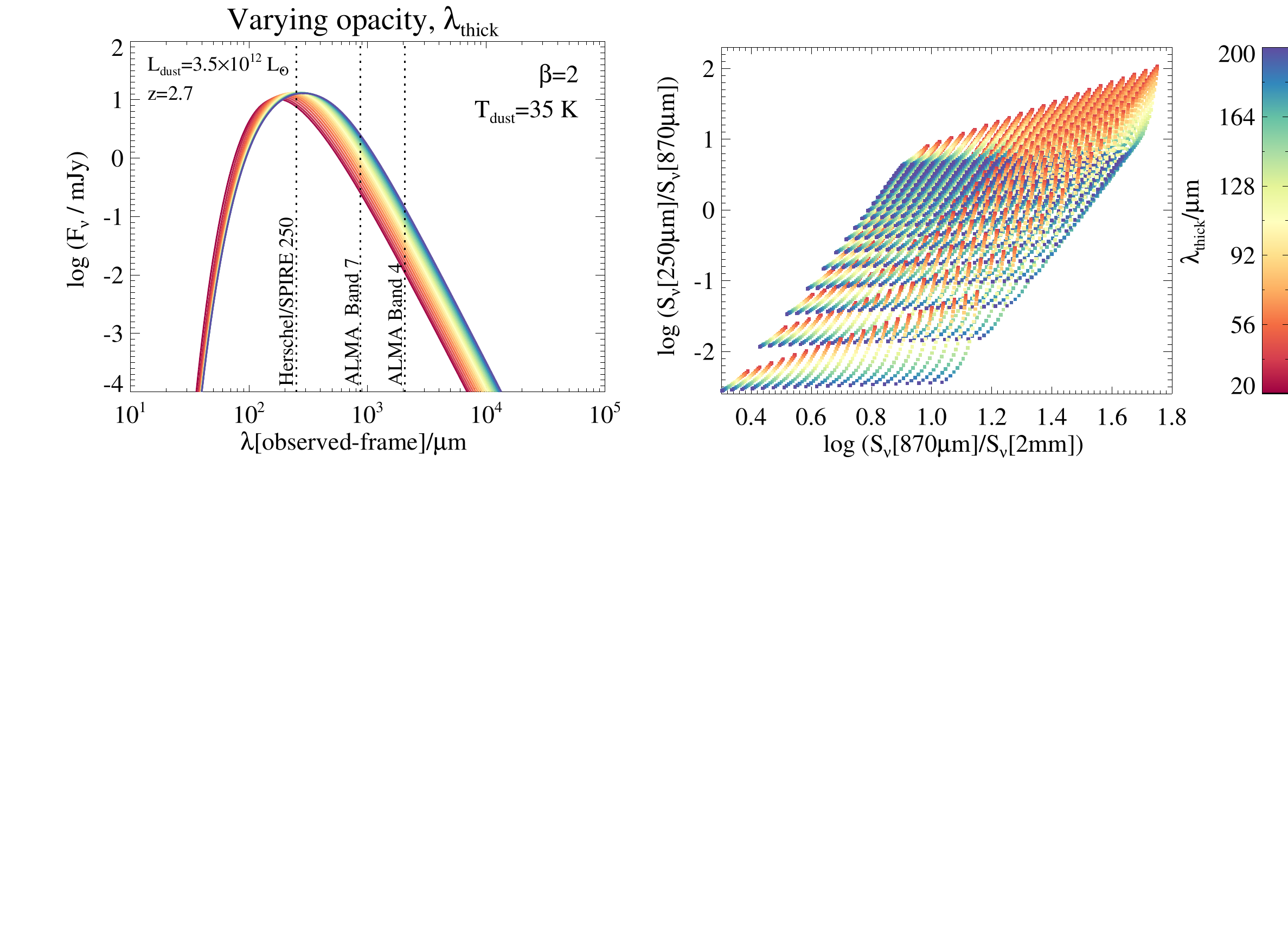}
\vspace{-6.3cm}
\caption{The predicted far-infrared colors of the dust emission in a galaxy at the median redshift of the ALESS sources ($z=2.7$) and with their median dust luminosity ($L_\mathrm{dust}=3.5\times10^{12}L_\odot$; \citealt{daCunha2015}), showing the effect of varying the dust temperature \tdust\ (top panels), emissivity index $\beta$ (middle panels), and dust opacity, parameterized by the wavelength up until which dust remains optically-thick, $\lambda_\mathrm{thick}$ (bottom panels; note that $\lthick\lesssim50\mic$ approximates the optically-thin case for the wavelength range sampled by our observations). The left-hand panels show the effect of varying each parameter at a time on the global dust SED, while the right-hand panels show the parameter space of galaxy colors varying all parameters at once, where we color-code each model by a single parameter at a time. Using a combination of at least three bands to sample both the peak of the SED and the Rayleigh-Jeans tail of the dust emission, we can form a color space where the variation of \tdust\ and $\beta$ are approximately orthogonal at fixed \lthick, and therefore break the degeneracy between these two parameters. In the general opacity scenario where \lthick\ is allowed to vary, \tdust\ and \lthick\ remain degenerate with currently available observations.
}
\label{fig:seds}
\end{figure*}

Here we assume the simplest approximation for the dust emission in galaxies: optically-thin dust, which means $\tau_\nu\ll1$ at the observed frequencies, hence $[1-\exp(-\tau_\nu)]\simeq\tau_\nu$, and the radiative transfer solution (eq.~\ref{rt_solution}) becomes:
\begin{equation}
S_\nu \propto \Sigma_\mathrm{dust} \kappa_\nu B_\nu(T_\mathrm{dust})\,.
\label{mbb}
\end{equation}

In this case, the shape of the far-infrared/sub-millimeter dust SED, at fixed redshift, depends solely on the dust temperature \tdust\ and the emissivity index $\beta$. These two parameters can be strongly degenerate in the 870\mic-to-2mm ratio, as shown in Fig.~\ref{fig:colours}(d). To break this degeneracy, we require additional observations sampling the dust emission closer to its peak.
Fig.~\ref{fig:seds} shows that in order to be sensitive to variations of both \tdust\ and $\beta$, we need to sample the dust SED in at least three bands, from the peak of the emission towards higher frequencies (which depends to first order on \tdust), to the Rayleigh-Jeans tail at low frequencies (which mainly depends on $\beta$).
Therefore to sample the peak of dust emission, we use {\it Herschel} flux measurements when available \citep{Swinbank2014}.
Of our sample of ALESS sources with robust (i.e., $\ge4\sigma$) 2mm measurements and $\ge3.5\sigma$ 870\mic\ measurements, along with spectroscopic redshifts, 27 have at least one {\it Herschel}/SPIRE measurement at 250\mic: this constitutes our {\it well-sampled subset}. We focus on this subset in the remainder of the paper because for these sources we have the minimum set of bands needed to adequately sample the SEDs (Fig.~\ref{fig:seds}), and redshift uncertainties are not likely to affect our results.

\subsubsection{General dust opacity}
\label{general_opacity}

\begin{figure}
\centering
\includegraphics[trim={1.2cm 0.3cm 0cm 0.2cm},width=0.5\textwidth]{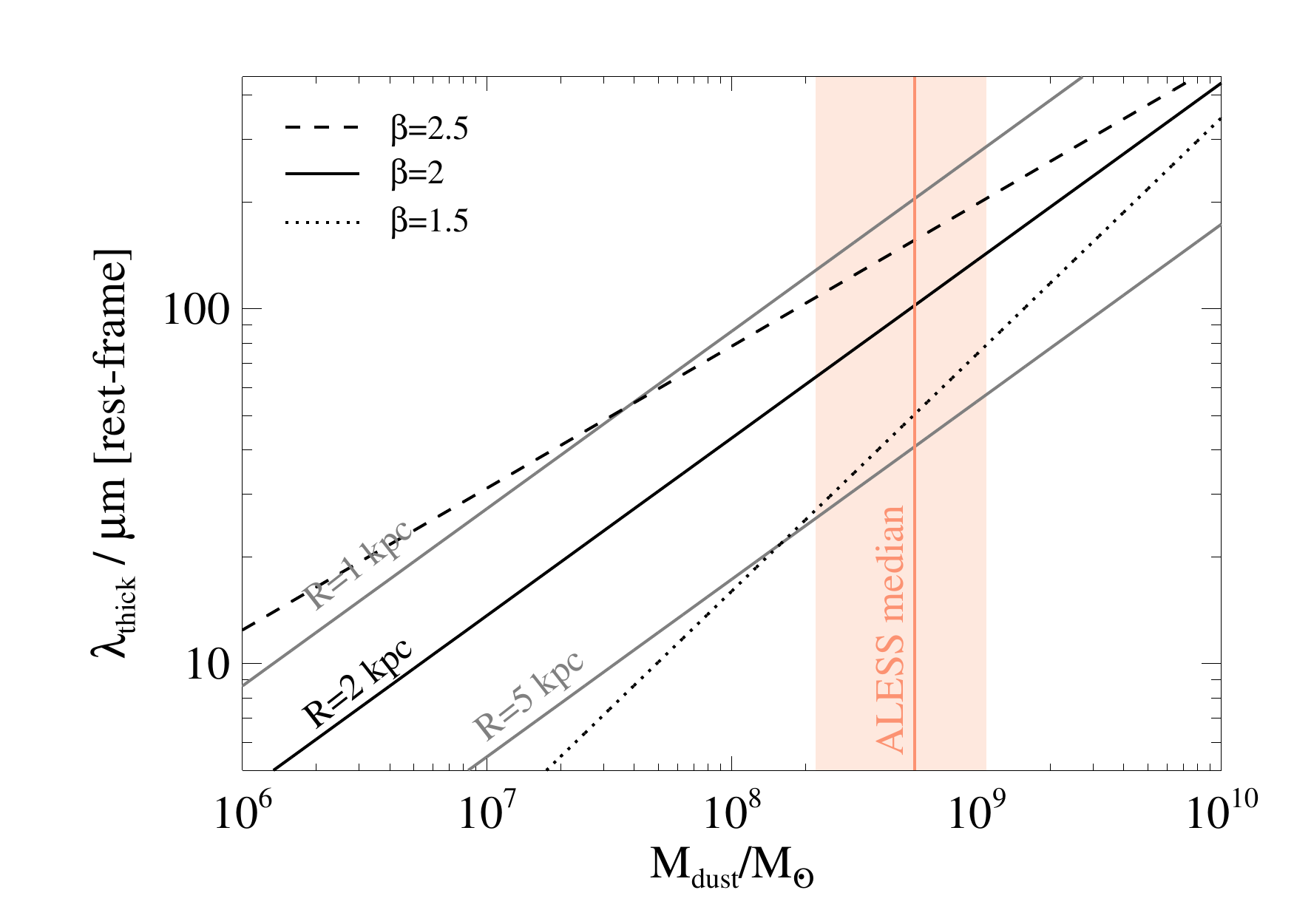}
\vspace{-0.4cm}
\caption{Wavelength up until which dust remains optically-thick, \lthick, as a function of total dust mass for different galaxy sizes and dust emissivity indexes (eq.~\ref{lthick}). The vertical orange line and shaded region shows the median dust mass of ALESS SMGs computed by \cite{daCunha2015}, and the 16th--84th percentile of the dust mass distribution for the sample ($\mdust=5.6^{+5.4}_{-3.4}\times10^8\msun$), which gives us a $\lambda_\mathrm{thick}\simeq100\pm40$\mic\ for the typical 2 kpc radius of the ALESS sources derived by \cite{Hodge2016}, assuming $\beta=2$.
}
\label{fig:opacity}
\end{figure}

For completeness, we also explore the more general scenario where dust may remain optically-thick towards far-infrared wavelengths. It is reasonable to consider that dust might be optically thick well into the far-infrared regime for very dusty sources such as SMGs  \citep[e.g.,][]{Conley2011,Simpson2017,Casey2019,Ugne2020,Cortzen2020}. In this case, we use eqs.~\ref{taunu} and \ref{kappa} to define an additional parameter, the wavelength up until which the dust remains optically thick, i.e., \lthick, such that $\tau_{\lambda_\mathrm{thick}}=1$. This \lthick\ becomes an additional free parameter of the model, along with \tdust\ and $\beta$, and it depends on the intrinsic properties of the dust through its opacity function $\kappa_\nu$, and on dust mass surface density.

In the bottom panels of Fig.~\ref{fig:seds}, we show how varying \lthick\ affects both the dust SEDs and the $S_\nu\mathrm{[870\mic]}/S_\nu\mathrm{[2mm]}$ vs $S_\nu\mathrm{[250\mic]}/S_\nu\mathrm{[870\mic]}$ color space. This introduces a clear degeneracy: increasing \lthick\ shifts the peak of the emission and affects the $S_\nu\mathrm{[250\mic]}/S_\nu\mathrm{[870\mic]}$ in the same direction as lowering the dust temperature. More detailed observations sampling the shape of the SED near its peak are needed to constrain all three parameters. Nevertheless, in order to account for the possibility that dust might not be optically thin at all wavelengths considered, we make an educated guess on \lthick. We assume a simple spherical shell geometry (e.g., \citealt{Inoue2020}), such that $\Sigma_\mathrm{dust}=\mdust/4\pi R^2$, to investigate how \lthick\ may vary as a function of dust mass and size for typical SMGs (Fig.~\ref{fig:opacity}). In this simple case, \lthick\ is given by:
\begin{equation}
\lthick=\lambda_0 \Big(\kappa_0\frac{\mdust}{4\pi R^2}\Big)^{1/\beta}\,.
\label{lthick}
\end{equation}
As expected, for a fixed galaxy size, dust remains optically-thick out to longer wavelengths as the total dust mass increases; at fixed dust mass, the dust column decreases and \lthick\ becomes shorter as the radius increases. If we take the median dust mass of ALESS SMGs (and the 16th--84th percentile range of the sample dust mass distribution) derived by \cite{daCunha2015} with MAGPHYS ($\mdust=5.6^{+5.4}_{-3.4}\times10^8\msun$), and the typical radius measured by \cite{Hodge2016} using high-resolution ALMA observations in Band 7 ($R\simeq2$~kpc), we obtain a typical $\lthick\simeq100\pm40\mic$ (assuming for now a dust emissivity index $\beta=2$; see also discussion in \citealt{Simpson2017}). This implies that, for the typical redshift of our sample ($z\simeq2.7$), observations shortwards of $\sim370\mic$ (observed-frame) are not necessarily in the optically-thin regime, which affects the SEDs (Fig.~\ref{fig:seds}). In order to investigate systematic effects of the optically thin versus general opacity assumptions on the derived dust parameters, we fit our ALESS SEDs both with optically-thin models and with a dust opacity model where we explore \lthick\ in the range 60\mic\ to 140\mic, as indicated by our simple calculation above.

\subsubsection{SED fitting method}
\label{fitting}

We use a Bayesian approach to fit the dust SEDs of our sources and recover posterior likelihood distributions for the free dust parameters: dust temperature (\tdust), dust emissivity index ($\beta$), dust luminosity (\ldust), dust mass (\mdust). Since we assume isothermal dust, for each galaxy we only include observations sampling the SED at wavelengths longer than 70\mic\ in the rest-frame, as the effects of warmer dust components from stochastically heated dust grains are likely to impact the SEDs at shorter wavelengths.
We generate a model library of dust SEDs with dust temperature \tdust\ uniformly distributed between 15 and 80~K, and emissivity index $\beta$ uniformly distributed between 1.0 and 3.0. For the general opacity case, we include \lthick\ as an additional free parameter and distribute it uniformly between 60\mic\ and 140\mic, based on the calculation in the previous section. 
For each source at redshift $z$, we place the model dust SEDs in the observed-frame and apply the appropriate cosmic microwave background (CMB) corrections as described in \cite{daCunha2013b}. We then compute the model fluxes in the {\it Herschel}/SPIRE bands at 250, 350, and 500\mic, and in the ALMA Bands 7 and 4 (we also include ALMA Band 3 at 3.3mm, available for 10 of our sources from \citealt{Wardlow2018}, and ALMA Band 8 at 0.63mm, available for two sources from \citealt{Rybak2019}). We compare these model fluxes with the observed fluxes (and upper limits, when available) by evaluating the $\chi^2$ goodness-of-fit of each model in the library. Then we obtain marginalized likelihood distributions for all the free parameters.

\subsection{Results: dust parameter constraints}
\label{results}

We start by focussing on the results obtained using the optically-thin assumption, and then compare with the results using the general opacity model using $\lthick=100\pm40\mic$.

\subsubsection{Results using the optically-thin approximation}
\label{results_thin}

\begin{figure*}
\begin{minipage}{\linewidth}
\centering
\includegraphics[width=0.95\textwidth]{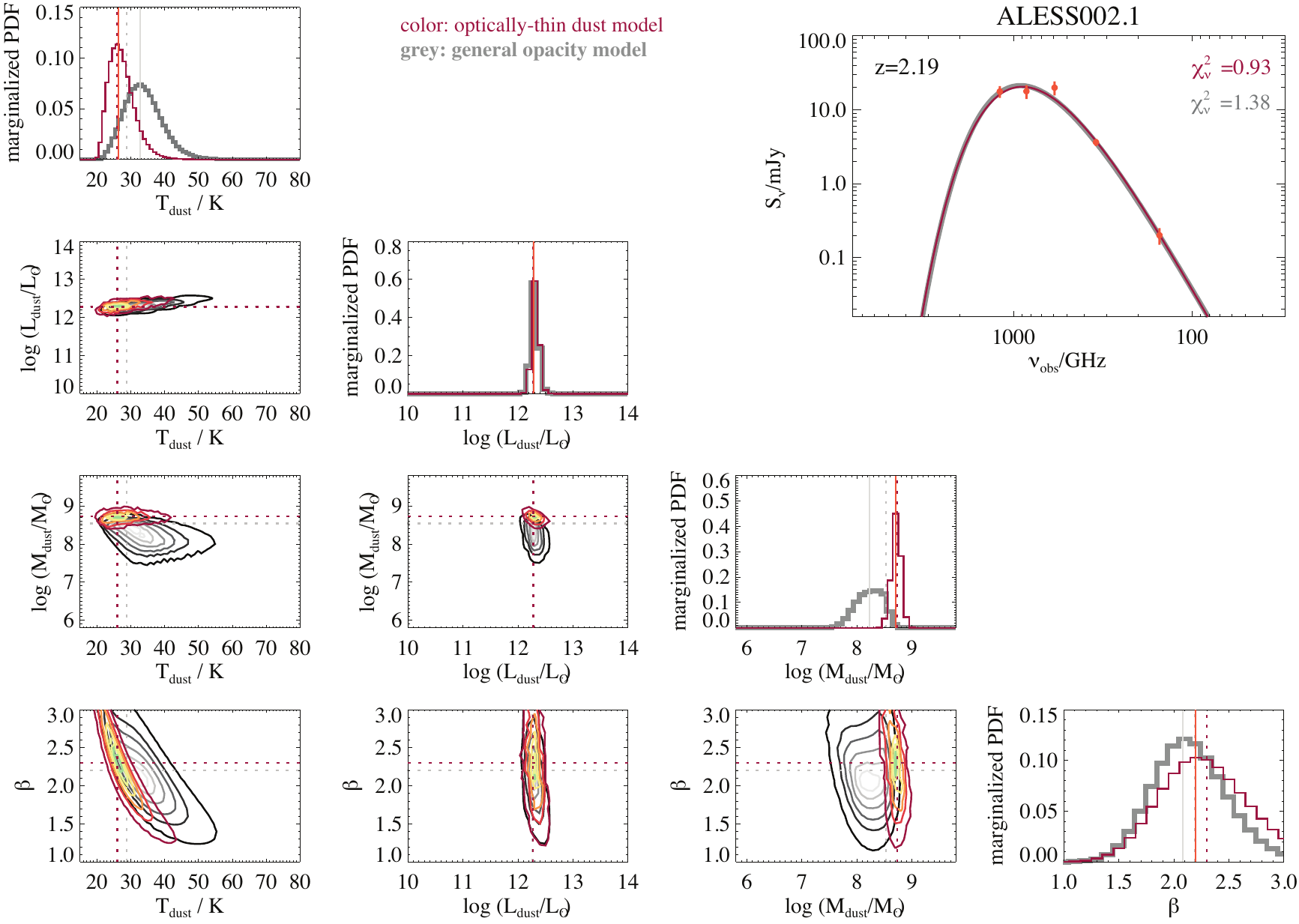}
\end{minipage}
\caption{Outputs of our Bayesian dust SED fitting for ALESS002.1, the brightest source in our well-sampled subset. The top right-hand panel shows the best-fit dust models (solid lines) and the observed fluxes (orange points). The remaining panels from top-left to bottom-right show the marginalized likelihood distributions for all the parameters in the fit: dust temperature (\tdust), luminosity (\ldust), mass (\mdust), and emissivity index ($\beta$), as well as the two-dimensional likelihood distributions of all combinations of these parameters. The contours show levels of equal probability: 1\%, 10\%, 25\%, 50\%, 75\%, and 90\% of the maximum probability value. The dotted lines indicate the best-fit parameters (i.e., corresponding to the minimum $\chi^2$), and the vertical solid lines show the medians of the likelihood distributions.
We plot the results for two fitting runs: the colored SED, histograms, and lines correspond to the optically-thin dust model; the grey scale ones correspond to the general opacity scenario. For ALESS002.1, the best-fit SEDs in these two scenarios are virtually indistinguishable from each other, and the reduced $\chi^2$ are very close ($\chi^2_\nu$ is actually smaller in the optically-thin case because it produces a similarly good fit with fewer free parameters than the general opacity model). In the general opacity case, \tdust\ and \mdust\ are less well-constrained due to the additional degeneracy with \lthick, and systematic offsets with the optically-thin model are clear; the \ldust\ and $\beta$ posteriors are very similar in the two cases.}
\label{fig:aless002p1}
\end{figure*}

In Fig.~\ref{fig:aless002p1}, we show an example of the outputs of our dust SED fitting for ALESS002.1, the brightest 870\,\mic\ source in our well-sampled subset. For this galaxy, the SED is sufficiently well sampled that both the dust temperature and emissivity index show well-constrained posterior likelihood distributions. Thanks to these two parameters being well constrained, the total dust luminosity and the dust mass are constrained to very small uncertainties as indicated by the narrow posterior distributions. The two-dimensional likelihood distributions allow us to explore parameter degeneracies in the fitting. As expected, and discussed in the previous section and in numerous works in the literature (e.g., \citealt{Shetty2009,Juvela2013}), there is a strong degeneracy between \tdust\ and $\beta$ which explains why while these parameters are well constrained, the likelihood distributions are relatively wide.

\begin{figure*}
\begin{minipage}{\linewidth}
\centering
\includegraphics[width=0.95\textwidth]{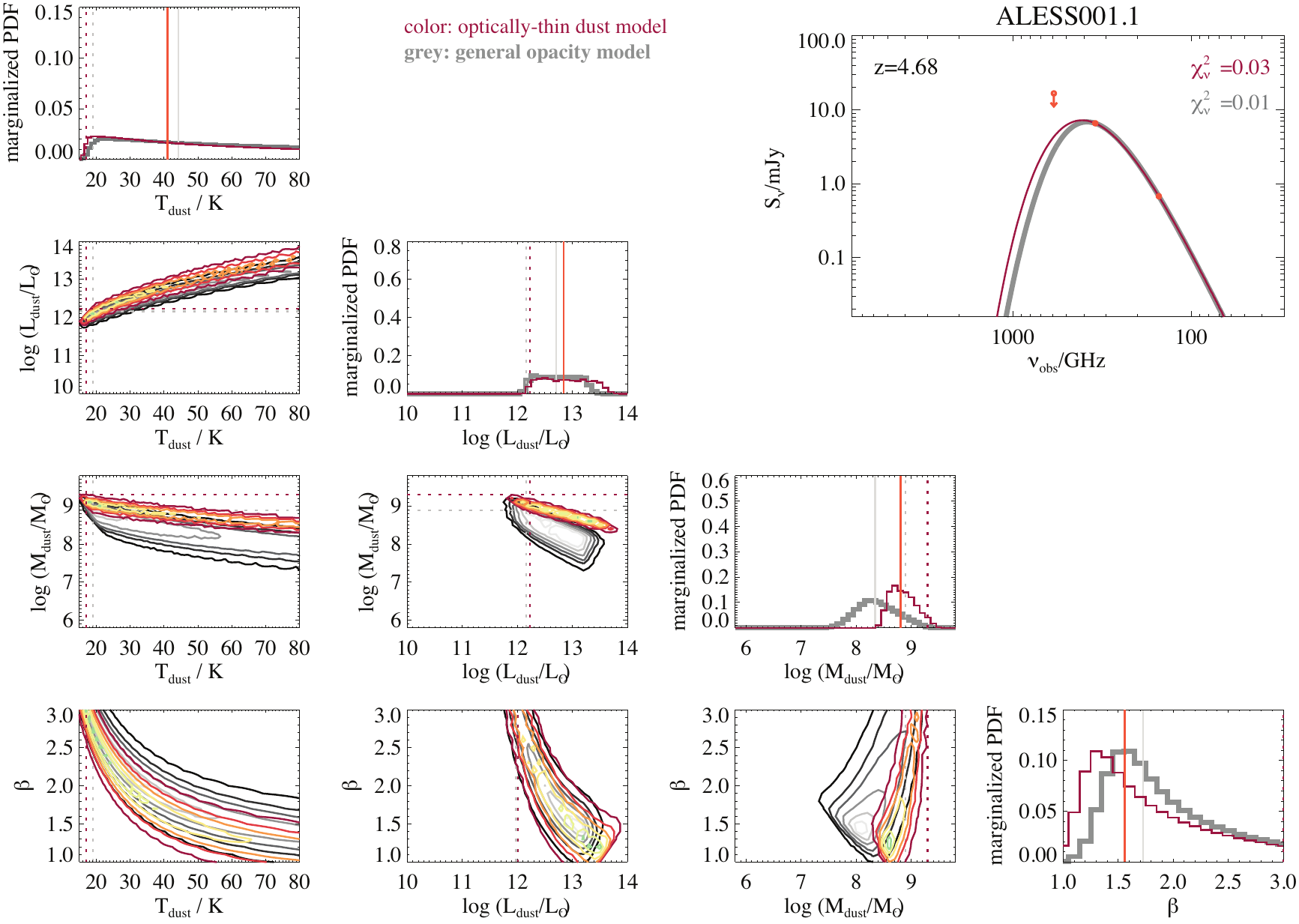}
\end{minipage}
\caption{Similar outputs of our fitting as in Fig.~\ref{fig:aless002p1}, but for ALESS001.1, a source without {\it Herschel} measurements (only upper limits). Since there are no measurements at frequencies higher than Band 7 to sample the peak of the dust emission, \tdust\ and $\beta$ are poorly constrained, and consequently the errors on \ldust\ and \mdust\ are also much larger (wider likelihood distributions) compared to ALESS002.1. Because of the lack of data, strong degeneracies between the model parameters become evident in the two-dimensional likelihood distributions, even in the optically-thin case.}
\label{fig:aless001p1}
\end{figure*}

In Fig.~\ref{fig:aless001p1} we show, for comparison, the results of fitting the dust SED of ALESS001.1, the brightest 870\,\mic\ source, which also has a robust 2\,mm detection but only {\it Herschel} limits. In this case, both \tdust\ and $\beta$ are severely unconstrained, resulting in much broader likelihood distributions for the dust masses and luminosities. The strong degeneracies between the various dust model parameters become clear in the two-dimensional likelihood distributions. This demonstrates that the parameters derived from this fitting for galaxies with only ALMA Band 7 and Band 4 detections are not reliable, particularly \tdust, $\beta$, and \mdust\ (see also Appendix~\ref{app_accuracy}). Therefore, in the remainder of this section, we focus solely on the well-sampled subset of 27 ALESS SMGs, accepting that these sources are not necessarily representative of the whole population (see Section~\ref{sec:selection}). In Table~\ref{table_properties}, we present the median-likelihood estimates of the dust temperatures, emissivity indexes, luminosities and masses, and respective confidence ranges, obtained using the optically-thin approximation for those 27 galaxies.

\begin{figure*}
\centering
\includegraphics[width=0.75\textwidth]{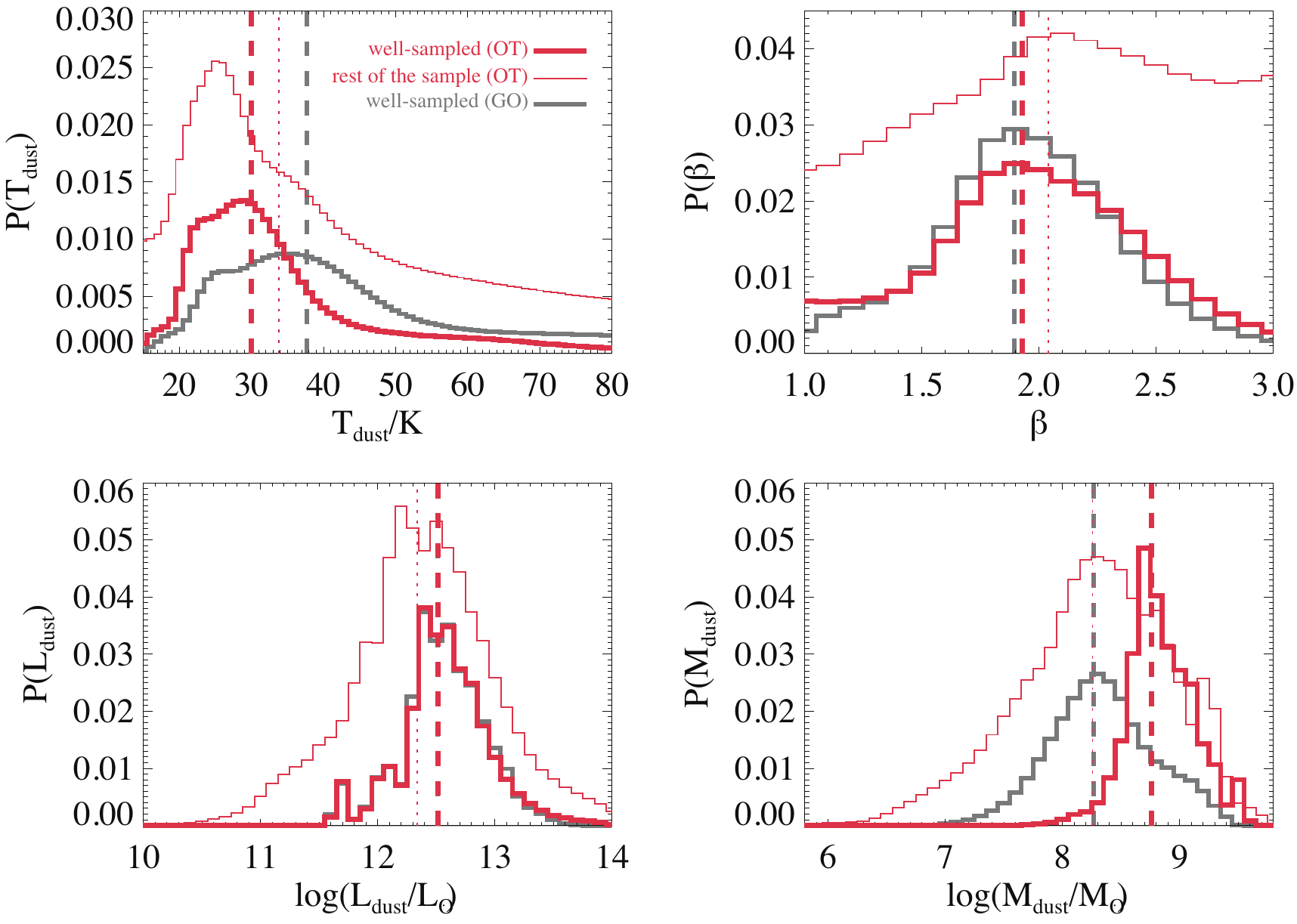}
\caption{Stacked likelihood distributions of the dust parameters for the ALESS sample. The red histograms show the results of fitting the SEDs using the optically-thin (OT) assumption: the thick line corresponds to the 27 sources in our well-sampled subset, and the thin line show the stacked PDFs of the remaining 72 sources. For the well-sampled subset, we find cold dust temperatures $\tdust\simeq30^{+14}_{-8}~K$, emissivity indexes $\beta\simeq1.9\pm0.4$, dust luminosities $\log(\ldust/\lsun)=12.5^{+0.4}_{-0.3}$, and dust masses $\log(\mdust/\msun)=8.8^{+0.3}_{-0.2}$ (medians of the stacked likelihood distributions, errors given by the 16th--84th percentiles). In grey we show, for comparison, the stacked likelihood distributions for the well-sampled sources using the general opacity (GO) model, which shows offsets towards higher \tdust\ and lower \mdust. However, \ldust\ and $\beta$ remain practically the same, meaning that the estimates of these parameters are robust against dust opacity assumptions.}
\label{fig:stacked}
\end{figure*}

\begin{figure*}
\centering
\includegraphics[trim={0cm 6.5cm 0cm 0cm},width=0.85\textwidth]{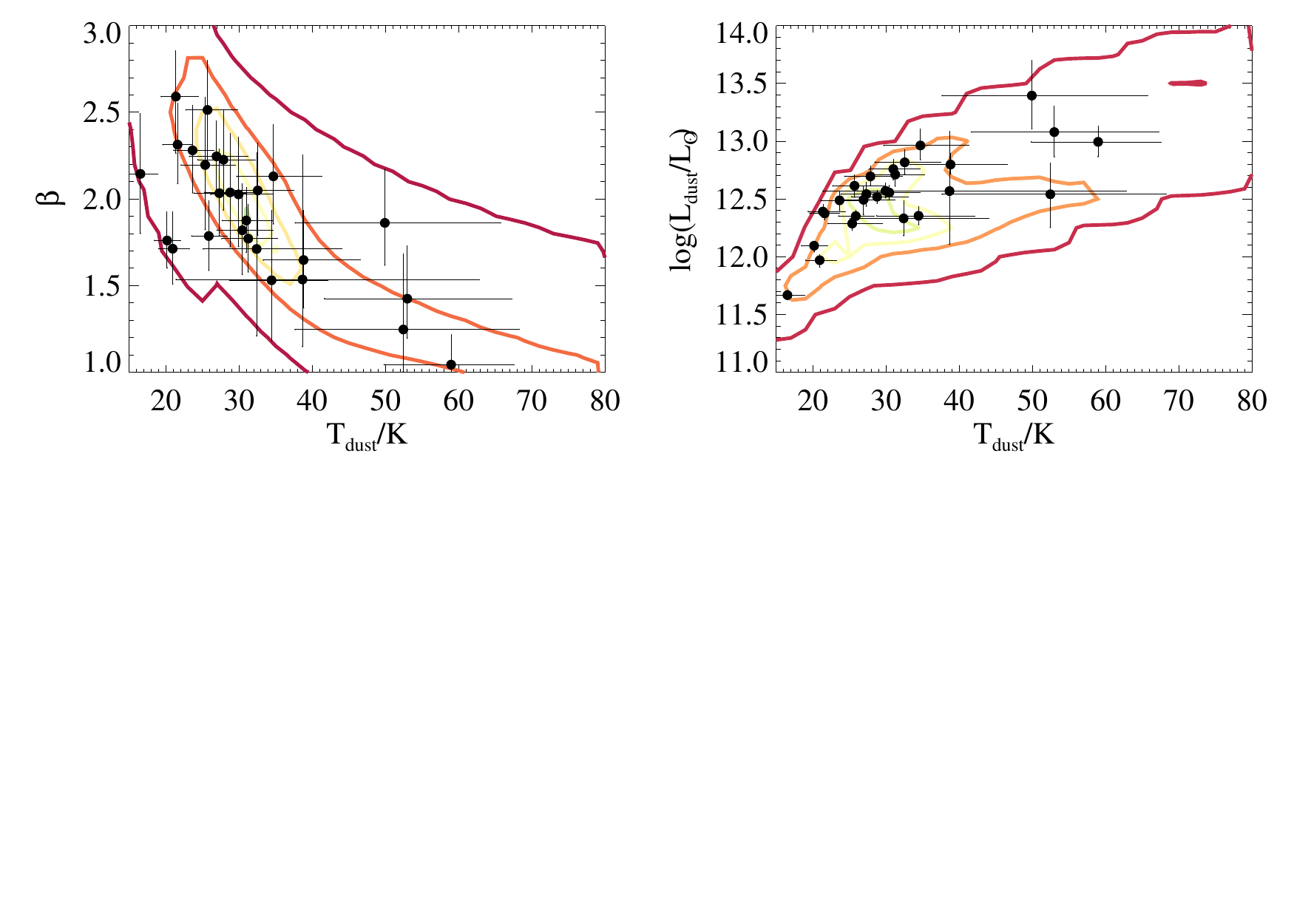}
\caption{Relation between dust temperature and emissivity index (left-hand panel), and dust luminosity (right-hand panel) for the 27 sources in our well-sampled subset (optically-thin case). The median-likelihood estimates for each source are plotted as black circles, with the errors given by the 16th--84th percentile ranges of the likelihood distributions. The contours show areas of equal probability of the stacked joint likelihood distributions. We note that we find similar correlations in the general opacity scenario (though the correlations are shifted due to the systematic offsets in \tdust\ described in Section~\ref{results_genopacity}).}
\label{fig:2dstacked}
\end{figure*}

The precision of our constraints can be quantified by the width (i.e., the 16th-84th percentile range) of the posterior likelihood distributions. For our well-sampled subset, we constrain the dust emissivity index to within $\pm0.25$. This error on $\beta$ depends strongly on the S/N in Band 4 and also to some extent in Band 7. The dust temperature \tdust\ is constrained to $\pm5$~K, and we achieve median precisions in dust luminosity and dust mass of $\pm0.10$ and $\pm0.08$~dex, respectively. We note that these are the median statistical errors on the fits within the context of assuming an optically-thin model, and systematics associated with the choice of dust opacity are not included. We discuss possible systematic uncertainties by comparing with results from the general opacity fits in Section~\ref{results_genopacity}.

To analyze the distribution of physical properties of our sources, we show, in Fig.~\ref{fig:stacked}, the stacked posterior likelihood distributions of $\beta$, \tdust, \ldust, and \mdust. We compare the results obtained for the well-sampled subset with those for the rest of the sample, which we include to check if we can conclude something about those sources in a statistical sense. The individual posterior distributions for the sources that are not in the well-sampled subset may not contain much information, however stacking them may reveal if any regions of the parameter space are preferred (as opposed to the completely unconstrained case where we would retrieve our flat priors). These sources seem to peak at slightly lower temperatures, however the probability extends to higher dust temperatures. Their dust luminosities and masses seem to be typically lower than for the well-sampled subset, as expected given their fainter sub-millimeter fluxes, and the emissivity indexes peak at similar values, although we note that the stacked posterior is much flatter.

For our well-sampled subset, we find median values of $\tdust=30^{+14}_{-8}$~K, and $\beta=1.9\pm0.4$ for the stacked likelihood distributions (the errors are the 16th-84th percentile ranges of the stacked distributions). The emissivity indexes are consistent with typical values $\beta=1.5-2.0$ measured in local galaxies and predicted by dust models (see Section~\ref{sec:comparison} for a more detailed discussion).
Fig.~\ref{fig:2dstacked} shows that there is a negative correlation between the dust temperatures and emissivity indexes of our well-sampled sources ($r_\mathrm{S}=-0.66$, $3.4\sigma$), and a positive correlation between dust luminosity and temperature ($r_\mathrm{S}=0.73$, $3.7\sigma$; a similar correlation is also found, e.g., by \citealt{daCunha2015}). To check the robustness of these correlations, we also stack the joint posterior likelihood distributions, shown as contours. The fact that the peak of the stacked likelihood distributions follows the observed correlations between the median-likelihood estimates shows that these correlations are robust\footnote{The median-likelihood estimate can be deceiving if a parameter is unconstrained: in that case the median will be the median of the prior; but the posterior would show that the parameter is unconstrained because it would resemble the prior, in our case, a flat distribution.}. This correlation between \tdust\ and $\beta$ could be caused to some extent by the intrinsic degeneracy between these two parameters, however we show in Section~\ref{sec:accuracy} that our parameter estimates are accurate enough for these sources (because the data we use break the degeneracy), so it is likely that the correlation is real (see also Section~\ref{sec:comparison}).

\subsubsection{Comparison with the general opacity model results}
\label{results_genopacity}

In Figs.~\ref{fig:aless002p1} and \ref{fig:aless001p1}, we also plot the results of our Bayesian fitting when using the general dust opacity model described in Section~\ref{general_opacity}, allowing \lthick\ to vary between 60 and 140\mic. In this case, the dust temperatures and dust masses are unconstrained even for our well-sampled subset, due to the strong degeneracy between \tdust\ and \lthick\ (Fig.~\ref{fig:seds}): the widths of the likelihood distributions for these parameters are much larger than in the optically-thin case. The best-fit model and median of the likelihood distribution of these parameters do not converge. This is another indication that the current observations are not sufficient to constrain a general opacity model where \lthick\ is allowed to vary. More observations sampling the SED peak and measurements of the size of the dust emission region would help constrain this parameter more precisely. With the current data, while the quality of the SED fits is comparable between the optically-thin approximation and the general opacity model, these two different modeling assumptions can lead to large differences in the inferred dust temperatures and, consequently, on the dust masses. 

We also compare the stacked likelihood distributions obtained using the general opacity model for our well-sampled subset in Fig.~\ref{fig:stacked}. In this case, the dust temperatures are more poorly constrained even for the ensemble of 27 well-sampled galaxies (due to the degeneracy with \lthick), but tend to peak at higher values. The dust masses tend to be lower than in the general opacity case, due to the higher \tdust. The recovered dust luminosities are very similar to those in the optically-thin scenario, and, importantly, so are the dust emissivity indexes. This shows that our estimates of $\beta$ for this subsample are robust regardless of whether an optically-thin or general opacity scenario are adopted.

\begin{figure*}
\centering
\includegraphics[trim={0cm 0cm 0cm 0cm},width=0.75\textwidth]{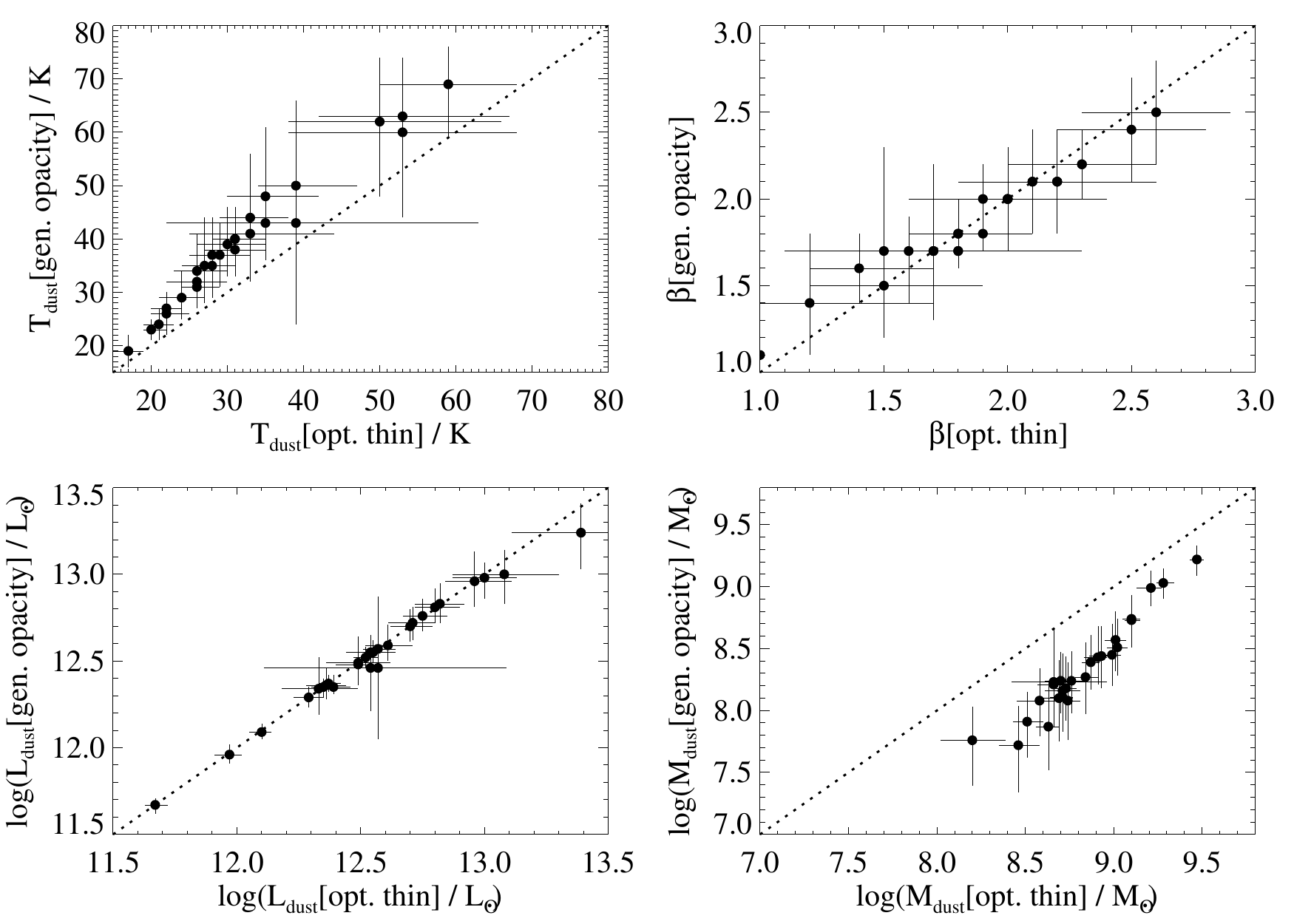}
\caption{Comparison between the median-likelihood estimates of dust temperatures (\tdust), emissivity indexes ($\beta$), luminosities (\ldust), and masses (\mdust) of our well-sampled subset of 27 ALESS SMGs when fitting their dust emission using optically-thin dust models and a general opacity model with $\lthick=60-140\mic$ (Section~\ref{general_opacity}). In each panel, the dotted line represents the identity. The error bars are the 16th-84th percentiles of the posterior likelihood distribution for each source. The parameters derived using different model assumptions are well correlated, however, we find strong systematic differences in \tdust\ and \mdust. \ldust\ and $\beta$ are robust against model assumptions on the dust opacity, as also shown in Fig.~\ref{fig:stacked}.}
\label{fig:parameters_ot_go}
\end{figure*}

In Fig.~\ref{fig:parameters_ot_go}, we compare the median-likelihood estimates of dust physical parameters derived using the optically-thin assumption with those derived using the general opacity model. The parameters are well-correlated, however, we find strong systematic offsets in the derived dust temperatures: the general opacity scenario produces \tdust\ typically $\sim$10 K warmer than the optically-thin case (see also \citealt{Simpson2017}). This leads to a strong offset in the inferred dust masses, which are typically 0.5 dex (i.e., a factor of 3) lower in the general opacity scenario than in the optically-thin case (the same offsets are also seen in the medians of the stacked likelihood distributions in Fig.~\ref{fig:stacked}). These differences can have very strong implications when interpreting inferred dust masses in the context of chemical evolution and dust production models, especially at high-redshift, where current models required substantial ISM growth to account for large inferred dust masses (e.g., \citealt{Rowlands2014,Mancini2015,Ugne2020,Ugne2021}). Nevertheless, we find that the inferred dust luminosities and emissivity indexes seem quite robust against dust model assumptions, with no systematic offsets. We checked that these differences would be more pronounced if we chose to include models where the dust remains optically thick beyond 140\mic. However, in that case, the parameters with the highest systematic offsets, \tdust\ and \mdust\ would remain almost unconstrained with the current data due to the strong degeneracy with \lthick, therefore the systematic offsets would be mostly a result of the prior. Given our calculation in Section~\ref{general_opacity}, very high values of \lthick\ seem unlikely (though see \citealt{Riechers2013}, who claim $\lthick\simeq200\mic$ in high-redshift, lensed SMGs). We caution that very optically-thick dust could lead to even more significant differences in the inferred dust masses (factors of 10 and more).

We note that, with the available data, we cannot distinguish which of these scenarios, optically-thin dust or general opacity with $\lthick=60-140\mic$, is more likely: the best-fit model probabilities of the optically thin to general opacity scenario are very close. Given the expected dust masses and sizes \citep{Hodge2016,Gullberg2019}, we expect the optically-thick models to be appropriate at least for some of the sources. Nevertheless, in the following sections, unless otherwise stated, we will focus on the optically-thin scenario, since these models are better constrained, and they are more widely used, thus facilitating comparisons with the literature.

\subsection{Accuracy of our fitting method}
\label{sec:accuracy}

Before we move on to interpreting our results in a broader context, we must first establish the accuracy of our derived dust parameters.
Previous studies focussing on modified black body fitting of the dust emission in compact galactic dust cores (e.g., \citealt{Shetty2009,Juvela2012,Juvela2013}) show that a correlation between \tdust\ and $\beta$ can be introduced artificially to some extent by performing $\chi^2$ minimization on noisy data. Bayesian methods such as ours are shown to produce more robust results (see also, e.g., \citealt{Kelly2012}) because they treat uncertainties rigorously and self-consistently, and as a result they do not produce spurious correlations between the parameters due to measurement uncertainties.

To address these issues in the context of this work, in this section we generate a suite of mock dust emission models to quantify the accuracy to which we expect the dust properties to be recovered from our fits. Again, we assume that dust emission in galaxies is isothermal and optically thin, and it can be described by simple modified black bodies with dust temperature \tdust\ and emissivity index $\beta$ (eq.~\ref{mbb}). We generate a library of $5,000$ models with dust temperatures uniformly distributed between 15 and 80 K, $\beta$ between 1 and 3, dust luminosity $\log(\ldust/\lsun)$ between 11.3 and 13.5 (a luminosity range similar to that of our SMGs; \citealt{daCunha2015}). To simulate our observables, we place these models at different redshifts using a Gaussian distribution centered at $z=2.7$, similar to the redshift distribution of our ALESS sources \citep{Simpson2014,daCunha2015}. For each model, we randomly draw a set of \tdust, \ldust, $\beta$ and $z$ from these distributions, and we compute the predicted (`observed') flux of each model in the same bands as for our observations, i.e., the {\it Herschel}/SPIRE bands, and ALMA Band 7 at 870\mic\ and Band 4 at 2mm. We include the effects of the CMB in the observed fluxes as prescribed in \cite{daCunha2013b}. We then perturb these observed fluxes by $\pm15\%$ to mimic our typical observational errors, and assign observational uncertainties to each flux that are similar to those of our real observed galaxies. That is, we assume: (i) a random signal-to-noise ratio drawn between 4 and 6 in the SPIRE 500- and 350-\mic\ bands, and between 4 and 10 in the 250-\mic\ band \citep{Swinbank2014}; (ii) a random signal-to-noise ratio drawn between 3 and 15 in ALMA Band 4, and
(iii) an ALMA Band 7 signal-to-noise that is correlated with the band 4 S/N in the same way as our observations (which yields a distribution between 4 and 40).\footnote{We note that, strictly speaking, the signal-to-noise should correlate with the actual model fluxes however, we choose to set our simulation up this way because it allows us to perform the test with realistic errors but at the same time without limiting the parameter space of our models.} We then fit our mock observations in the same way as we fit the actual observations in Section~\ref{modeling}.

\begin{figure*}
\centering
\includegraphics[width=0.6\textwidth,angle=90]{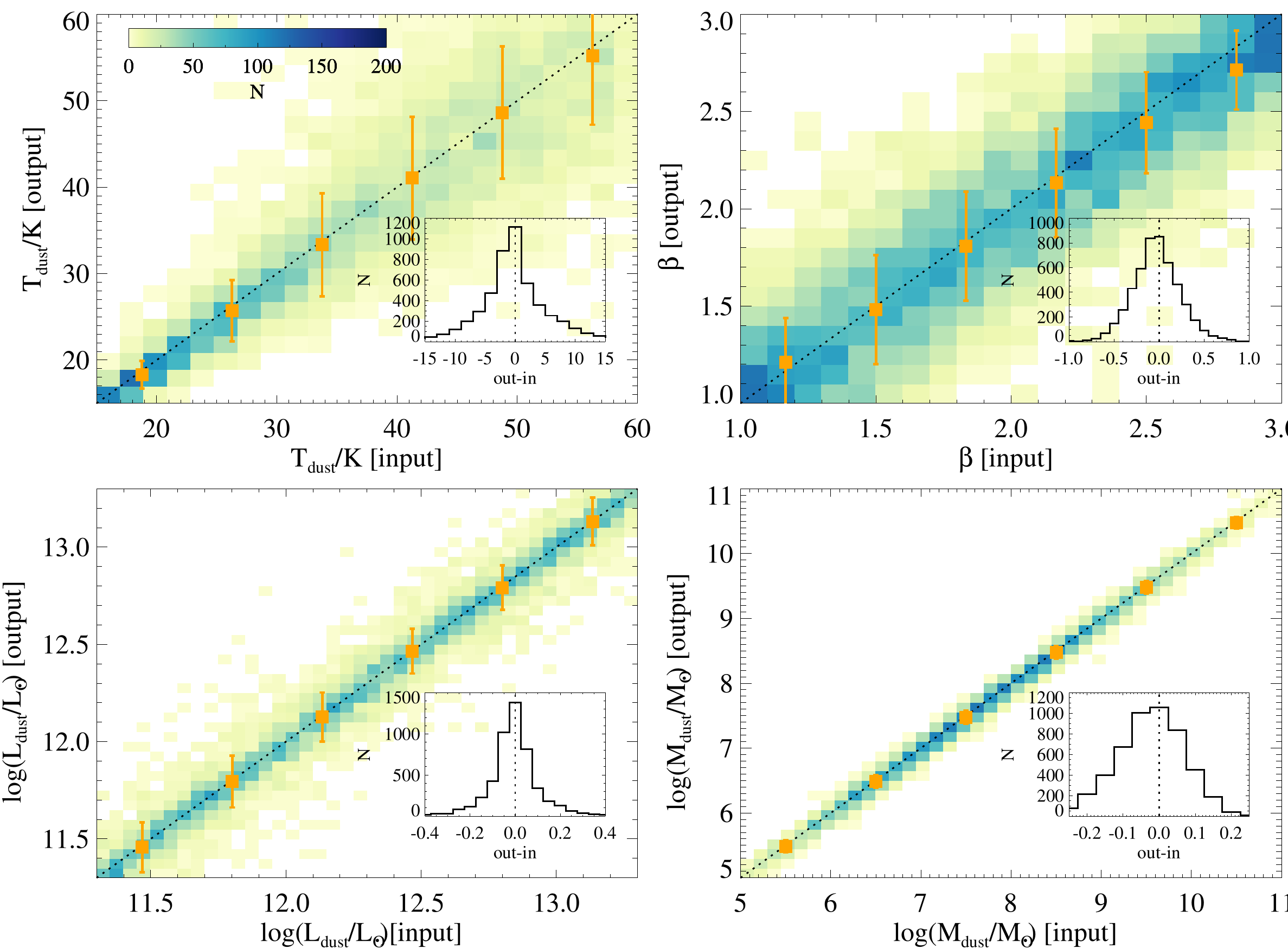}
\caption{Accuracy of our fitting method in constraining the dust temperature (top left), emissivity index (top right), luminosity (bottom left) and mass (bottom right). For each of our $5,000$ mock models (described in Section~\ref{sec:accuracy}), we plot the density of input parameter versus output median-likelihood estimate from our fitting. The black dotted line is the identity line, and the orange squares show the median in different bins, with the error bar showing the standard deviation in each bin. The inset plots show the overall distribution of the output value minus the input.}
\label{fig:accuracy}
\end{figure*}

Fig.~\ref{fig:accuracy} shows the accuracy of the derived dust properties of our mock observations.
The input parameters are well recovered by our method for SEDs that have as many observational constraints as our well-sampled subset of ALESS sources. Our method typically recovers the input dust luminosities to $\pm0.13$\,dex, the dust temperatures to within $\pm6$\,K, the emissivity indexes are recovered within $\pm0.27$, and the dust masses to $\pm0.1$\,dex (these values are the standard deviations of the difference between input and output values). The systematics are minimal, with median offsets (i.e., difference between output and input values) of $-0.007$\,dex for \ldust, $-0.5$\,K for \tdust, $-0.04$ for $\beta$, and $-0.017$\,dex for \mdust. It is worth noting that the accuracy of \tdust\ estimates decreases for higher temperatures, which is expected because the peak of the dust SED shifts to lower wavelengths and is less well sampled by the SPIRE bands. However even in that regime the output values are still distributed around the input values (i.e., no significant systematics). The results of this test allow us to conclude that the dust parameters obtained for our well-sampled subset are robust, at least if assumptions about dust opacity are correct (see discussion below). We check that when we perform this test, we start with uncorrelated \tdust\ and $\beta$, and the results are also uncorrelated, therefore we conclude that our found correlation between \tdust\ and $\beta$ is not likely a result of fitting noisy data using our method.
In Appendix~\ref{app_accuracy_data}, we use similar simulations to show that our dust parameters would not be accurate enough if the fits did not include Band 4 data or {\it Herschel} data, which is why we chose to focus mainly on our well-sampled subset for which both are available.

It has to be noted that the self-consistency check described above assumes that we are using the correct model for the dust emission, but that may not be the case if, for example, dust is more optically-thick than assumed. Therefore, in Appendix~\ref{app_accuracy_genopacity}, we test the accuracy of our derived parameters in the case where the input mock observations are generated using the general opacity model, but the models used to fit those observations include only optically-thin models. Fig.~\ref{fig:accuracy_go} shows that, at least for the range of \lthick\ explored (between 60\mic\ and 140\mic), no significant biases are found in $\beta$ and \ldust\ when using the incomplete assumption of optically-thin dust. However, not surprisingly, systematic offsets arise for \tdust\ (because this parameter depends on the peak of SED, which is most affected by the optical depth effects), and \mdust\ (because it depends strongly on \tdust). The offsets correlate strongly with \lthick, in the sense that the longer \lthick, the more the model deviates from the optically-thin assumption, as well as with \tdust, since hotter dust peaks at shorter wavelengths, and therefore is more affected by the dust opacity assumptions.

\section{Discussion}
\label{discussion}

\subsection{Selection effects}
\label{sec:selection}

In the previous section, we show that our dust parameters are robust for our well-sampled subset (bar systematics due to opacity modeling assumptions), and therefore the measured correlation between \tdust\ and $\beta$ is not likely introduced by our fitting method. In this section, we explore the impact of selection effects on our derived dust properties and their correlations. We use the library of (optically-thin) dust models from the previous section, for which we have, for each dust model at a given redshift, the predicted fluxes in the {\it Herschel} and ALMA bands. Then, we apply the same flux selections to those models as in our observations. We apply two selections: (1) all models that would be detected above $4\sigma$ in our 2mm observations ($\sigma=0.053$~mJy~beam$^{-1}$), plus $\ge3.5\sigma$ detection in Band 7 (870\mic), where $\sigma=0.4$~mJy~beam$^{-1}$ (this is the original ALESS selection; \citealt{Hodge2013}), and (2) all models that would obey our `well-sampled' subset detection criteria, i.e., models that obey the previous criterion and that would have a $\ge4\sigma$ detection in {\it Herschel}/SPIRE at 250\mic, where $\sigma=3$~mJy \citep{Swinbank2014}.

\begin{figure*}
\centering
\includegraphics[trim={0cm 0cm 0cm 0cm},width=0.72\textwidth,angle=90]{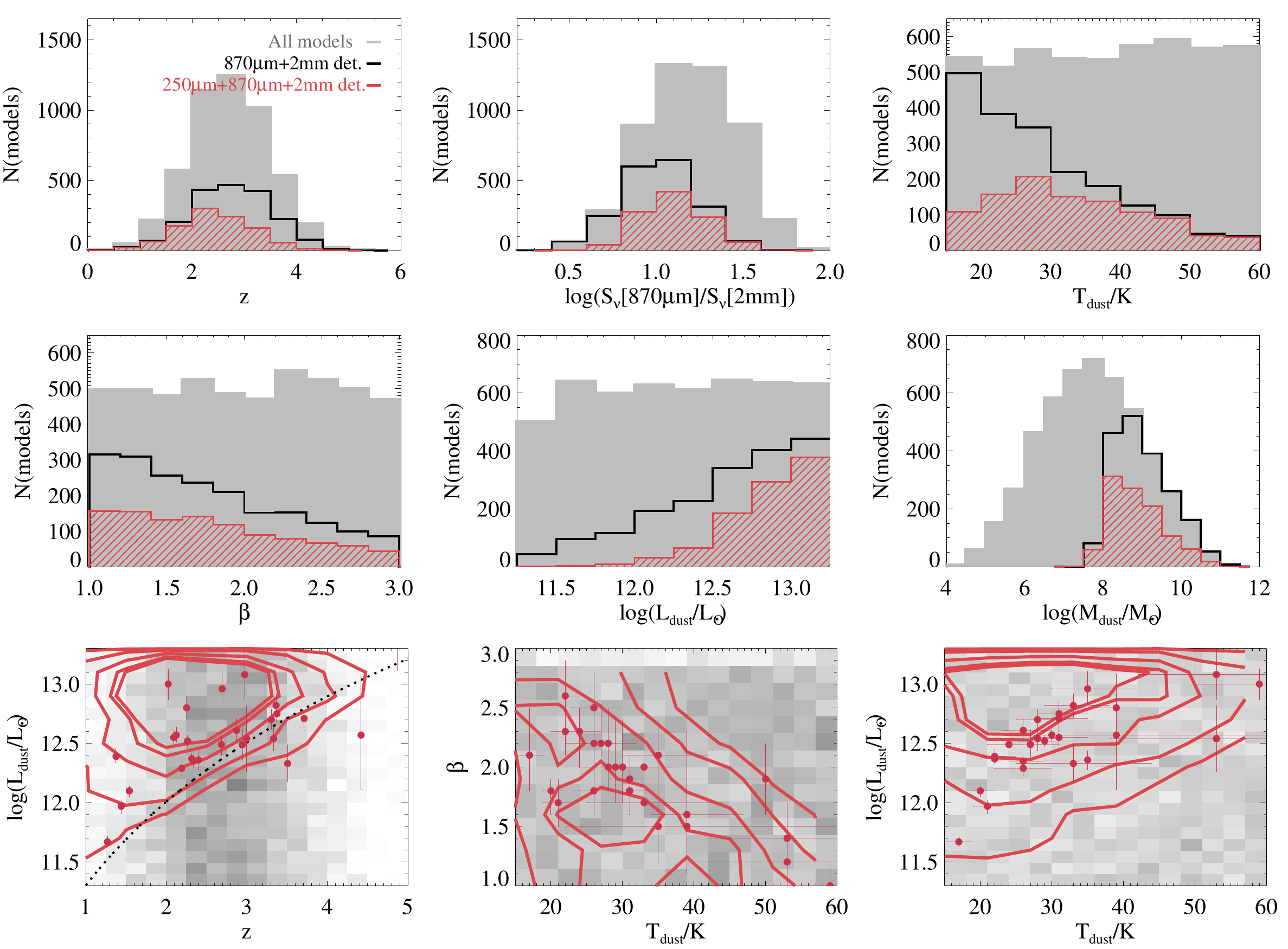}
\caption{Distribution of physical parameters and observables in our library of dust models depending on our selection criteria. The model library was set up to have a similar redshift distribution to that of our ALESS SMGs, and uniformly sample the dust temperature, emissivity index, and dust luminosity. {\it Top and middle panels:}  The grey histograms show the full library of models, while the other two histograms show the effect of applying flux cuts similar to the ones in our ALESS observations: {\em black}: $\ge3.5\sigma$ detection in Band 7 (870\mic), where $\sigma=0.4$~mJy~beam$^{-1}$ \citep{Hodge2013} and $\ge4\sigma$ detection in Band 4 (2mm), where $\sigma=0.053$~mJy~beam$^{-1}$ (this paper); {\em red}: same as previous, i.e. strong detections in Bands 7 and 4, plus $\ge4\sigma$ detection in {\it Herschel}/SPIRE at 250\mic, where $\sigma=3$~mJy \citep{Swinbank2014}: this criterion is similar to the selection of our 27 `well-sampled' ALESS sources. {\it Bottom panels:} Relation between physical properties in our library of dust emission models. The grey shading shows the density of models in the full library, showing there are no {\it a priori} correlations between the physical parameters. The red contours show the distribution of models to which flux cuts mimicking the selection of our `well-sampled' subset have been applied. For comparison, we also plot the median-likelihood estimates of the properties of our well-sampled subset as red circles. The dotted line in the first panel shows a $\sim(1+z)^4$ selection on luminosity; this is mostly imposed by the 250-\mic\ flux cut.}
\label{fig:histsim}
\end{figure*}

Figure~\ref{fig:histsim} shows that the first selection criterion (Band 4 and 7 detections) selects sources with higher dust luminosities, lower dust temperatures, and higher emissivity indexes: all these tend to boost the 2\,mm flux. The result is that the models with the highest dust masses are selected (the 2\,mm flux cut is effectively a dust mass selection; 870\mic\ is also close to a dust mass selection, as shown by \citealt{Ugne2020}). This flux cut also reproduces the typical 870\mic-to-2mm flux ratio of our observations and the fact that we preferentially detect the brightest 870\,\mic\ sources. When the second criterion is applied (i.e., detections in Bands 4 and 7, plus at 250\mic), we tend to select models at lower redshifts, we lose a large fraction of low \tdust\ models, and the $\beta$ distribution becomes flatter. These changes are mainly caused by the 250-\mic\ flux cut, which imposes a luminosity limit with redshift and selects preferentially warmer dust. The resulting \tdust\ distribution looks similar to the stacked \tdust\ posterior for our well-sampled subset (Fig.~\ref{fig:stacked}), peaking at about 30~K.

The bottom panels of Fig.~\ref{fig:histsim} show the relations between several physical properties in our model library, and how they change when we apply our selection criteria for the well-sampled subset. These panels clearly show that our flux cuts exclude some regions of the parameter space, making {\it a priori} uncorrelated properties appear correlated for SMG samples when applying certain multi-wavelength flux cuts, as illustrated by the red contours. Most notably we would not detect high-$\beta$, high-\tdust\ sources according to our selection, which could be affecting our derived $\beta$-\tdust\ correlation, at least to some extent. This is driven by the ALMA selection at 870\mic\ and 2mm: imposing solely a 250-\mic\ flux cut would retrieve models spanning the full \tdust-$\beta$ parameter space of our library. We note that we would still detect low $\beta$ and low \tdust\ sources, but the 250-\mic\ flux cut means that it would be less likely, and indeed such sources do not seem to exist in our sample. Nevertheless our models are more strongly correlated than the distribution of models shown by the red contours (with Spearman rank correlation coefficients $r_\mathrm{S}=-0.69$ and $r_\mathrm{S}=-0.47$ for our well-sampled subset and the models, respectively), indicating that the correlation between \tdust\ and $\beta$ is not necessarily caused only by the sample selection (the same can be said of the correlation between \tdust\ and \ldust, with Spearman rank correlation coefficients of $r_\mathrm{S}=0.73$ and $r_\mathrm{S}=0.17$ for our sources and the models shown in red, respectively.)


\subsection{Evolution in $S_\nu\mathrm{[870\mic]}/S_\nu\mathrm{[2mm]}$?}
\label{ratio_evolution}

\begin{figure}
\centering
\includegraphics[trim={0cm 0cm 0cm 0cm},width=0.5\textwidth]{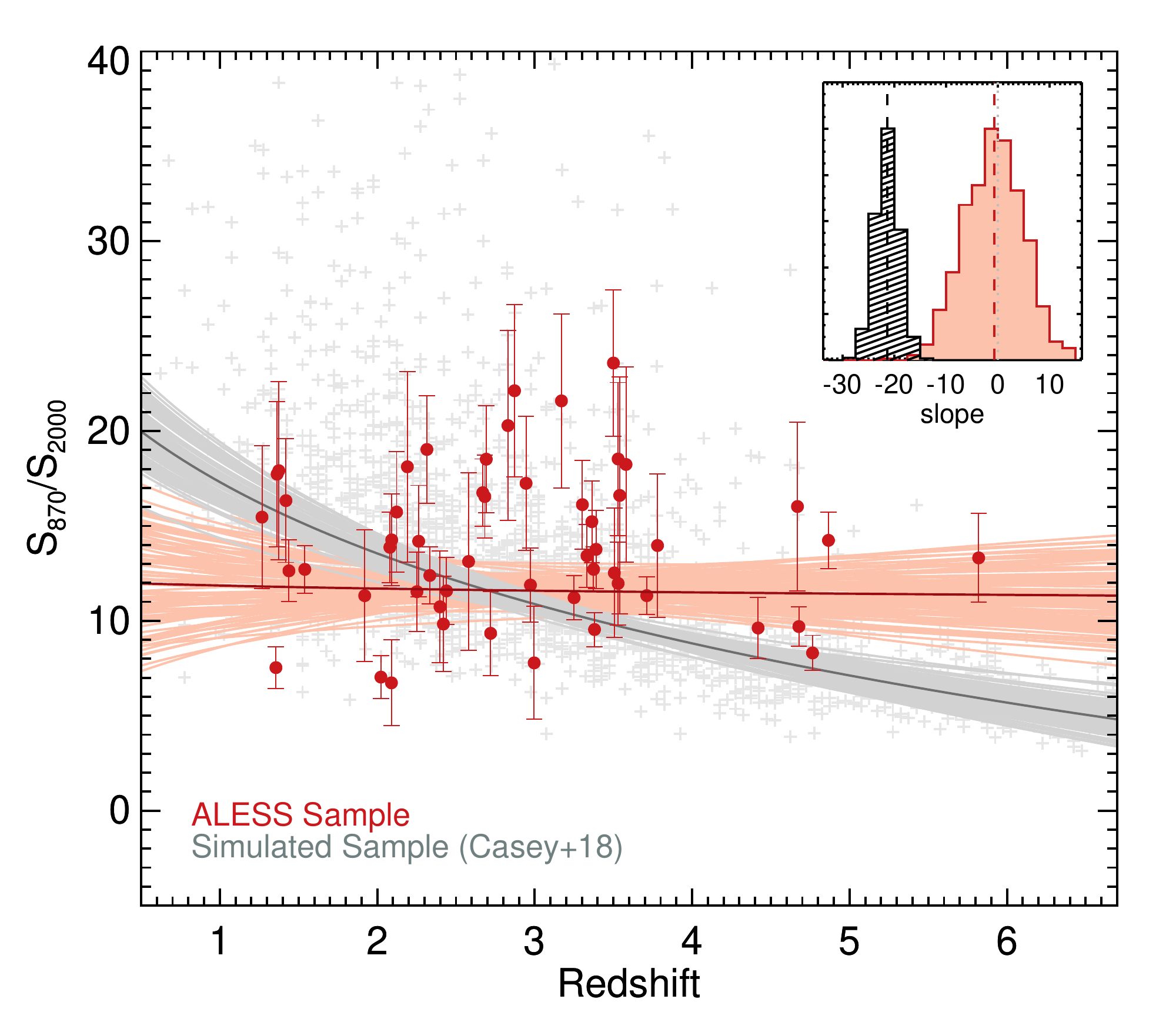}
\caption{A comparison of the measured slope in $\log(1+z)$ versus $S_\nu\mathrm{[870\mic]}/S_\nu\mathrm{[2mm]}$ between the ALESS sample (we include only sources with $\sn\ge4$ here) and simulated sources
from \citet{Casey2018a} as projected in $z$ versus $S_\nu\mathrm{[870\mic]}/S_\nu\mathrm{[2mm]}$.  The light gray points show the distribution of extracted
flux ratios for simulated sources after accounting for the statistical
noise of observations of the ALESS sample both at 870\mic\ and 2mm.
The gray lines represent best fits to Monte Carlo draws from the
simulated sources, chosen to have similar signal-to-noise and redshift
distribution to the ALESS data.  The best-fit line to ALESS data is
shown in dark red, with orange lines representing the fits to
bootstrapped subsamples of ALESS.  The inset plot shows the
distribution of best-fit slopes to the simulated data (black) in
relation to the measured slope (orange, with bootstrapped
uncertainties); the deviation between the two is significant at the $\sim3.3\sigma$ level.}
\label{fig:cmcmodel}
\end{figure}

In Section~\ref{colours}, we find very little evidence for evolving $S_\nu\mathrm{[870\mic]}/S_\nu\mathrm{[2mm]}$ with redshift for the sources with robust 2\,mm fluxes.
Here we employ the empirical backward evolution model of \citet{Casey2018a}
to test whether or not a non-evolving ratio of 870\mic-to-2mm ratio would be expected in this ALESS dataset.  For example,
if the signal-to-noise ratio of individual detections are low, then an
evolution in $S_\nu\mathrm{[870\mic]}/S_\nu\mathrm{[2mm]}$ might not be observable unless
the sample is sufficiently large.  Alternatively, if there is
substantial evolution in the average dust temperature of SMGs towards
higher redshifts (as suggested by recent theoretical works,
e.g., \citealt{Behrens2018,Ma2019,Liang2019,Sommovigo2020}; though we note such evolution is not seen when comparing similar luminosity samples, e.g., \citealt{Ugne2020}), then the degeneracy between
redshift and dust temperature could result in a non-evolving $S_\nu\mathrm{[870\mic]}/S_\nu\mathrm{[2mm]}$ ratio.

To test the ability with which our sample can constrain
the evolution of this ratio, we draw on $\sim$2,000 simulated sources extracted
from output photometry of the \citet{Casey2018a} Model B. This model produces a `dust-rich early Universe', where dusty star-forming galaxies dominate 
the cosmic star formation history from at $1.5\lesssim z\lesssim6.5$; however we note
that the luminosity function model does not affect the
redshift-dependence of (sub-)millimeter colors.  An implicit assumption of
this model is that there is no redshift evolution in \tdust\, though there is a non-evolving luminosity-dependence of
infrared luminosity with some intrinsic scatter.  The emissivity spectral
index is assumed to be fixed at $\beta=1.8$. In other
words, redshift evolution of $S_\nu\mathrm{[870\mic]}/S_\nu\mathrm{[2mm]}$ is a fundamental assumption of
the model, with scatter caused by variation in dust temperature
and observational noise.  Simulated sources take into account the effect of the CMB as described
in \cite{daCunha2013b}, and have analogous flux
densities and detection signal-to-noise ratios as the ALESS sample at 870\mic.

We downsample the $\sim$2,000 simulated sources to the sample size and redshift distribution of the ALESS
sources in Monte Carlo trials.  Within the redshift range of the majority of the ALESS data
($1.2<z<5.0$), the redshift evolution of the millimeter color should
be linear in $\log(1+z)$ versus $S_\nu\mathrm{[870\mic]}/S_\nu\mathrm{[2mm]}$. Thus, we
fit linear relationships, weighted by signal-to-noise in $S_\nu\mathrm{[870\mic]}/S_\nu\mathrm{[2mm]}$, between these quantities both for our simulated
Monte Carlo samples from the \citet{Casey2018a} model and the ALESS
data, bootstrapping the latter to constrain the uncertainty in the
inferred relationship.  We show the results in Fig.~\ref{fig:cmcmodel}, where the
measured slope of the relation deviates from the median
relation of the model Monte Carlo trials at the $3.3\sigma$ level (inset histogram showing
bootstrapped slope measured for ALESS sample versus slope of
simulated sources).  The flatter measured slope in ALESS
hints at a possible breakdown in the assumptions of the model, i.e., that there may be real evolution in dust temperatures and/or
$\beta$ from $1\lesssim z\lesssim 5$. This is consistent with the fact that we find a correlation between 
\tdust\ and redshift for our well-sampled subset ($r_\mathrm{S}=0.51$, $2.6\sigma$), though that correlation is likely to be due, at least in part,
to selection effects (see also, \citealt{Ugne2020}). We find no correlation between $\beta$ and redshift for the
well-sampled subset ($r_\mathrm{S}=-0.13$, $<1\sigma$).

What is clear from this test is that the statistical significance with which the relationship between redshift and  $S_\nu\mathrm{[870\mic]}/S_\nu\mathrm{[2mm]}$ is
flatter than expected is low, due in part to our relatively small
sample size and low S/N on individual sources.  A robust
characterization of the evolution of this ratio, and its implications
on measured dust temperatures and emissivity indexes will require
samples $\sim5\times$ larger than currently exist.

\subsection{Comparison with other measurements and theoretical predictions}
\label{sec:comparison}

\begin{figure*}
\centering
\includegraphics[width=0.65\textwidth]{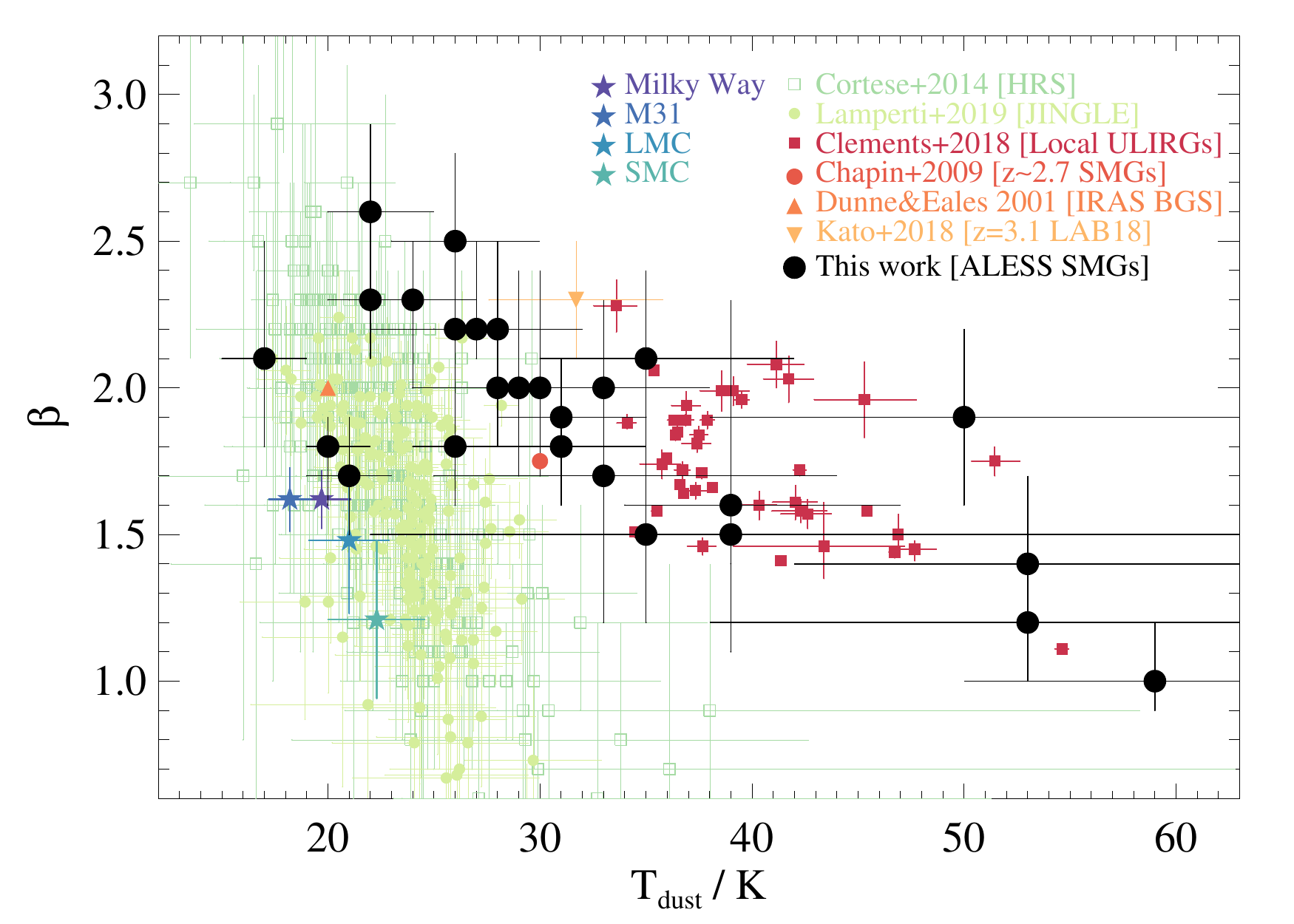}
\caption{Dust temperature against dust emissivity index for our SMGs and other samples in the literature: local star-forming galaxies from the {\it Herschel} Reference Sample (HRS; \citealt{Cortese2014}) and the JCMT dust and gas In Nearby Galaxies Legacy Exploration (JINGLE) sample \citep{Lamperti2019}; local ULIRGs from the {\it Herschel} ULIRG Survey (HERUS; \citealt{Clements2018}). The remaining symbols show either single-source measurements (e.g., a $z=3.1$ Lyman-$\alpha$ blob from \citealt{Kato2018}) or average values for galaxy samples (average value for a sample of 5 SMGs at $z\sim2.7$ observed with AzTEC at 1.1mm from \citealt{Chapin2009}; average value for a sample of 32 local infrared galaxies from the {\it IRAS} Bright Galaxy Sample observed with SCUBA in the sub-millimeter by \citealt{Dunne2001}). The stars show measurements from {\it Planck} for the Milky Way \citep{Planck2014}, M31 \citep{Planck2015}, and the Magellanic Clouds \citep{Planck2011}.}
\label{fig:samples}
\end{figure*}

In Fig.~\ref{fig:samples}, we compare our derived dust temperatures and emissivity indexes with those obtained for other samples of galaxies, mostly in the local Universe where such measurements have been possible with {\it Herschel}, SCUBA and {\it Planck}. The exact values of these parameters, especially of the dust temperatures, might not be directly comparable between samples, as they can depend strongly on the SED fitting method and wavelength coverage of the data. However, some general conclusions may be derived from this comparison. Both local ULIRGs and high-redshift SMGs show on average warmer dust temperatures, presumably because of their intense star formation activities which produce stronger interstellar radiation fields heating the dust grains. Both moderately star-forming local galaxies from the {\it Herschel} Reference sample \citep{Cortese2014} and the JCMT dust and gas In Nearby Galaxies Legacy Exploration (JINGLE) sample \citep{Lamperti2019}, as well as local ultra-luminous infrared galaxies (ULIRGs) from the HERUS sample \citep{Clements2018} show a negative correlation between \tdust\ and $\beta$, and the slope of this correlation is similar to ours in the case of the HERUS sample, though we note that our ALESS SMGs extend to lower temperatures than local ULIRGs (a well-known difference between local ULIRGs and high-redshift SMGs; e.g., \citealt{Symeonidis2013,Swinbank2014}). More importantly, our SMGs seem to span a similar range in dust emissivity index as those local samples, with values ranging between $\beta\simeq1.0$ and 2.5.

Our results are broadly consistent with the best previous constraints of the dust emissivity index in $z\sim2.7$ SMGs using AzTEC 1.1mm data, which yielded an average value of $\beta\simeq1.75^{+0.25}_{-0.75}$ for a sample of 5 SMGs \citep{Chapin2009}. \cite{Magnelli2012} also found on average $\beta=2.0\pm0.2$ for a sample of 19 SMGs observed with all {\it Herschel}/PACS and SPIRE bands, plus at least one detection longwards of 1mm, from modeling their SEDs with more complex, multi-temperature models. Their quoted error on $\beta$ is smaller than ours, likely due to better sampling of the SEDs. With our sample we not only confirm this value to be a reasonable average value for high-redshift dusty star-forming galaxies ($\beta\simeq1.5-2$ is often assumed when modeling sources at both low and high redshift; e.g., \citealt{Scoville2016,Galliano2018}), but we also show that there can be significant variation from galaxy to galaxy. We note that the recent ALMA measurement of $\beta=2.3$ in a $z=3.1$ galaxy by \cite{Kato2018} is entirely consistent with the range of values we find for our SMGs, and therefore that source is not necessarily an outlier in dust properties.

\cite{Aravena2016} derive $\beta=1.3\pm0.2$ (using the Rayleigh-Jeans approximation) for a source individually detected at 1.3mm and 3mm in the ASPECS ALMA spectroscopic deep field pilot \citep{Walter2016}, and even lower values ($\beta=0.9\pm0.4$) for a stacked sample. Taken at face value these results interestingly could indicate different dust emissivity indexes (and therefore different dust grain properties) in samples of galaxies of lower infrared luminosity than SMGs that are also selected a different wavelengths. However, in Appendix~\ref{appendix_rj}, we demonstrate that the Rayleigh-Jeans approximation is invalid even for the $\lambda>1$mm ASPECS bands at $z>1$, and that could lead to the emissivity index being underestimated by between 0.2 and 1.0, depending on the exact redshifts and dust temperatures of the sources (Fig.~\ref{rj_test_longwave}).

The average dust emissivity index of our SMGs, $\beta=1.9\pm0.4$, is entirely consistent with the predictions of theoretical dust models of interstellar dust (e.g., \citealt{Draine1984,Draine2011,Koehler2015}), which predict values typically between 1 and 2.5, depending on grain composition. For example, \cite{Koehler2015} predict $\beta\sim1.5$ for core mantle grains in the diffuse ISM, and an increase to $\beta\sim1.8-2.0$ towards denser environments due to coagulation and accretion onto the dust grains which change their optical properties (see also, e.g., \citealt{Jones2017}). 

The variation in $\beta$ is connected with a variation in temperature given the strong correlation between these two properties. A negative correlation between $\beta$ and \tdust\ is found not only from global SED fits of galaxies but it is also well known from fits to the dust emission of galactic molecular clouds and cold cores (e.g., \citealt{Dupac2003,Desert2008,Paradis2010,Juvela2013}). The existence of this correlation is robust against uncertainties introduced by SED fitting methods, wavelength coverage of the data, etc., and is thought to be a result of a change in the intrinsic emissivity properties of dust grains with temperature. Indeed, laboratory measurements of interstellar dust grain analogues such as amorphous silicate grains also show a negative correlation between $\beta$ and temperature (\citealt{Agladze1996, Mennella1998,Boudet2005,Meny2007,Coupeaud2011}; see also Appendix B of \citealt{Inoue2020}). Additional radiative transfer effects could also contribute to this correlation by introducing departures to the isothermal approximation, in the sense that a mix of dust temperatures along the line of sight can contribute to broaden the SEDs and hence lower the inferred $\beta$ values (e.g., \citealt{Shetty2009,Coupeaud2011,Koehler2015}).  

We conclude that while it is challenging to connect our measured emissivity indexes directly to evolutionary grain models, the typical values found for our SMGs are broadly consistent with local measurements, theoretical dust models, and laboratory measurements. Therefore there does not seem to be a strong evolution of the dust properties in dusty star-forming galaxies between $z\sim2.7$ and $z\sim0$, at least for massive, chemically-evolved galaxies such as SMGs.

\section{Summary \& Conclusions}
\label{conclusions}

In this paper we present new ALMA 2\,mm continuum observations of the 99 870\mic-selected SMGs in the ALESS sample. We find that at the sensitivity of our observations (53\textmu Jy~beam$^{-1}$ on average), we detect 70 sources (i.e., 71\% of our sample), including 53 above $4\sigma$.
We model the dust emission in the ALESS SMGs using isothermal, optically-thin models of varying dust temperatures and emissivity indexes, and we also explore more general opacity models where we vary the wavelength at which the dust becomes optically-thin. In order to break degeneracies in the models, we include {\it Herschel} fluxes sampling near the peak of the dust emission. This allows us to derive robust dust properties for a subset of 27 ALESS SMGs with well-sampled SEDs and spectroscopic redshifts.
The main conclusions of our work are the following:
\begin{itemize}
\item At the depth of our observations, the detection rate of our SMGs at 2mm is practically independent of redshift, and brighter SMGs (with higher 870\mic\ fluxes and stellar masses) are the most likely to be detected. 29 of the ALESS sources remain undetected in our 2\,mm observations; this could be due in part because of the depth of our observations, or they may have peculiar dust emission properties.

\item For the sources for which we measure a 2mm flux density (i.e., $\sn>1.5$), the median 870\mic-to-2mm flux ratio is $14\pm5$. The median flux ratio for the entire ALESS sample, including upper 2mm upper limits, is $17\pm9$. For our detected sources, we find that this flux ratio does not depend on redshift, which could point to an evolution of dust temperatures and/or emissivity indexes with redshift. This needs to be further explored with larger samples.

\item We demonstrate that the 870\mic-to-2mm flux ratio alone cannot be used to derive the emissivity index of the dust $\beta$ using the Rayleigh-Jeans approximation. This approximation can lead to a severe underestimation of the true value of $\beta$.

\item For a subset of 27 SMGs for which we have well-sampled SEDs and spectroscopic redshifts, we estimate $\beta$, \tdust, \ldust, and \mdust\ using isothermal, optically-thin dust models. Including 2\,mm observations allows us to constrain $\beta$ very precisely, to within $\pm0.25$ for each source, which leads to \mdust\ being constrained to $\pm0.08$~dex, a two-fold improvement in precision when compared to not including 2\,mm fluxes in the fitting (in which case $\beta$ remains unconstrained). The median dust mass of our well-sampled SMGs from is $M_\mathrm{dust}=5.8^{+5.9}_{-2.4}\times10^8\,M_\odot$, consistent with previous estimates using MAGPHYS which also assume optically-thin dust and $\beta=1.5-2$ (though multiple temperature components), and no fluxes beyond 870\mic\ \citep{daCunha2015}. We note however, that this value and the quoted precision on individual \mdust\ measurements were obtained using the optically-thin dust assumption, and additional systematics need to be taken into account, as these values depend strongly on the assumed dust opacity. We find that in our general opacity scenario we would obtain dust masses that are typically three times lower. Our current observations are not sufficient to distinguish between optically-thin and optically-thick dust. Better sampling of the dust emission near its peak, specifically through high-\sn\, high-frequency ALMA observations, could help, as well as more measurements of the dust emission sizes of the sources. Uncertainties in the normalization of the dust emissivity per unit mass can lead to further $\sim$ factor of 3 systematic uncertainties in the dust masses (e.g.,~\citealt{Galliano2018,Bianchi2019,Inoue2020}).

\item We measure a median value of $\beta=1.9\pm0.4$ for the dust emissivity index of our subset of 27 well-sampled SMGs. Contrary to the dust mass estimates, this result is robust against dust opacity assumptions in the models, and it is consistent with previous estimates for local galaxies and with expectations from theoretical modeling and laboratory measurements of interstellar dust grain analogs.

\item We find a negative correlation between $\beta$ and \tdust\ that is not introduced by our method, but could be introduced, to some extent, by selection effects, since sources with simultaneously high $\beta$ and high \tdust\ would not be selected in our well-sampled dataset. However, this correlation is also found for local galaxy samples and Milky Way dust clouds selected in different ways, and is predicted by theoretical dust models, therefore selection effects might not be the only cause for the correlation found in our sample.

\end{itemize}

This work confirms the dust emissivity index between 1.5 and 2.0 that is typically assumed in most high-redshift studies. This implies that the properties of dust at $z\simeq1-3$ are similar to the properties of local galaxies. This is true at least for SMGs, which based on their relatively high stellar masses ($M_\ast\gtrsim10^{10}\,M_\odot$; \citealt{daCunha2015,Ugne2020}) are likely to have already reached solar metallicities in their ISM. We speculate that therefore they are likely to have reached a critical metallicity for their dust grain evolution to be happening mainly through ISM growth (see \citealt{Ugne2021})  which is also thought to be the dominant mechanism in the Milky Way and other present-day galaxies (e.g., \citealt{Asano2013}). A larger number of robust measurements of the emissivity index of high-redshift galaxies of lower masses/metallicities with ALMA would help establish if there is an evolution of the dust grain properties in galaxies with a less chemically-evolved ISM.

\section*{Acknowledgements}
We thank the anonymous referee for their detailed reading and comments which helped us improve this paper. We thank Isabella Lamperti for sending us the JINGLE emissivity indexes and temperatures.
EdC gratefully acknowledges the Australian Research Council as the recipient of a Future Fellowship (project FT150100079) and the ARC Centre of Excellence for All Sky Astrophysics in 3 Dimensions (ASTRO 3D; project CE170100013).
JH and HA acknowledge support of the VIDI research program with project number 639.042.611, which is (partly) financed by the Netherlands Organisation for Scientific Research (NWO).
MK acknowledges support from the International Max Planck Research School for Astronomy and Cosmic Physics at Heidelberg University (IMPRS-HD).
IRS and AMS acknowledge support from STFC (ST/T000244/1).
KK acknowledge support from the Knut and Alice Wallenberg Foundation and from the Swedish Research Council (2015-05580).
HD acknowledges financial support from the Spanish Ministry of Science, Innovation and Universities (MICIU) under the 2014 Ramon y Cajal program RYC-2014-15686 and AYA2017-84061-P, the later one co-financed by FEDER (European Regional Development Funds).
This paper makes use of the following ALMA data: ADS/JAO.ALMA\#2015.1.00948.S. ALMA is a partnership of ESO (representing its member states), NSF (USA) and NINS (Japan), together with NRC (Canada), NSC and ASIAA (Taiwan), and KASI (Republic of Korea), in cooperation with the Republic of Chile. The Joint ALMA Observatory is operated by ESO, AUI/NRAO and NAOJ.

\appendix

\section{Source positions and ALMA fluxes}
\label{app_sources}

\startlongtable 
\begin{deluxetable*}{lcccccCC}
\tablecaption{Positions and measured and deboosted ALMA fluxes in Band 7 (\citealt{Hodge2013}) and Band 4 (this work) of our 99 ALESS sources. \label{table_sources} }\tablehead{
\colhead{Source} & \colhead{R.A.} & \colhead{Declination} & \colhead{$S_\nu^\mathrm{measured}$[$870\mic$]} & \colhead{$S_\nu^\mathrm{measured}$[$2$mm]} & S/N[2mm] & \colhead{$S_\nu^\mathrm{deboosted}$[$870\mic$]} & \colhead{$S_\nu^\mathrm{deboosted}$[$2$mm]}\\
 \colhead{ID} & \colhead{(J2000)} & \colhead{(J2000)} & \colhead{(mJy)} & \colhead{(mJy)} & & \colhead{(mJy)} & \colhead{(mJy)}}
\startdata
  ALESS001.1 & 03:33:14.46 & $-$27:56:14.5 & $6.7\pm0.5$ & $0.69\pm0.05$ & 13.8 & 6.6\pm0.5 & 0.68\pm0.05\\
  ALESS001.2 & 03:33:14.41 & $-$27:56:11.6 & $3.5\pm0.4$ & $0.25\pm0.05$ & 5.0 & 3.3\pm0.4 & 0.21\pm0.05\\
  ALESS001.3 & 03:33:14.18 & $-$27:56:12.3 & $1.9\pm0.4$ & $<0.05$ & $<1.5$ & 1.6\pm0.4 & -\\
  ALESS002.1 & 03:33:02.69 & $-$27:56:42.8 & $3.8\pm0.4$ & $0.24\pm0.05$ & 4.8 & 3.6\pm0.4 & 0.20\pm0.05\\
  ALESS002.2 & 03:33:03.07 & $-$27:56:42.9 & $4.2\pm0.7$ & $0.30\pm0.05$ & 6.0 & 3.7\pm0.7 & 0.27\pm0.05\\
  ALESS003.1 & 03:33:21.50 & $-$27:55:20.3 & $8.3\pm0.4$ & $0.66\pm0.05$ & 13.2 & 8.2\pm0.4 & 0.65\pm0.05 \\
  ALESS005.1 & 03:31:28.91 & $-$27:59:09.0 & $7.8\pm0.7$ & $0.49\pm0.05$ & 9.8 & 7.6\pm0.7 & 0.47\pm0.05\\
  ALESS006.1 & 03:32:56.96 & $-$28:01:00.7 & $6.0\pm0.4$ & $0.49\pm0.05$ & 9.8 & 5.9\pm0.4 & 0.48\pm0.05 \\
  ALESS007.1 & 03:33:15.42 & $-$27:45:24.3 & $6.1\pm0.3$ & $0.35\pm0.05$ & 7.0 & 6.1\pm0.3 & 0.33\pm0.05\\
  ALESS009.1 & 03:32:11.34 & $-$27:52:11.9 & $8.8\pm0.5$ & $0.63\pm0.05$ & 12.6 & 8.7\pm0.5 & 0.61\pm0.05\\
  ALESS010.1 & 03:32:19.06 & $-$27:52:14.8 & $5.2\pm0.5$ & $0.34\pm0.05$ & 6.8 & 5.0\pm0.5 & 0.31\pm0.05\\
  ALESS011.1 & 03:32:13.85 & $-$27:56:00.3 & $7.3\pm0.4$ & $0.46\pm0.05$ & 9.2 & 7.2\pm0.4 & 0.44\pm0.05\\
  ALESS013.1 & 03:32:48.99 & $-$27:42:51.8 & $8.0\pm0.6$ & $0.71\pm0.05$ & 14.2 & 7.8\pm0.6 & 0.70\pm0.05\\
  ALESS014.1 & 03:31:52.49 & $-$28:03:19.1 & $7.5\pm0.5$ & $0.79\pm0.05$ & 15.8 & 7.4\pm0.5 & 0.77\pm0.05\\
  ALESS015.1 & 03:33:33.37 & $-$27:59:29.6 & $9.0\pm0.4$ & $0.55\pm0.05$ & 11.0 & 8.9\pm0.4 & 0.53\pm0.05\\
  ALESS015.3 & 03:33:33.59 & $-$27:59:35.4 & $2.0\pm0.5$ & $<0.05$ & $<1.5$ & 1.5\pm0.5 & -\\
  ALESS017.1 & 03:32:07.30 & $-$27:51:20.8 & $8.4\pm0.5$ & $0.67\pm0.05$ & 13.4 & 8.3\pm0.5 & 0.65\pm0.05\\
  ALESS018.1 & 03:32:04.88 & $-$27:46:47.7 & $4.4\pm0.5$ & $0.39\pm0.05$ & 7.8 & 4.2\pm0.5 & 0.36\pm0.05\\
  ALESS019.1 & 03:32:08.26 & $-$27:58:14.2 & $5.0\pm0.4$ & $0.30\pm0.05$ & 6.0 & 4.9\pm0.4 & 0.26\pm0.05\\
  ALESS019.2 & 03:32:07.89 & $-$27:58:24.1 & $2.0\pm0.5$ & $0.15\pm0.05$ & 3.0 & 1.5\pm0.5 & -\\
  ALESS022.1 & 03:31:46.92 & $-$27:32:39.3 & $4.5\pm0.5$ & $0.33\pm0.05$ & 6.6 & 4.3\pm0.5 & 0.30\pm0.05\\
  ALESS023.1 & 03:32:12.01 & $-$28:05:06.5 & $6.7\pm0.4$ & $0.51\pm0.05$ & 10.2 & 6.6\pm0.4 & 0.49\pm0.05\\
  ALESS023.7 & 03:32:11.92 & $-$28:05:14.0 & $1.8\pm0.5$ & $<0.05$ & $<1.5$ & 1.3\pm0.5 & -\\
  ALESS025.1 & 03:31:56.88 & $-$27:59:39.3 & $6.2\pm0.5$ & $0.31\pm0.05$ & 6.2 & 6.1\pm0.4 & 0.27\pm0.05\\
  ALESS029.1 & 03:33:36.90 & $-$27:58:09.3 & $5.9\pm0.4$ & $0.48\pm0.05$ & 9.6 & 5.8\pm0.4 & 0.46\pm0.05\\
  ALESS031.1 & 03:31:49.79 & $-$27:57:40.8 & $8.1\pm0.4$ & $0.72\pm0.05$ & 14.4 & 8.0\pm0.4 & 0.71\pm0.05\\
  ALESS035.1 & 03:31:10.51 & $-$27:37:15.4 & $4.4\pm0.3$ & $0.39\pm0.05$ & 7.8 & 4.3\pm0.3 & 0.36\pm0.05\\
  ALESS035.2* & 03:31:10.22 & $-$27:37:18.1 & $1.4\pm0.4$ & $<0.05$ & $<1.5$ & 1.0\pm0.4 & -\\
  ALESS037.1 & 03:33:36.14 & $-$27:53:50.6 & $2.9\pm0.4$ & $0.32\pm0.05$ & 6.4 & 2.7\pm0.4 & 0.29\pm0.05\\
  ALESS037.2 & 03:33:36.36 & $-$27:53:48.3 & $1.6\pm0.4$ & $<0.05$ & $<1.5$ & 1.2\pm0.4 & -\\
  ALESS039.1 & 03:31:45.03 & $-$27:34:36.7 & $4.3\pm0.3$ & $0.39\pm0.05$ & 7.8 & 4.2\pm0.3 & 0.36\pm0.05\\
  ALESS041.1 & 03:31:10.07 & $-$27:52:36.7 & $4.9\pm0.6$ & $0.20\pm0.05$ & 4.0 & 4.6\pm0.6 & -\\
  ALESS041.3 & 03:31:10.30 & $-$27:52:40.8 & $2.7\pm0.8$ & $0.10\pm0.05$ & 4.0 & 1.8\pm0.8 & -\\
  ALESS043.1 & 03:33:06.64 & $-$27:48:02.4 & $2.3\pm0.4$ & $0.14\pm0.05$ & 2.8 & 2.0\pm0.4 & -\\
  ALESS045.1 & 03:32:25.26 & $-$27:52:30.5 & $6.0\pm0.5$ & $0.31\pm0.05$ & 6.2 & 5.8\pm0.5 & 0.27\pm0.05\\
  ALESS049.1 & 03:31:24.72 & $-$27:50:47.1 & $6.0\pm0.7$ & $0.36\pm0.05$ & 7.2 & 5.7\pm0.7 & 0.33\pm0.05\\
  ALESS049.2 & 03:31:24.47 & $-$27:50:38.1 & $1.8\pm0.5$ & $<0.06$ & $<1.5$ & 1.3\pm0.5 & -\\
  ALESS051.1 & 03:31:45.06 & $-$27:44:27.3 & $4.7\pm0.4$ & $0.29\pm0.05$ & 5.8 & 4.6\pm0.4 & 0.26\pm0.05\\
  ALESS055.1 & 03:33:02.22 & $-$27:40:35.5 & $4.0\pm0.4$ & $0.53\pm0.05$ & 10.6 & 3.8\pm0.4 & 0.51\pm0.05\\
  ALESS055.2* & 03:33:02.16 & $-$27:40:41.3 & $2.4\pm0.6$ & $<0.05$ & $<1.5$ & 1.8\pm0.6 & - \\
  ALESS055.5 & 03:33:02.35 & $-$27:40:35.4 & $1.4\pm0.4$ & $0.10\pm0.05$ & 2.0 & 1.0\pm0.4 & - \\
  ALESS057.1 & 03:31:51.92 & $-$27:53:27.1 & $3.6\pm0.6$ & $<0.05$ & $<1.5$ & 3.2\pm0.6 & -\\
  ALESS059.2 & 03:33:03.82 & $-$27:44:18.2 & $1.9\pm0.4$ & $0.27\pm0.05$ & 5.4 & 1.6\pm0.4 & 0.23\pm0.05\\
  ALESS061.1 & 03:32:45.87 & $-$28:00:23.4 & $4.3\pm0.5$ & $0.44\pm0.05$ & 8.8 & 4.1\pm0.4 & 0.42\pm0.05\\
  ALESS063.1 & 03:33:08.45 & $-$28:00:43.8 & $5.6\pm0.3$ & $0.42\pm0.05$ & 8.4 & 5.5\pm0.3 & 0.40\pm0.05\\
  ALESS065.1 & 03:32:52.27 & $-$27:35:26.3 & $4.2\pm0.4$ & $0.17\pm0.05$ & 3.4 & 4.1\pm0.4 & -\\
  ALESS066.1 & 03:33:31.93 & $-$27:54:09.5 & $2.5\pm0.5$ & $0.08\pm0.05$ & 1.6 & 2.1\pm0.5 & -\\
  ALESS067.1 & 03:32:43.20 & $-$27:55:14.3 & $4.5\pm0.4$ & $0.31\pm0.05$ & 6.2 & 4.4\pm0.4 & 0.28\pm0.05\\
  ALESS067.2 & 03:32:43.02 & $-$27:55:14.7 & $1.7\pm0.4$ & $<0.05$ & $<1.5$ & 1.3\pm0.4 & -\\
  ALESS068.1 & 03:32:33.33 & $-$27:39:13.6 & $3.7\pm0.6$ & $0.30\pm0.05$ & 6.0 & 3.3\pm0.6 & 0.27\pm0.05\\
  ALESS069.1 & 03:31:33.78 & $-$27:59:32.4 & $4.9\pm0.6$ & $0.26\pm0.05$ & 5.2 & 4.6\pm0.6 & 0.23\pm0.05\\
  ALESS069.2 & 03:31:34.13 & $-$27:59:28.9 & $2.4\pm0.6$ & $0.10\pm0.05$ & 2.0 & 1.8\pm0.6 & -\\
  ALESS069.3* & 03:31:33.97 & $-$27:59:38.3 & $2.1\pm0.6$ & $<0.05$ & $<1.5$ & 1.4\pm0.6 & -\\
  ALESS070.1 & 03:31:44.02 & $-$27:38:35.5 & $5.2\pm0.4$ & $0.39\pm0.05$ & 7.8 & 5.1\pm0.4 & 0.36\pm0.05\\
  ALESS071.1 & 03:33:05.65 & $-$27:33:28.2 & $2.9\pm0.6$ & $0.17\pm0.05$ & 3.4 & 2.4\pm0.6 & -\\
  ALESS071.3 & 03:33:06.14 & $-$27:33:23.1 & $1.4\pm0.4$ & $<0.05$ & $<1.5$ & 1.0\pm0.4 & -\\
  ALESS072.1 & 03:32:40.40 & $-$27:37:58.1 & $4.9\pm0.5$ & $0.38\pm0.05$ & 7.6 & 4.7\pm0.5 & 0.35\pm0.05\\
  ALESS073.1 & 03:32:29.29 & $-$27:56:19.7 & $6.1\pm0.5$ & $0.73\pm0.05$ & 14.6 & 5.9\pm0.5 & 0.72\pm0.05\\
  ALESS074.1 & 03:33:09.15 & $-$27:48:17.2 & $4.6\pm0.7$ & $0.20\pm0.05$ & 4.0 & 4.2\pm0.7 & -\\
  ALESS075.1 & 03:31:27.19 & $-$27:55:51.3 & $3.2\pm0.4$ & $0.14\pm0.05$ & 2.8 & 3.0\pm0.4 & -\\
  ALESS075.4 & 03:31:26.57 & $-$27:55:55.7 & $1.3\pm0.4$ & $<0.06$ & $<1.5$ & 0.8\pm0.4 & -\\
  ALESS076.1 & 03:33:32.34 & $-$27:59:55.6 & $6.4\pm0.6$ & $0.57\pm0.05$ & 11.4 & 6.2\pm0.6 & 0.45\pm0.05\\
  ALESS079.1 & 03:32:21.14 & $-$27:56:27.0 & $4.1\pm0.4$ & $0.36\pm0.05$ & 7.2 & 3.9\pm0.4 & 0.33\pm0.05\\
  ALESS079.2 & 03:32:21.60 & $-$27:56:24.0 & $2.0\pm0.4$ & $<0.06$ & $<1.5$ & 1.7\pm0.4 & -\\
  ALESS079.4* & 03:32:21.18 & $-$27:56:30.5 & $1.8\pm0.5$ & $<0.05$ & $<1.5$ & 1.3\pm0.5 & -\\
  ALESS080.1 & 03:31:42.80 & $-$27:48:36.9 & $4.0\pm0.9$ & $0.28\pm0.05$ & 5.6 & 3.2\pm0.9 & 0.24\pm0.05 \\
  ALESS080.2 & 03:31:42.62 & $-$27:48:41.0 & $3.5\pm0.9$ & $<0.05$ & $<1.5$ & 2.6\pm0.9 & -\\
  ALESS082.1 & 03:32:54.00 & $-$27:38:14.9 & $1.9\pm0.5$ & $0.10\pm0.05$ & 2.0 & 1.4\pm0.5 & -\\
  ALESS083.4 & 03:33:08.71 & $-$28:05:18.5 & $1.4\pm0.4$ & $0.54\pm0.05$ & 10.8 & 0.96 & -\\
  ALESS084.1 & 03:31:54.50 & $-$27:51:05.6 & $3.2\pm0.6$ & $0.28\pm0.05$ & 5.6 & 2.7\pm0.6 & 0.24\pm0.05\\
  ALESS084.2 & 03:31:53.85 & $-$27:51:04.4 & $3.2\pm0.8$ & $<0.06$ & $<1.5$ & 2.4\pm0.8 & -\\
  ALESS087.1 & 03:32:50.88 & $-$27:31:41.5 & $1.3\pm0.4$ & $0.17\pm0.05$ & 3.4 & 0.8\pm0.4 & -\\
  ALESS087.3* & 03:32:51.27 & $-$27:31:50.7 & $2.4\pm0.6$ & $<0.06$ & $<1.5$ & 1.8\pm0.6 & -\\
  ALESS088.1 & 03:31:54.76 & $-$27:53:41.5 & $4.6\pm0.6$ & $0.32\pm0.05$ & 6.4 & 4.3\pm0.6 & 0.28\pm0.05\\
  ALESS088.2* & 03:31:55.39 & $-$27:53:40.3 & $2.1\pm0.5$ & $<0.06$ & $<1.5$ & 1.6\pm0.5 & -\\
  ALESS088.5 & 03:31:55.81 & $-$27:53:47.2 & $2.9\pm0.7$ & $<0.08$ & $<1.5$ & 2.2\pm0.7 & -\\
  ALESS088.11 & 03:31:54.95 & $-$27:53:37.6 & $2.5\pm0.7$ & $<0.06$ & $<1.5$ & 1.7\pm0.7 & -\\
  ALESS092.2 & 03:31:38.14 & $-$27:43:43.4 & $2.4\pm0.7$ & $<0.05$ & $<1.5$ & 1.6\pm0.7 & -\\
  ALESS094.1 & 03:33:07.59 & $-$27:58:05.8 & $3.2\pm0.5$ & $0.19\pm0.05$ & 3.8 & 2.9\pm0.5 & -\\
  ALESS098.1 & 03:31:29.92 & $-$27:57:22.7 & $4.8\pm0.6$ & $0.30\pm0.06$ & 5.0 & 4.5\pm0.6 & 0.25\pm0.06\\
  ALESS099.1* & 03:32:51.82 & $-$27:55:33.6 & $2.1\pm0.4$ & $<0.05$ & $<1.5$ & 1.8\pm0.4 & -\\
  ALESS102.1 & 03:33:35.60 & $-$27:40:23.0 & $3.1\pm0.5$ & $0.09\pm0.05$ & 1.8 & 2.8\pm0.5 & -\\
  ALESS103.3* & 03:33:25.04 & $-$27:34:01.1 & $1.4\pm0.4$ & $<0.05$ & $<1.5$ & 1.0\pm0.4 & -$\\
  ALESS107.1 & 03:31:30.50 & $-$27:51:49.1 & $1.9\pm0.4$ & $0.25\pm0.06$ & 4.2 & 1.6\pm0.4 & 0.20\pm0.06\\
  ALESS107.3 & 03:31:30.72 & $-$27:51:55.7 & $1.5\pm0.4$ & $0.08\pm0.06$ & $<1.5$ & 1.1\pm0.4 & -\\
  ALESS110.1 & 03:31:22.66 & $-$27:54:17.2 & $4.1\pm0.5$ & $0.26\pm0.05$ & 5.2 & 3.9\pm0.5 & 0.21\pm0.05\\
  ALESS110.5 & 03:31:22.96 & $-$27:54:14.4 & $2.4\pm0.6$ & $<0.06$ & $<1.5$ & 1.8\pm0.6 & -\\
  ALESS112.1 & 03:32:48.86 & $-$27:31:13.3 & $7.6\pm0.5$ & $0.42\pm0.05$ & 8.4 & 7.5\pm0.5 & 0.39\pm0.05\\
  ALESS114.1 & 03:31:50.49 & $-$27:44:45.3 & $3.0\pm0.8$ & $<0.06$ & $<1.5$ & 2.2\pm0.8 & -\\
  ALESS114.2 & 03:31:51.11 & $-$27:44:37.3 & $2.0\pm0.5$ & $<0.07$ & $<1.5$ & 1.5\pm0.5 & -\\
  ALESS115.1 & 03:33:49.70 & $-$27:42:34.6 & $6.9\pm0.4$ & $0.48\pm0.06$ & 8.0 & 6.8\pm0.4 & 0.45\pm0.06\\
  ALESS116.1 & 03:31:54.32 & $-$27:45:28.9 & $3.1\pm0.5$ & $0.22\pm0.05$ & 4.4 & 2.8\pm0.5 & 0.17\pm0.05\\
  ALESS116.2 & 03:31:54.44 & $-$27:45:31.5 & $3.4\pm0.6$ & $0.13\pm0.05$ & 2.6 & 3.0\pm0.6 & -\\
  ALESS118.1 & 03:31:21.92 & $-$27:49:41.4 & $3.2\pm0.5$ & $0.31\pm0.06$ & 5.2 & 2.9\pm0.5 & 0.27\pm0.06\\
  ALESS119.1 & 03:32:56.64 & $-$28:03:25.2 & $8.3\pm0.5$ & $0.38\pm0.05$ & 7.6 & 8.2\pm0.5 & 0.35\pm0.05\\
  ALESS122.1 & 03:31:39.54 & $-$27:41:19.7 & $3.7\pm0.4$ & $0.53\pm0.06$ & 8.8 & 3.5\pm0.4 & 0.50\pm0.06\\
  ALESS124.1 & 03:32:04.04 & $-$27:36:06.4 & $3.6\pm0.6$ & $0.36\pm0.06$ & 6.0 & 3.2\pm0.6 & 0.32\pm0.06\\
  ALESS124.4 & 03:32:03.89 & $-$27:36:00.1 & $2.2\pm0.6$ & $<0.06$ & $<1.5$ & 1.6\pm0.6 & -\\
  ALESS126.1 & 03:32:09.61 & $-$27:41:07.7 & $2.2\pm0.5$ & $0.1\pm0.06$ & 1.7 & 1.7\pm0.5 & -\\
\enddata
\tablecomments{For the 2mm sources with $1.5\le\sn<4$, we use the measured fluxes; there is no need to correct these for flux boosting because we measure the flux directly at the known position of the 870\mic\ source.
}
\end{deluxetable*}

\begin{deluxetable*}{lccccccccc}
\tablecaption{Significant detections in our 2\,mm maps that do not have a counterpart in the ALESS main catalog. \label{extra_sources}}
\tablehead{
\colhead{Field} & \colhead {ID} & \colhead{R.A.} & \colhead{Declination} & \colhead{$S_\nu$[$870\mic$]} & \colhead{$S^\mathrm{measured}_\nu$[$2$mm]} & S/N[2mm] & PB[2mm] & \colhead{$S^\mathrm{deboosted}_\nu$[$2$mm]} & IRAC?$^a$ \\
 & & \colhead{(J2000)} & \colhead{(J2000)} &  \colhead{(mJy)} & \colhead{(mJy)} & & & \colhead{(mJy)} &
 }
\startdata
LESS 5   & ALESS005\_2mm.1 & 03:31:29.96 & $-$27:58:47.04 & PB$<0.2$ & $0.22\pm0.05$ & 4.4 & 0.319 & $0.17\pm0.05$ &N\\
LESS 59 & ALESS059\_2mm.1 & 03:33:02.89 & $-$27:44:33.27 & PB$<0.2$ & $0.45\pm0.05$ & 9.0 & 0.548 & $0.43\pm0.05$ & Y \\
LESS 59 & ALESS059\_2mm.2 & 03:33:03.37 & $-$27:44:26.20 & PB$<0.2$ & $0.23\pm0.05$ & 4.6 & 0.857 & $0.19\pm0.05$ & N \\
LESS 76 & ALESS076\_2mm.1 & 03:33:34.13 & $-$27:59:48.75 & PB$<0.2$ & $0.25\pm0.05$ & 5.0 & 0.356 & $0.21\pm0.05$ & N \\
LESS 83$^b$ & ALESS083.1 & 03:33:09.41 & $-$28:05:30.90 & $1.4\pm0.4$ & $0.32\pm0.05$ & 6.4 & 0.666 & $0.29\pm0.05$ & Y \\
LESS 92 & ALESS092\_2mm.1 & 03:31:37.00 & $-$27:43:41.26 & PB$<0.2$ & $0.26\pm0.05$ & 5.2 & 0.710 & $0.21\pm0.05$  & Y \\
LESS 114$^c$ & ALESS114\_2mm.1 & 03:31:50.30 & $-$27:44:46.84 & $<3.9$ & $0.30\pm0.05$ & 6.0 & 0.995 & $0.25\pm0.05$ &Y \\
LESS 114 & ALESS114\_2mm.2 & 03:31:48.85 & $-$27:44:29.80 & PB$<0.2$ & $0.25\pm0.06$ & 4.2 & 0.303 & $0.20\pm0.06$ & N \\
\enddata
\tablecomments{$^a$ {\it Spitzer}/IRAC 3.6\mic\ counterpart within 1-arcsec in the ECDFS catalog of \cite{Damen2011}; $5\sigma$ limiting magnitude is $m_\mathrm{AB}=23.8$.
$^b$ This is one of the supplementary ALESS sources that had been excluded from the main sample for being outside the primary beam of the Band 7 map \cite{Hodge2013}. However, this source does have a measured 870\mic\ flux in ALESS, and our detection at 2mm confirms its SMG status. \cite{Simpson2014} estimate a photometric redshift of $z=2.36^{+0.67}_{-0.22}$ for ALESS083.1 based on SED modeling of its optical/near-IR counterpart, consistent with the average redshift of the ALESS main sample.
$^c$ This source is spatially offset by $\sim3$~arcsec from ALESS114.1, and we confirm that it is undetected in the 870\mic\ observations despite being within the primary beam.
}
\end{deluxetable*}

\section{Testing the Rayleigh-Jeans approximation}
\label{appendix_rj}

In the Rayleigh-Jeans (RJ) regime, two flux densities at different frequencies should be perfectly correlated independently of the intrinsic dust temperature and redshift of the source, and indeed the ratio of two (sub-)millimeter fluxes is often used to estimate the dust emissivity index (e.g., \citealt{Aravena2016}). Here we demonstrate that the 870\mic-to-2mm flux density ratios measured for our sources cannot be used to obtain the dust emissivity index using the RJ approximation.

In the Rayleigh-Jeans regime, i.e., at sufficiently low frequencies, the Planck function can be approximated as:
\begin{equation}
B_\nu(T_\mathrm{dust}) \approx \frac{2 k_B \tdust}{c^2} \nu^2 \,,
\label{rjbb}
\end{equation}
where $k_B$ is the Boltzmann constant, and $c$ is the speed of light.
If we combine equations (\ref{mbb}), (\ref{kappa}), and (\ref{rjbb}) to obtain the dust emission in the RJ regime, the ratio of any two flux densities depends only on the ratio of their frequencies and on the RJ dust emissivity index $\beta_\mathrm{RJ}$:
\begin{equation}
S_{\nu_1}=S_{\nu_2} \left( \frac{\nu_1}{\nu_2}\right)^{2+\beta_\mathrm{RJ}}\,.
\label{beta_rj}
\end{equation}

\begin{figure*}
\begin{minipage}{\linewidth}
\centering
\includegraphics[angle=90,width=0.45\textwidth]{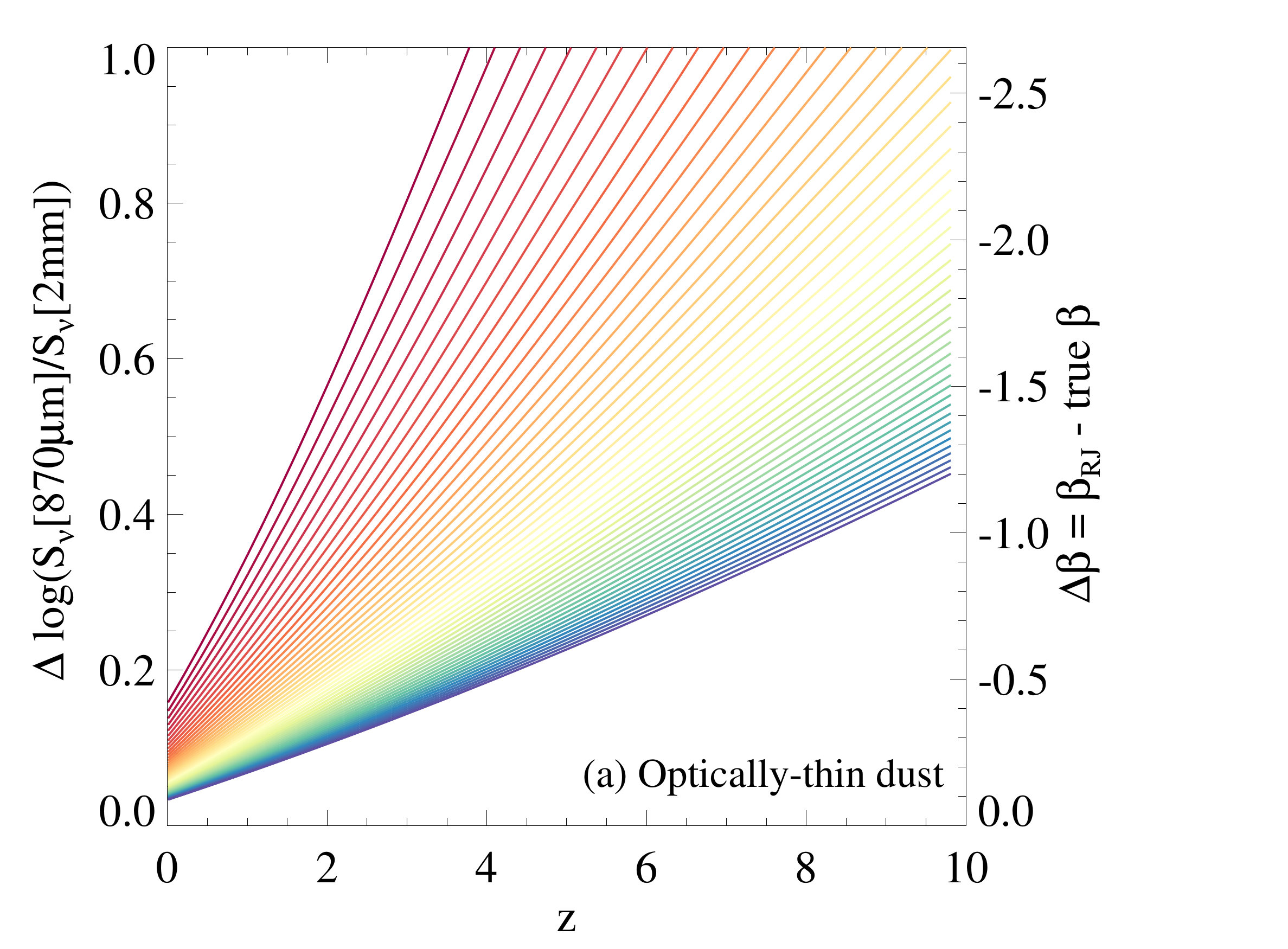}
\includegraphics[angle=90,width=0.45\textwidth]{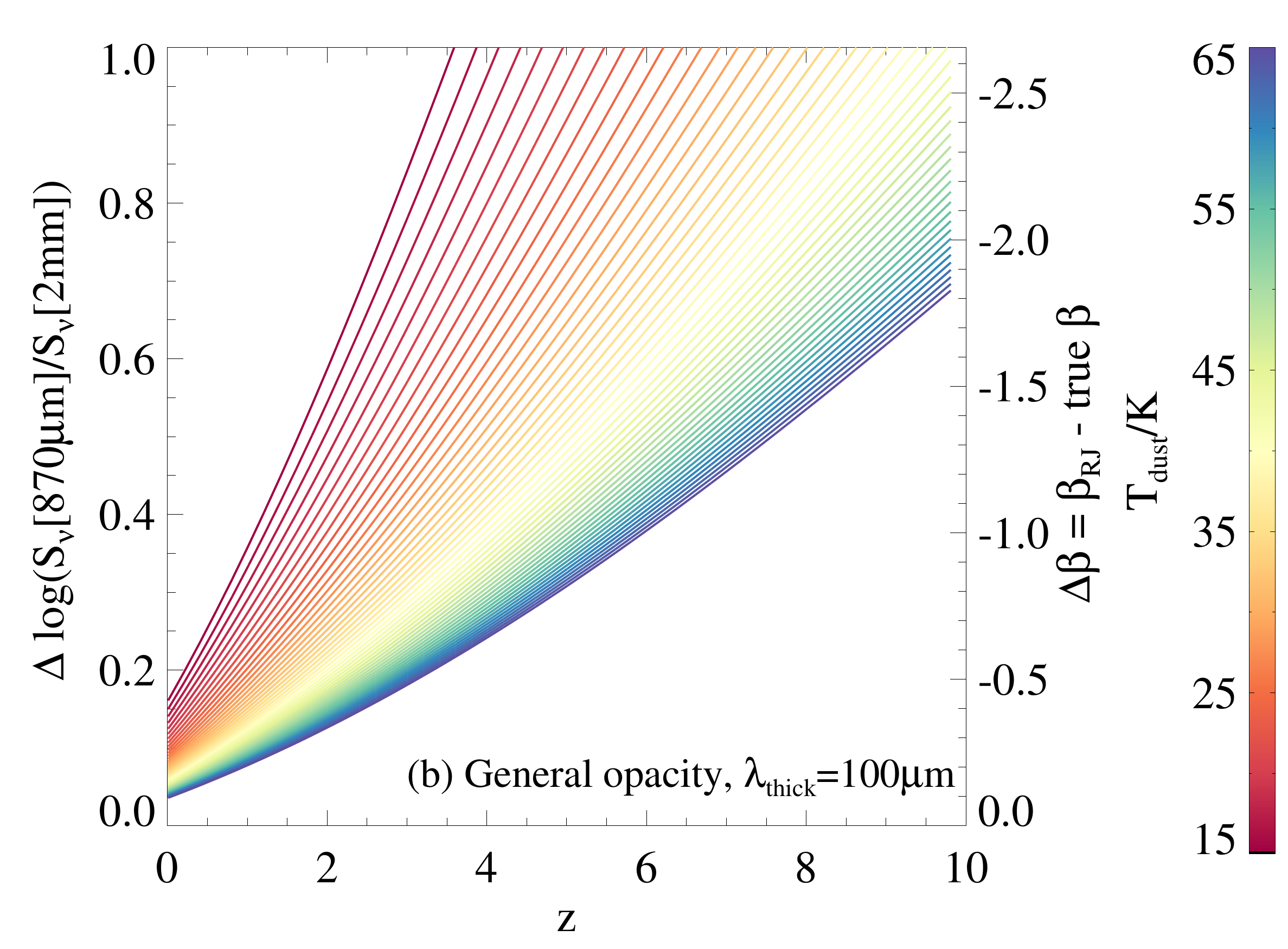}
\end{minipage}
 \caption{Deviation from the Rayleigh Jeans approximation as a function of redshift and temperature. The y-axis shows the difference between the ratio of Band 7 to Band 4 flux densities computed using the RJ approximation (eq.\,\ref{beta_rj}), and the true ratio, i.e., $\Delta\log(S_\nu[870\mic]/S_\nu[2\mathrm{mm}])=\log(S_\nu[870\mic]/S_\nu[2\mathrm{mm}])_\mathrm{RJ}-\log(S_\nu[870\mic]/S_\nu[2\mathrm{mm}])_\mathrm{true}$, computed using (a) the optically-thin approximation (Section~\ref{optically_thin}), and (b) the general opacity model, where we fix $\lthick=100\mic$ (Section~\ref{general_opacity}). This translates linearly to a difference $\Delta\beta$ between the inferred RJ emissivity index $\beta_\mathrm{RJ}$ and the true emissivity index (right-hand y-axes). } 
 \label{rj_test}
\end{figure*}

\begin{figure*}
\begin{minipage}{\linewidth}
\centering
\includegraphics[angle=90,trim={0cm 0cm 0cm 0cm},width=0.45\textwidth]{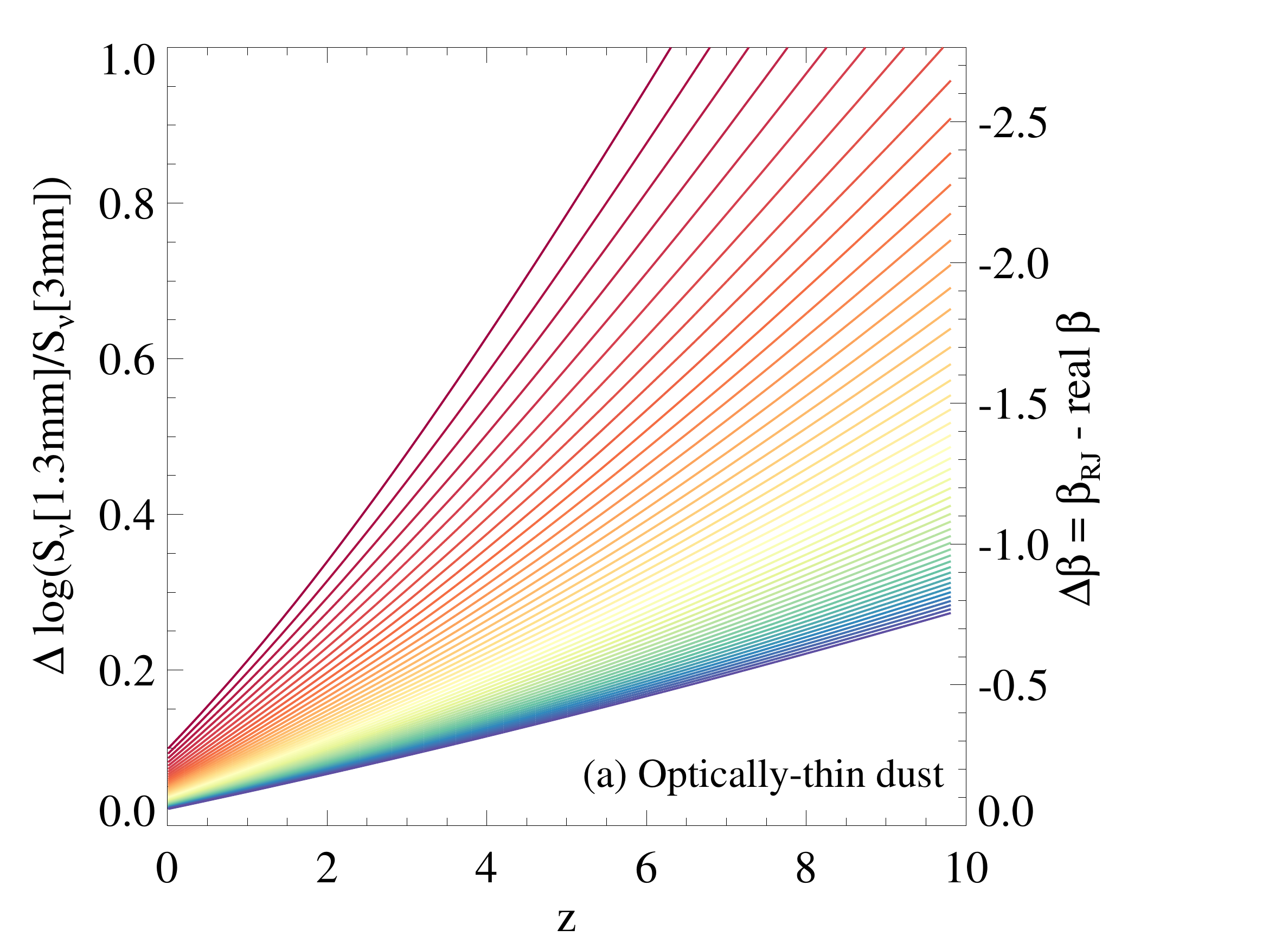}
\includegraphics[angle=90,trim={0cm 0cm 0cm 0cm},width=0.45\textwidth]{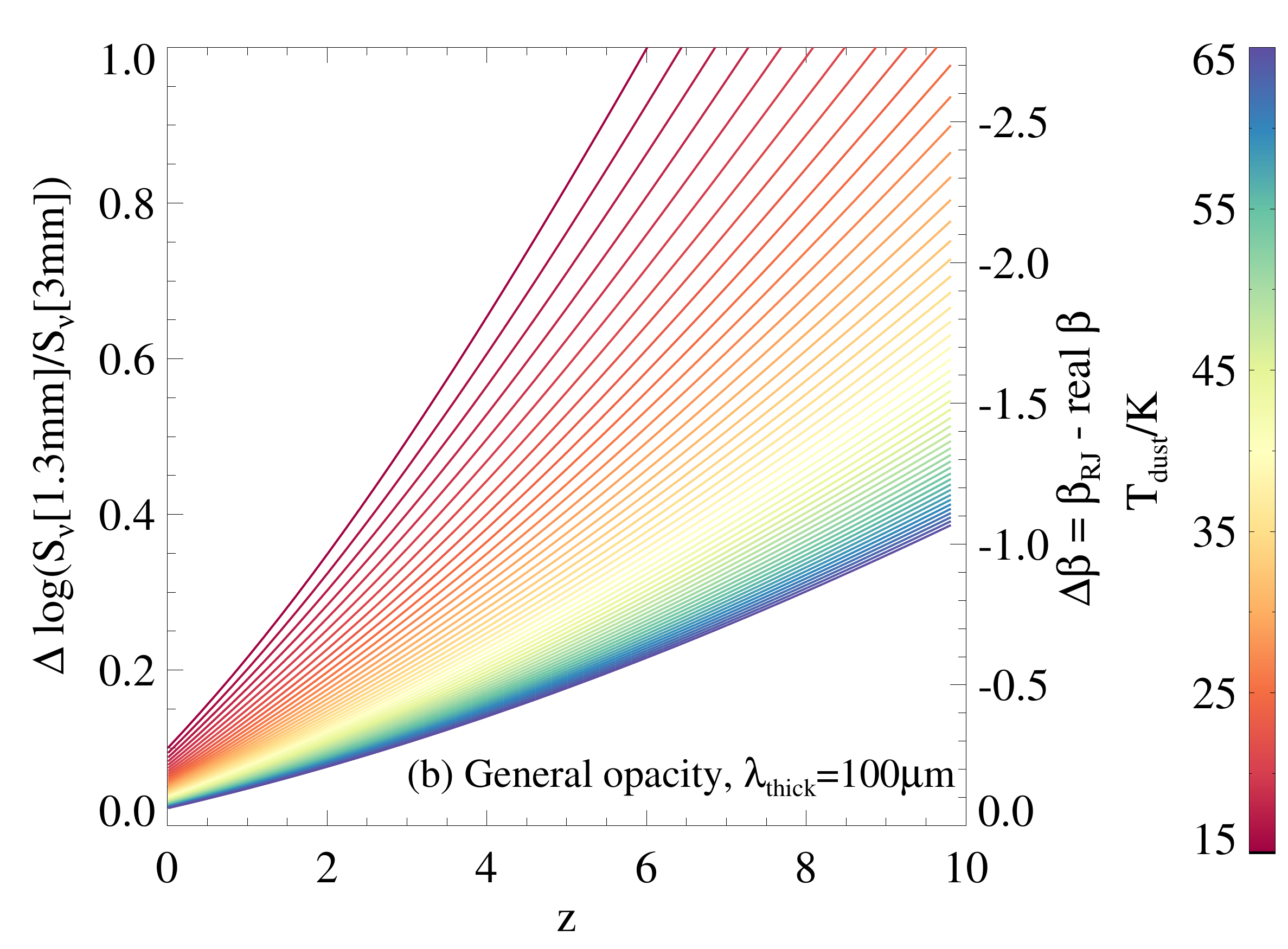}
\end{minipage}
 \caption{Same as Fig.~\ref{rj_test}, but using the ALMA Band 6 and Band 3 fluxes (at 1.3mm and 3mm).} 
 \label{rj_test_longwave}
\end{figure*}

In Fig.~\ref{b4_pred_b7}(b), we use eq.~(\ref{beta_rj}) with $\nu_1=345$~GHz (the frequency of our Band 7 observations) $\nu_2=145$~GHz (the frequency of our Band 4 observations) to plot the predicted relation between the 870\,\mic\ and 2\,mm flux densities in the Rayleigh-Jeans approximation assuming different values for $\beta_\mathrm{RJ}$. Our observed correlation between the fluxes has a similar slope and seems to be consistent with $\beta_\mathrm{RJ}\simeq1$. The dispersion could be explained using different values of $\beta_\mathrm{RJ}$, and could go as low as $\beta_\mathrm{RJ}=0.5$, as shown in the figure. At face value, it seems that the majority of our sources would have dust emissivity index values of less than 1.5, which is very low compared to the typically assumed values between 1.5 and 2.0. This could be either because the ALESS SMGs indeed have low dust emissivity indexes, or because the Rayleigh-Jeans approximation is not appropriate at our observed frequencies (for the redshifts and dust temperatures of our sample).

To test this, we generate both optically-thin and general opacity (with $\lthick=100\mic$, for reference) dust emission models (Section~\ref{modeling}) with fixed $\beta=2$ and \tdust\ between 15 and 80 K. Then, we place each model at different redshifts ($z=0-10$), and compute the ratio of their flux densities $S_\nu\mathrm{[870\mic]}/S_\nu\mathrm{[2mm]}$ at each redshift. In Fig.~\ref{rj_test}, we show how this ratio deviates from the ratio computed assuming the RJ approximation (eq.~\ref{beta_rj}) as a function of the model temperature \tdust\ and observed redshift. Not surprisingly, the deviation from RJ increases with redshift, and at fixed redshift it is larger for lower dust temperatures, because both higher redshift and cooler dust temperatures will shift the peak of the dust emission closer to the observed frequencies. The flux ratio overestimation in the RJ approximation leads directly to a systematic underestimation of $\beta$ in both dust modelling cases. Fig.~\ref{rj_test}(a) shows that even at $z=0$, Bands 7 and 4 sample the RJ regime (i.e., the deviation is close to zero) only for hot temperatures. At the typical redshift of our SMG sample ($z\simeq2.7$), the RJ approximation over-predicts the flux ratios by at least 0.15~dex, which translates to an underestimation of the emissivity index by at least $\Delta\beta=0.5$ (but the difference can be much larger for the cooler dust temperatures). The effect is similar but slightly more pronounced in the general opacity case.

We conclude that the Rayleigh-Jeans approximation cannot be used to constrain the dust emissivity indexes of our galaxies using our ALMA observations at 870\mic\ and 2mm. We also note that caution must be taken when adopting the Rayleigh Jeans approximation to derive the dust emissivity index of high-redshift galaxies observed even at lower frequencies with ALMA.
We demonstrate this by repeating the test above using the two frequencies used in the ASPECS survey \citep{Walter2016,Aravena2016,Gonzalez2019}: 100~GHz (3mm) and 230~GHz (1.3mm).
As expected, the deviation from the RJ regime is less significant in this case. However, adopting the RJ approximation can still lead to underestimating $\beta$ by at least $\Delta\beta=0.5$ at $z\simeq2.7$ for both the optically-thin and general opacity scenarios (see Fig.~\ref{rj_test_longwave}).

\section{Dust parameters derived for our well-sampled subset}
\label{app_parameters}

In Table~\ref{table_properties}, we list the median-likelihood estimates of the dust physical properties of our 27 sources with well-sampled SEDs and spectroscopic redshifts (and their respective confidence ranges) obtained using the isothermal, optically-thin dust assumption.

\begin{deluxetable*}{lcccc}
\tablecaption{Dust physical properties of the 27 galaxies in our well-sampled subset derived using the optically-thin dust approximation (Section~\ref{optically_thin}). For each property, we indicate our best estimate and confidence range, given by the median and 16th-to-84th percentile range of the posterior likelihood distribution, respectively. \label{table_properties} }
\tablehead{
\colhead{Source} & $\tdust\ / K$ & $\beta$ & $\log(\ldust/\lsun)$ & $\log(\mdust/\msun)$
}
\startdata
ALESS002.1	& $26^{+4}_{-3}$ & $2.2^{+0.4}_{-0.4}$ & $12.29^{+0.06}_{-0.06}$ & $8.70^{+0.08}_{-0.09}$ \\
ALESS003.1	 & $31^{+4}_{-3}$ & $1.9^{+0.2}_{-0.2}$ & $12.76^{+0.09}_{-0.09}$ & $8.99^{+0.05}_{-0.06}$ \\
ALESS005.1	 & $28^{+5}_{-3}$ & $2.2^{+0.3}_{-0.3}$ & $12.70^{+0.09}_{-0.09}$ & $8.84^{+0.06}_{-0.06}$ \\
ALESS006.1	  & $26^{+3}_{-2}$ & $1.8^{+0.2}_{-0.2}$ & $12.35^{+0.05}_{-0.05}$ & $9.10^{+0.04}_{-0.04}$ \\
ALESS007.1	 & $35^{+7}_{-5}$ & $2.1^{+0.3}_{-0.2}$ & $12.96^{+0.15}_{-0.12}$ & $8.63^{+0.06}_{-0.07}$ \\
ALESS009.1	 & $50^{+16}_{-12}$ & $1.9^{+0.3}_{-0.3}$ & $13.39^{+0.31}_{-0.28}$ & $8.46^{+0.12}_{-0.11}$ \\
ALESS011.1	  & $24^{+3}_{-2}$ & $2.3^{+0.2}_{-0.3}$ & $12.49^{+0.08}_{-0.09}$ & $9.00^{+0.08}_{-0.06}$ \\
ALESS017.1	& $21^{+2}_{-1}$ & $1.8^{+0.1}_{-0.2}$ & $12.10^{+0.04}_{-0.06}$ & $9.46^{+0.05}_{-0.03}$ \\
ALESS018.1	  & $39^{+8}_{-5}$ & $1.6^{+0.3}_{-0.2}$ & $12.80^{+0.09}_{-0.08}$ & $8.71^{+0.06}_{-0.07}$ \\
ALESS022.1	 &$29^{+5}_{-3}$ & $2.0^{+0.4}_{-0.3}$ & $12.52^{+0.06}_{-0.06}$ & $8.76^{+0.07}_{-0.07}$ \\
ALESS023.1	 & $28^{+4}_{-3}$ & $2.0^{+0.3}_{-0.2}$ & $12.54^{+0.11}_{-0.10}$ & $8.93^{+0.08}_{-0.07}$ \\
ALESS025.1	  & $26^{+4}_{-3}$ & $2.5^{+0.3}_{-0.3}$ & $12.61^{+0.10}_{-0.09}$ & $8.68^{+0.09}_{-0.08}$ \\
ALESS029.1	 & $21^{+3}_{-1}$ & $1.7^{+0.2}_{-0.2}$ & $11.97^{+0.05}_{-0.06}$ & $9.28^{+0.05}_{-0.04}$ \\
ALESS031.1	  & $32^{+4}_{-4}$ & $1.8^{+0.2}_{-0.2}$ & $12.71^{+0.08}_{-0.10}$ & $9.02^{+0.06}_{-0.06}$ \\
ALESS035.1	  & $53^{+15}_{-11}$ & $1.4^{+0.3}_{-0.2}$ & $13.08^{+0.22}_{-0.21}$ & $8.51^{+0.10}_{-0.08}$ \\
ALESS049.1	  & $27^{+5}_{-3}$ & $2.2^{+0.3}_{-0.1}$ & $12.49^{+0.13}_{-0.13}$ & $8.71^{+0.10}_{-0.10}$ \\
ALESS061.1	  & $39^{+24}_{-17}$ & $1.5^{+0.8}_{-0.4}$ & $12.57^{+0.52}_{-0.46}$ & $8.66^{+0.30}_{-0.24}$ \\
ALESS067.1	  & $30^{+5}_{-3}$ & $2.0^{+0.4}_{-0.3}$ & $12.57^{+0.07}_{-0.06}$ & $8.73^{+0.06}_{-0.06}$ \\
ALESS068.1	 & $33^{+12}_{-7}$ & $1.7^{+0.6}_{-0.5}$ & $12.33^{+0.16}_{-0.15}$ & $8.59^{+0.12}_{-0.14}$ \\
ALESS070.1	  & $31^{+4}_{-3}$ & $1.8^{+0.3}_{-0.2}$ & $12.56^{+0.06}_{-0.07}$ & $8.87^{+0.05}_{-0.05}$ \\
ALESS088.1	 & $17^{+3}_{-2}$ & $2.1^{+0.4}_{-0.3}$ & $11.67^{+0.05}_{-0.04}$ & $9.21^{+0.07}_{-0.09}$ \\
ALESS098.1	 & $22^{+3}_{-2}$ & $2.6^{+0.3}_{-0.3}$ & $12.39^{+0.06}_{-0.05}$ & $8.91^{+0.05}_{-0.08}$ \\
ALESS107.1	 & $53^{+16}_{-15}$ & $1.2^{+0.5}_{-0.2}$ & $12.54^{+0.28}_{-0.29}$ & $8.20^{+0.19}_{-0.19}$ \\
ALESS112.1	 & $22^{+2}_{-2}$ & $2.3^{+0.3}_{-0.2}$ & $12.38^{+0.04}_{-0.06}$ & $9.10^{+0.05}_{-0.05}$ \\
ALESS115.1	 & $33^{+5}_{-4}$ & $2.0^{+0.3}_{-0.2}$ & $12.82^{+0.11}_{-0.10}$ & $8.74^{+0.07}_{-0.07}$ \\
ALESS118.1	 & $35^{+8}_{-6}$ & $1.5^{+0.4}_{-0.3}$ & $12.35^{+0.09}_{-0.07}$ & $8.66^{+0.09}_{-0.09}$ \\
ALESS122.1	  & $59^{+9}_{-9}$ & $1.0^{+0.2}_{-0.1}$ & $12.99^{+0.14}_{-0.12}$ & $8.66^{+0.08}_{-0.08}$ \\
 \enddata
\end{deluxetable*}

\section{Further tests of the accuracy of our fitting method}
\label{app_accuracy}

\subsection{Effect of unavailable Band 4 or Herschel data}
\label{app_accuracy_data}

\begin{figure*}
\centering
\includegraphics[width=0.53\textwidth,angle=90]{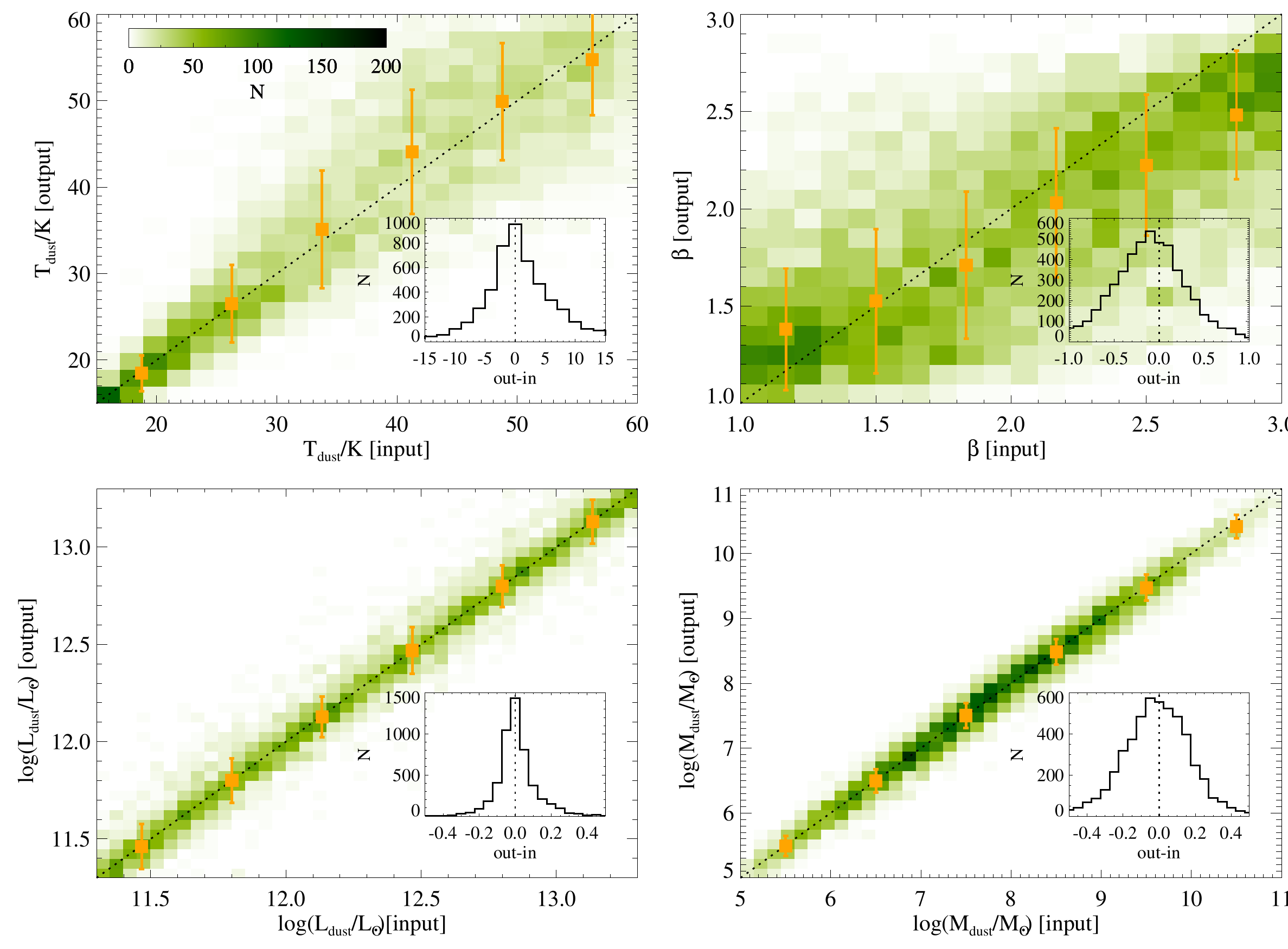}
\caption{Same as Fig.~\ref{fig:accuracy}, but excluding Band 4 from the fits. This shows that \ldust\ is still robust without low frequency data, because the fits include data sampling the peak of the SED. The main effect of not having Band 4 is that the accuracy in $\beta$ decreases significantly, which affects the accuracy in \tdust\ to some extent (because of the degeneracy between these two parameters), and most importantly affects \mdust\ estimates.}
\label{fig:accuracy_noBand4}
\end{figure*}

\begin{figure*}
\centering
\includegraphics[width=0.53\textwidth,angle=90]{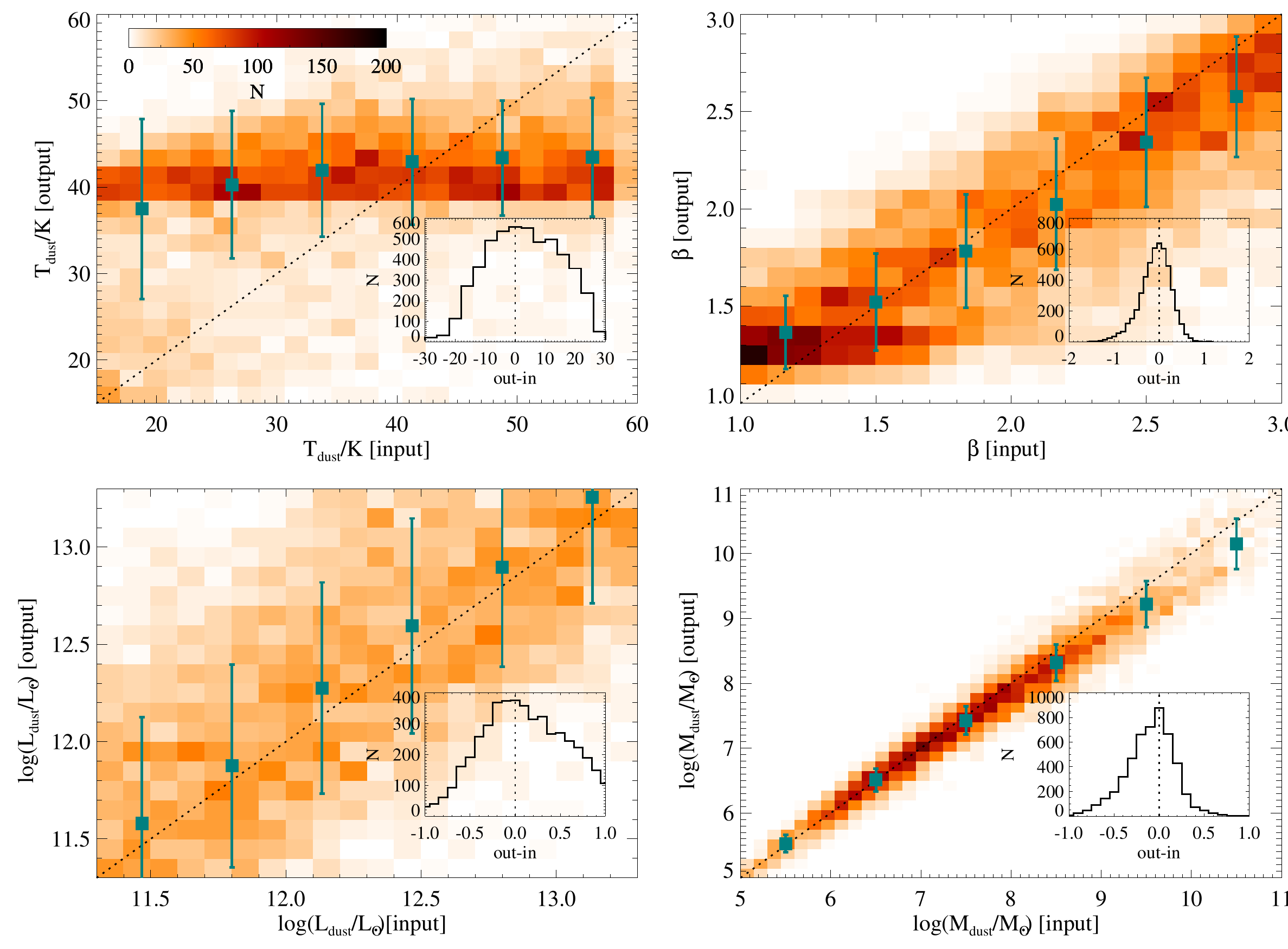}
\caption{Same as Fig.~\ref{fig:accuracy}, but excluding {\it Herschel} data from the fits (i.e, only fitting fluxes in ALMA Bands 7 and 4 here). This shows that we need to sample the peak of the SED to get reliable \tdust, hence also \ldust\ and \mdust. $\beta$ is relatively well constrained, but with larger errors.}
\label{fig:accuracy_noHerschel}
\end{figure*}

Here we use mock dust SED fits similar to the ones presented in Section~\ref{sec:accuracy} to test the effect of excluding fluxes in certain bands from our fits. Figs.~\ref{fig:accuracy_noBand4} and \ref{fig:accuracy_noHerschel} show cases where the SED is not as well-sampled as in the simulation shown in Fig.~\ref{fig:accuracy}, i.e. excluding the 2\,mm data and the {\it Herschel} data, respectively. Fig.~\ref{fig:accuracy_noBand4}  shows that not extending the observations into wavelengths longer than 870\mic\ impacts the estimates of $\beta$ significantly, however the accuracy of the remaining parameters is not significantly affected. That is, if no observations sampling the Rayleigh-Jeans tail of the dust emission (e.g., at 2mm) are available, we can still recover the temperature, and consequently the total luminosity and dust mass, albeit with lower average accuracy. On the other hand, if the peak of the SED is not well sampled, as shown in Fig.~\ref{fig:accuracy_noHerschel} where only the 870\mic\ and 2mm fluxes are included in the fits, \ldust\ and \tdust\ become very hard to constrain and inaccurate. Surprisingly, the dust masses are still reasonably accurate (within about 0.5 dex) even in this case. We attribute this to the fact that the parameter priors are realistic (at least in the simulation, since the mock SEDs parameters are drawn from the sample distribution as the priors used in the fitting). When applying this to real galaxies, the effects of lacking data could be much worse if the real distribution of parameters differs significantly from the priors.

\subsection{Effect of using optically-thin models to fit general opacity dust}
\label{app_accuracy_genopacity}

\begin{figure*}
\centering
\includegraphics[width=0.53\textwidth,angle=90]{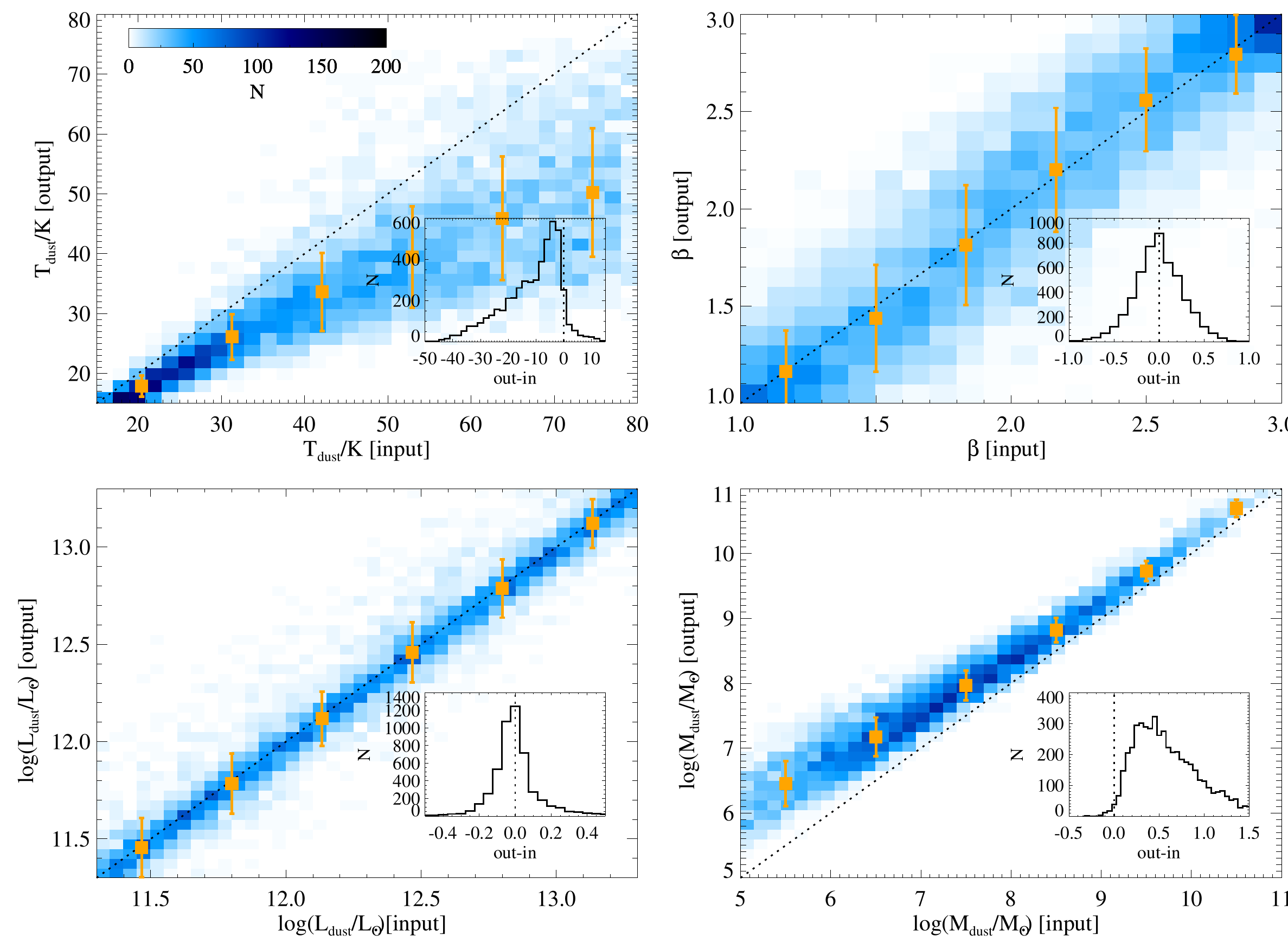}
\caption{Same as Fig.~\ref{fig:accuracy}, but using general opacity models to generate the mock observations (input), and then fitting those observations with optically-thin models (output). This shows that, for the range of general opacity models used (\lthick\ varying between 60 and 140\mic), the constraints on $\beta$ and \ldust\ are robust, however important systematic offsets are seen in \tdust\ and \mdust.}
\label{fig:accuracy_go}
\end{figure*}

Here we test the effect of using the wrong assumption regarding dust optical depth on the accuracy of our results. We generate a suite of dust emission models using our general opacity scenario (Section~\ref{general_opacity}), with a uniform prior on \lthick\ varying between 60\mic\ and 140\mic. Then, we use our Bayesian fitting routine to fit the mock SEDs produced by these models, but assuming only optically-thin dust. Fig.~\ref{fig:accuracy_go} shows the results of that exercise. We find that our constraints on $\beta$ and \ldust\ are robust against dust optical depth assumptions, however significant systematics may arise in \tdust\ and, consequently, the inferred dust masses. These systematics are larger when the input models have higher \lthick\ (i.e., they deviate more from the optically-thin assumption), and hotter dust temperatures, because both of these will affect the peak of the dust emission SED more significantly.

\def\aj{AJ}
\def\araa{ARA\&A}
\def\apj{ApJ}
\def\apjl{ApJ}
\def\apjs{ApJS}
\def\apss{Ap\&SS}
\def\aap{A\&A}
\def\aapr{A\&A~Rev.}
\def\aaps{A\&AS}
\def\mnras{MNRAS}
\def\pasp{PASP}
\def\pasj{PASJ}
\def\qjras{QJRAS}
\def\nat{Nature}

\def\aplett{Astrophys.~Lett.}
\def\aas{AAS}
\let\astap=\aap
\let\apjlett=\apjl
\let\apjsupp=\apjs
\let\applopt=\ao

\bibliographystyle{apj}
\bibliography{bib_dacunha}

\end{document}